\documentclass[twocolumn,epjc3]{svjour3}
\journalname{CERN-TH}
\usepackage[utf8]{inputenc} % allow utf-8 input
\usepackage[T1]{fontenc}    % use 8-bit T1 fonts
\usepackage{subcaption}
\usepackage{cite}
\usepackage{amsmath}
\usepackage{xspace}
\usepackage{enumitem}
\usepackage{float}
\usepackage{stfloats}
\usepackage[T1]{fontenc}
\usepackage{color}
\usepackage{graphicx}
\usepackage[utf8]{inputenc}
\usepackage[colorlinks = true,
            linkcolor = blue,
            urlcolor  = blue,
            citecolor = blue,
            anchorcolor = blue]{hyperref}
\usepackage{appendix}
\usepackage{multirow}
\usepackage{multicol}
\usepackage[export]{adjustbox}

\newcommand{\NC}{N_{\mathrm{c}}}

\newcommand{\herwig}{\textsf{Herwig}\xspace}
\newcommand{\herwigv}[1]{\textsf{Herwig~#1}}
\newcommand{\pythia}{\textsf{Pythia}\xspace}
\newcommand{\pythiav}[1]{\pythia~\textsf{#1}}
\newcommand{\sherpa}{\textsf{Sherpa}\xspace}

\newcommand{\instruction}[1]{}

\newcommand{\ThePEG}{\textsf{ThePEG}\xspace}
\newcommand{\ThePI}{\textsf{TheP8I}\xspace}

\newcommand{\rivet}{\textsf{Rivet}\xspace}
\newcommand{\professor}{\textsf{Professor}\xspace}
\newcommand{\autotunes}{\textsf{Autotunes}\xspace}
\newcommand{\eetune}{\textsc{LH Tune}\xspace}
\newcommand{\empt}[1]{}

\newcommand{\UE}{Underlying Event\xspace}
\newcommand{\MB}{Minimum Bias\xspace}
\newcommand{\ISR}{Initial-State Radiation\xspace}
\newcommand{\FSR}{Final-State Radiation\xspace}
\newcommand{\AOPS}{Angular Ordered Parton Shower\xspace}

\newcommand{\energymomentum}{energy--momentum\xspace}
\graphicspath{ {./figures/} }

\begin{document}
\begin{sloppypar}

\title{Herwig 7 with the Lund String Model:\\ Tuning and Comparative Hadronization Studies}

\author{Michaela Divisova\thanksref{e1,addr1}
        \and
        Miroslav Myska\thanksref{e2,addr1}
        \and
        Pratixan Sarmah\thanksref{e3,addr2}
        \and
        Andrzej~Si\'{o}dmok\thanksref{e4,addr2, addr3}
}
\institute{Czech Technical University in Prague, Brehova 7, 115 19, Prague, Czech Republic
        \label{addr1}
        \and
        Jagiellonian University, 
        ul. prof. Stanislawa \L{}ojasiewicza 11, 30-348 Krak\'{o}w, Poland
        \label{addr2}
        \and
        Theoretical Physics Department, CERN, 1211 Geneva 23, Switzerland
        \label{addr3}
}
\thankstext{e1}{e-mail: divismi5@fjfi.cvut.cz }
\thankstext{e2}{e-mail: miroslav.myska@fjfi.cvut.cz }
\thankstext{e3}{e-mail: pratixan.sarmah@doctoral.uj.edu.pl}
\thankstext{e4}{e-mail: andrzej.siodmok@uj.edu.pl}

\maketitle
    
\begin{abstract}
The modelling of the formation of colour-singlet hadrons from coloured partons, known as Hadronization, is crucial for generating realistic events in Monte Carlo Event Generators. Due to limited understanding of the non-perturbative regime, physically motivated phenomenological hadronization models with tunable parameters are used and later tuned to the experimental data. 
Modern Monte Carlo generators primarily employ one of two hadronization models: the Lund string model, which is the default in \pythia, and the cluster model, which is the default in \herwig{} and \sherpa{}.  
In this work, we combine the Lund string hadronization model, as implemented in \pythiav{8}, with \herwigv{7} using \ThePI interface. We tune the string model with \herwigv{7}'s Angular Ordered Parton Shower (AOPS) to lepton and hadron collision data, resulting in the Les Houches Tune (\eetune), which shows good performance across a wide range of observables. The \eetune will be included in the \herwigv{7.4} release\footnote{It can also be used with \herwigv{7.3}, provided that the relevant patch is applied to the \texttt{herwig-bootstrap} installation script which can be provided by the authors.}. 
This development enables a direct comparative study of the two hadronization models within \herwig{}, both interfaced with the \AOPS, which serves as the main motivation behind this work.
\end{abstract}

\section{Introduction}
General-purpose Monte Carlo generators (GPMC), \herwig~\cite{Bellm:2015jjp,Bellm:2019zci,Bewick:2023tfi}, \pythia~\cite{Sjostrand:2014zea,Bierlich:2022pfr} and \sherpa~\cite{Gleisberg:2008ta,Sherpa:2019gpd, Sherpa:2024mfk}, play a key role in high-energy physics. 
These tools provide fully exclusive simulations of particle collisions, allowing direct comparisons with experimental data even when complicated experimental cuts have been applied.  Consequently, GPMC generators have been an integral part of many measurements and discoveries in modern high-energy physics. Furthermore, they are widely used by both theorists and experimentalists to simulate the physics of future experiments and guide the development of their operational requirements. Therefore, it is important to increase their precision as well as to improve the uncertainty estimates associated with these complex simulations.

Over the past few years, significant progress has been made in improving the perturbative aspects of GPMCs. 
In particular, steps forward have been made in developing new parton showers~\cite{Hoche:2024dee, FerrarioRavasio:2023kyg, Schumann:2007mg, Nagy:2007ty, Nagy:2015hwa, Bewick:2021nhc, Lee:2023hef,Fischer:2016vfv,Hoang:2018zrp,Holguin:2020joq, Frixione:2023ssx, Forshaw:2025bmo} and improving their hard (perturbative) limit by matching them to next-to-leading order (NLO) matrix elements
~\cite{Frixione:2002ik, Nason:2004rx,Frixione:2007vw,Alioli:2010xd, Alioli:2008gx,Alioli:2008tz,Alioli:2009je,Hamilton:2010mb,Platzer:2011bc,Hoeche:2011fd,Nason:2012pr,Frederix:2012dh,Heinrich:2017kxx,Jones:2017giv,Cormier:2018tog,ATL-PHYS-PUB-2023-029, Jadach:2011cr,Jadach:2015mza,Jadach:2016qti, Sarmah:2024hdk, Frixione:2023hwz,Giele:2007di, Chen:2021phj, Bellm:2017ktr}. 
The automatised matching advancements are often called the ``NLO revolution'' and show that GPMCs have entered the precision era. This progress also triggered work on the estimation of perturbative uncertainties of GPMC predictions. As a result, significant activities have been seen in this field within all GPMC projects\cite{Bothmann:2016nao,Mrenna:2016sih,Bellm:2016voq,Bellm:2016rhh}. 

Increased control of perturbative corrections has led to a situation where the precision of LHC measurements is increasingly limited by the non-perturbative components of GPMCs, such as hadronization. 
The modelling of non-perturbative quantum chromodynamics (QCD) has already become one of the dominant sources of systematic uncertainty in precision measurements of important Standard Model parameters such as the top-mass~\cite{Argyropoulos:2014zoa} and the $W$-mass measurement using a new method proposed in~\cite{Freytsis:2018cev} (where hadronization effects are as big as $\sim 10\%$).
Moreover, hadronization remains the main bottleneck for thrust measurement in Higgs decays~\cite{Hou:2016sho} or the extraction of the strong coupling~\cite{Johnson:2017ttl}, where the authors write in the summary that:
``a better understanding of non-perturbative effects will improve the accuracy of the extraction significantly''.
Therefore, work on improving the modelling of hadronization phenomena has recently 
intensified~\cite{Fischer:2016zzs,Ferreres-Sole:2018vgo,Gieseke:2018gff,Platzer:2022jny, Bierlich:2020naj, Bierlich:2022vdf, Chakraborty:2022nui,Bierlich:2023okq, Bierlich:2014xba,Kerbizi:2023cde,Hoang:2024nqi, Masouminia:2023zhb,Chahal:2022rid,Gieseke:2025mcy} 
including novel ideas of using Machine Learning techniques~\cite{Ghosh:2022zdz,Ilten:2022jfm,Chan:2023ume,Chan:2023icm,Bierlich:2023fmh,Heller:2024onk,Bierlich:2024xzg,Bierlich:2023zzd, Assi:2025avy}. 
Nevertheless, there are still just two physically motivated hadronization models employed in Monte Carlo generators: the Lund string model~\cite{Andersson:1983ia}, used in \pythia{}, and the cluster model~\cite{Webber:1983if}, employed by default in \herwig{} and \sherpa.

Historically, \pythia{} has put a strong emphasis on soft physics, such as hadronization (i.e. it was created as a result of research in Lund on the structure of the fragmentation of a single quark into a stream of hadrons~\cite{Sjostrand:2019zhc}), \MB (MB) physics and the soft \UE (UE). 
On the other hand, \herwig{} has historically put its emphasis on the perturbative description of an event and has
originated in the coherence studies leading to the \AOPS~\cite{Marchesini:1987cf,Catani:1990rr}.
While all modern GPMCs have well-developed simulation components for both perturbative and non-perturbative physics, it would be beneficial to be able to independently combine different components from different generators to examine their effects.
The use of different ``mixtures'' of models to describe perturbative and non-perturbative quantum chromodynamics will not only potentially improve the description of the data but will also allow us to study the uncertainties associated with these effects.

Thus, in this article, we present a combination of the different components of the \herwig and \pythia{} event generators.
More specifically, we provide a full simulation of \herwigv{7} with a recent version of AOPS~\cite{Bewick:2021nhc} and a modern version of the Lund string model from \pythiav{8}, which is also interfaced with the colour reconnection (CR) model introduced in~\cite{Christiansen:2015yqa}.
\footnote{A combination of the AOPS with the Lund sting model was already attempted, see~\cite{Bellm:2019owc, LaCagnina:2023yvi}. However, the authors in their publications focused mainly on the development of a new method of tuning rather than obtaining an optimal result. For this reason, they focused on LEP data, and their results cannot be used in an optimal way for LHC physics. For example, they did not interface the string model with the colour reconnection model, which is an important component needed for the description of hadron collisions. They also used an older version of AOPS before major modifications reported in~\cite{Bewick:2021nhc}.}
We show that competitive predictions for both lepton-lepton and hadron-hadron collisions can be achieved with this combination.

This also allows us to study the uncertainties related to non-perturbative hadronization effects by fixing all the other components of the simulation in \herwig{} and switching the hadronization models.
Similarly, we can compare the results of \herwig{} and \pythia{} using the same hadronization model (i.e. the string model) and see how the other physics components within the generators change the simulation results.

This paper is structured as follows: in Section~\ref{hadronizationModels} we briefly summarise the different hadronization models; in Section~\ref{StringQCD} we present the technical details of the combination of the string hadronization and colour reconnection models with AOPS
in \herwig{} and the tuning procedure; in Section~\ref{results} we show the comparisons between \pythia{} and \herwig{} with the cluster and string hadronization models, and finally we summarise the study in Section~\ref{summary}.

%%%%%%%%%%%%%%%%%%%%%%%%%%%%%%%%%%%%%%%%%%%%%%%%%%%%%%%%%%%%%%%%%%%%%%%%%%%%%%%%%%%%%%%%%%%%%%%%%%%%%%%%%%%

\section{Hadronization models in Monte Carlo Event Generators}
\label{hadronizationModels}
Modern Monte Carlo event generators primarily employ one of two hadronization models: the Lund string model~\cite{Andersson:1983ia} which is the default in \pythia, and the cluster model~\cite{Webber:1983if} which is the default in \herwig{} and \sherpa{}.

The Lund string model is based on insights from lattice QCD simulations. The model represents the interaction between colour-charged quarks as a colour flux tube, behaving like a relativistic string. Lattice QCD calculations support the idea that the potential energy between a quark and an antiquark increases linearly with their separation $r$, following the relation
\begin{align}
\label{eq:V_kappa}
\hspace{1in} V(r) = \kappa r,
\end{align}
where $\kappa$ is the string tension, approximately $1$ GeV/fm. 
This linear confinement prevents quarks from existing as free particles. The string tension in the Lund string model is implicitly related to the fragmentation parameters but is not explicitly tunable as a single parameter.
As quarks move apart in high-energy collisions, the string stretches and accumulates energy until it reaches a critical threshold, making it energetically favourable to create new quark-antiquark pairs (or diquark pairs) from the vacuum.
This process of string breaking leads to the formation of hadrons, which inherit fractions of the initial energy and momentum from the string.
A key feature of the model is the Lund fragmentation function, which describes how energy and momentum are distributed among the produced hadrons. The probability of a hadron carrying a fraction $z$ of the remaining string energy, including the Bowler modification~\cite{Bowler:1981sb} follows
\begin{align}
    \hspace{0.5in} f(z) \propto \frac{1}{z^{1+r_Qbm^2_Q}}\left({1-z}\right)^{a}
 e^{-b m_\mathrm{T}^2 / z},
  \label{eq:fragz}
\end{align}
where $Q$ represents the heavy quark flavour, $m_Q$ is the quark mass, $r_Q$ is a tunable parameter that can be adjusted separately for bottom and charm quarks\footnote{The Bowler modification $z^{-r_Qbm^2_Q}$ in Eq.~\ref{eq:fragz} is only applied for heavy quarks, i.e. charm and bottom using corresponding parameters \texttt{rFactC} and \texttt{rFactB} in \pythia,  for light quarks $r_Q=0$.},
$a$ and $b$ are tunable parameters, corresponding to \texttt{aLund}\footnote{There is also a parameter \texttt{aExtraSQuark} which allows to increase  $a$ for $s$ quarks, with total $a~=~$\texttt{aLund}~+~\texttt{aExtraSQuark} and similar parameter  \texttt{aExtraDiquark} which does the same for diquarks.} 
and \texttt{bLund} in \pythia, respectively. The ``transverse mass'' $m_\mathrm{T}$ of the hadron is defined as $m_\mathrm{T}^2=m^2+p_\mathrm{T}^2$. 
During the fragmentation process, hadrons get Gaussian $p_\mathrm{T}$-kicks, modelled with the width parameter $\sigma$, and acquire a transverse momentum.
Additionally, gluons are modelled as kinks on the string, effectively splitting it into multiple segments, each of which undergoes independent fragmentation. 

On the other hand, the cluster model is based on `t Hooft's planar diagram theory \cite{tHooft:1973alw}: the dominant colour structure of QCD diagrams in the perturbation expansion in $1/\NC$ can be represented in a planar form using colour lines, which is commonly known as the $\NC \to \infty$ limit. 
The resulting colour topology in Monte Carlo events with partons in the final-state features open colour lines following the parton shower evolution. 
After the remaining gluons decay non-perturbatively into light quark–antiquark pairs, the resulting event consists of colour-connected partons in colour triplet or antitriplet states. These partons combine into colour-singlet groups called clusters.  
The phenomenon of the partons generated in parton showers to be arranged in colour-singlet clusters (pre-hadrons) within a limited range in both coordinate and momentum space is known as colour preconfinement~\cite{Amati:1979fg}.
The clusters then decay into hadrons, or, if they are too massive, into lighter clusters first, and then into hadrons.
As the cluster model has already been tuned and is used as the default within \herwig, we do not describe its parameters here and refer the interested reader to the \herwig{} manual~\cite{Bahr:2008pv}.

Both models have their advantages and disadvantages. The string model, for example, offers a very predictive framework for how its space--time motion and breakup translates into an \energymomentum{} distribution of the primary hadrons. 
The main weakness of the string model is that there are many parameters related to flavour properties. 
The cluster model, to some extent, has opposite features, i.e., a simpler \energymomentum{} picture but a better and more predictive flavour composition.
Therefore, a comparative study between these two different hadronization models, both with the \AOPS in \herwig{} serves as the motivation behind this study\footnote{It is worth adding that an interface to a machine learning hadronization model MLHAD~\cite{Ghosh:2022zdz} has also been tested within \herwigv{7}. However, it is too early for this model to be used in practice, so we are not considering it in this study.}.

%%%%%%%%%%%%%%%%%%%%%%%%%%%%%%%%%%%%%%%%%%%%%%%%%%%%%%%%%%%%%%%%%%%%%%%%%%%%%%%%%%%%%%%%%%%%%%%%%%%%%%%%%%%

\section{The String Hadronization and Colour Reconnection models in \herwigv{7}}
\label{StringQCD}
The string hadronization and colour reconnection models in \pythia{} can be interfaced with \herwigv{7} by exploiting the \ThePEG (Toolkit for High Energy Physics Event Generation)~\cite{Lonnblad:2006pt,Lonnblad:2009zz} framework.
\ThePEG is a general framework for implementing physics models for event generation in high-energy particle collisions. 
Its modular structure allows users to define different physics models independently and combine them to create customised event generation setups.
However, \ThePEG{} was only used as a backbone in one of the three existing Monte Carlo generators, \herwig{}. This meant that it was not often possible to exploit its full potential and use components from different Monte Carlo generators. 
Recently, it was shown, see~\cite{Bellm:2019owc}, that with some modifications to an interface between \ThePEG{}  and \pythiav{8} (called \ThePI~\cite{TheP8I}) one can achieve a working environment of \herwigv{7} with the Lund string model.
We have extended \ThePI{} to further allow the use of the so-called QCD-based colour reconnection model within \pythiav{8}~\cite{Christiansen:2015yqa}. These modifications are described in more detail in~\ref{appendix_thep8i} and will be made available as a patch in the \texttt{herwig-bootstrap} installation script of the \herwigv{7.4} release.

\subsection{Tuning}
With a combined working setup of \herwigv{7} using Leading Order (LO) matrix element, \AOPS and the string hadronization model, we fit selected free parameters of the models against a wide range of experimental data from LEP, the Tevatron, and the LHC to provide a general-purpose tune for collider physics.
To this end, we segregate the parameters into four subgroups and employ a systematic four-stage parameter optimisation approach, where we tune:
\begin{itemize}
    \item the string fragmentation parameters along with the \FSR (FSR) parameters subject to LEP measurements, as listed in Table~\ref{tab:frag_params}, in the first stage,
    \item the flavour composition governed by the flavour parameters, as listed in Table~\ref{tab:flavor_params}, subject to particle multiplicity data as shown in Fig.~\ref{fig:multiplicity_distributions} in the second stage,
    \item the \ISR (ISR) strong coupling $\alpha_{\mathrm S}^{\mathrm{ISR}}$ and intrinsic  $k_{\mathrm{T}}$, as listed in Table~\ref{tab:Z_tune}, subject to Drell-Yan measurements from ATLAS experiment shown in Fig.~\ref{fig:Z_distributions} in the third stage,
    \item and the Multiple Parton Interactions (MPI) and colour reconnection (CR) parameters subject to \MB and \UE observables from hadron collision data, as listed in Table~\ref{tab:UE_params}, in the fourth stage.
\end{itemize}
This is achieved using the well-established tools -- \rivet \cite{Buckley_Rivet:2010ar, 10.21468/SciPostPhys.8.2.026, Bierlich:2024vqo} and \professor \cite{Buckley:2009bj}. 
\rivet serves as a preservation tool that contains multiple sets of analyses from high-energy collider experiments over the years, along with measurement logics, which can be used for direct comparisons of our Monte Carlo predictions.
\professor is used to numerically optimise the parameter values of the model to obtain a better description of the experimental data in a two-step procedure --
at first, \professor makes a non-exhaustive scan to sample from the high-dimensional parameter space and generates an interpolating polynomial for each observable bin, called the response function, as a function of the parameters.
This is followed by a global optimisation of the response function yielding optimal parameter values. 

In the optimisation step, \professor{} performs a minimisation of the weighted Goodness-of-Fit function (GoF), a $\chi^2$ function in our case. The GoF measure as a function of the parameter vector $\boldsymbol{p}$ is defined as
\begin{align}
\label{eqn:professor_gof_eqn}
\hspace{0.5in}\chi^2(\boldsymbol{p}) = \sum_{\mathcal{O}} w_{\mathcal{O}} \sum_{b \in \mathcal{O}} \frac{\left(f^{(b)}(\boldsymbol{p})-\mathcal{R}_b\right)^2}{\Delta_b^2},
\end{align}
where $w_{\mathcal{O}}$ is the weight for each observable $\mathcal{O}$ adjusted to the tune, $f^{(b)}(\boldsymbol{p})$ represents the generator response function for each bin, $\mathcal{R}_b$ denotes the reference value for bin $b$ and $\Delta_b$ is the total uncertainty for the reference value in bin $b$. Other GoF functions can also be used in the newer versions of \professor.
As evident from Eq. \ref{eqn:professor_gof_eqn}, \professor offers the flexibility to assign arbitrary weights to each distribution or bin in the GoF function, allowing users to emphasise the observables that are physically most relevant. 
The weights used for tuning in this publication can be found in~\ref{appendix_tune}.

\subsubsection{Lepton collisions}
\label{sec:lepton_collisions_3_1_1}
We started tuning the string model with \herwigv{7} using $e^+e^-$ collision data at $91.2$ GeV and measured at the LEP collider. These events provide a clean final-state environment sensitive only to the final-state parton shower and hadronization model. 
Similarly to the \pythia{} tune~\cite{Buckley:2009bj}, we simplify interpolation and tuning by splitting the set of 18 primary parameters of the string model into two distinct subgroups. 
We adopt the same assumption as other \pythia{} tunes~\cite{Buckley:2009bj,Skands:2010ak,Skands:2014pea}, that these sets of parameters decouple and can be tuned separately in smaller groups without significantly affecting the respective tunes. We expect this assumption to hold for our setup as well without introducing significant correlations between the subset of fragmentation and flavour parameters.
 
In the first stage of tuning, seven string fragmentation parameters, see Table~\ref{tab:frag_params}, are tuned along with two \FSR parameters.
\herwigv{7.3} offers the separation of the effective values of the strong coupling for \ISR and \FSR~\cite{Bewick:2023tfi}, allowing us to focus only on $\alpha_{\mathrm{S}}^{\mathrm{FSR}}(M_Z)$ at this stage. 
We also consider the infrared shower cutoff $p_{\perp}^{\mathrm{min}}$ for the tuning keeping in mind its strong anti-correlation with the strong coupling. 
In the string hadronization model, we consider the parameters described in Section~\ref{hadronizationModels} during the tuning, which are -- the Gaussian width, $\sigma$, of the transverse momenta of the broken string components, the longitudinal momenta given by the Lund fragmentation function and its two parameters, $a$ and $b$. We also include four flavour-related parameters (\texttt{aExtraSQuark}, \texttt{aExtraDiquark}, \texttt{rFactC}, and \texttt{rFactB}) to help reproduce the flavour-sensitive observables.
The observables used in this stage of the tuning were event shapes, momentum spectra, and the mean charged multiplicities measured by the DELPHI Collaboration~\cite{DELPHI:1996sen} and the flavour-specific mean charged multiplicities measured by the OPAL Collaboration~\cite{OPAL:1998arz}. These sets of experimental observables along with their assigned weights are listed in Table~\ref{tab:LEPtuneweights1} of~\ref{appendix_tune}.
\begin{table}[]
    \centering
    \begin{tabular}{lcc}
        Parameter & Range & \eetune\\
        \hline
        \texttt{$\alpha_{\mathrm{S}}^{\mathrm{FSR}}$}  & 0.100 - 0.134 & 0.126 \\
        \texttt{$p_{\perp}^{\mathrm{min}}$} [GeV]    & 0.40 - 1.50     & 1.03 \\
        \texttt{$a$}            & 0.00 - 2.00     & 0.75 \\
        $b$ [GeV$^{-2}$]          & 0.20 - 2.00     & 0.90 \\
        $\sigma$ [GeV]       & 0.00 - 1.00     & 0.31 \\
        \texttt{aExtraSQuark}    & 0.00 - 2.00     & 0.18 \\
        \texttt{aExtraDiquark}   & 0.00 - 2.00     & 0.05 \\
        \texttt{rFactC}          & 0.00 - 2.00     & 0.68 \\
        \texttt{rFactB}          & 0.00 - 2.00     & 1.27 \\
    \end{tabular}
    \caption{Final-State AOPS and Lund string fragmentation parameters tuned in the first stage of tuning.}
    \label{tab:frag_params}
\end{table}
\begin{table}[]
    \centering
    \begin{tabular}{lcc}
        Parameter       & Range & \eetune\\
        \hline
        \texttt{probStoUD}       & 0.10 - 0.30     & 0.19 \\
        \texttt{probQQtoQ}       & 0.00 - 0.20     & 0.08 \\
        \texttt{probSQtoQQ}      & 0.80 - 1.00     & 0.99 \\
        \texttt{probQQ1toQQ0}     & 0.00 - 0.10     & 0.02 \\
        \texttt{etaSup}           & 0.20 - 0.70     & 0.51 \\
        \texttt{etaPrimeSup}      & 0.00 - 0.30     & 0.18 \\
        \texttt{popcornRate}      & 0.00 - 1.00     & 0.73 \\
        \texttt{mesonUDvector}    & 0.20 - 0.60     & 0.33 \\
        \texttt{mesonSvector}     & 0.45 - 1.00    & 0.68 \\
        \texttt{mesonCvector}     & 0.00 - 2.00     & 1.07 \\
        \texttt{mesonBvector}     & 1.40 - 3.00     & 1.85 \\
    \end{tabular}
    \caption{Lund string flavour parameters tuned in the second stage of tuning.}
    \label{tab:flavor_params}
\end{table}

\begin{figure}[htp]
\centering
    \begin{subfigure}{0.45\textwidth}
        \includegraphics[trim=0 340 0 0, clip,width=1.0\textwidth]{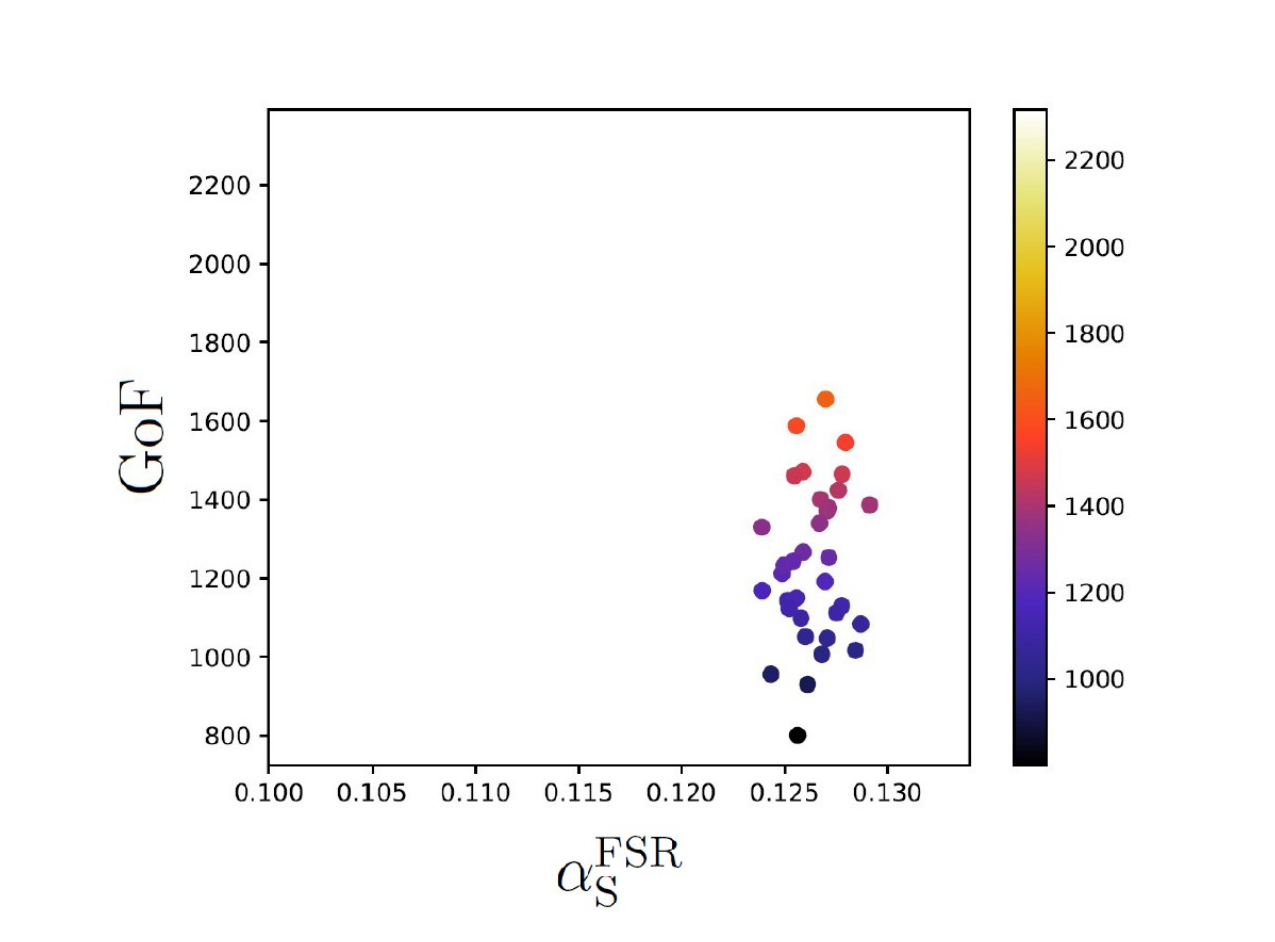}
        \caption{}
        \label{fig:chi2_LEP:a}
    \end{subfigure}\\
    \begin{subfigure}{0.45\textwidth}
        \includegraphics[trim=0 340 0 0, clip,width=1.0\textwidth]{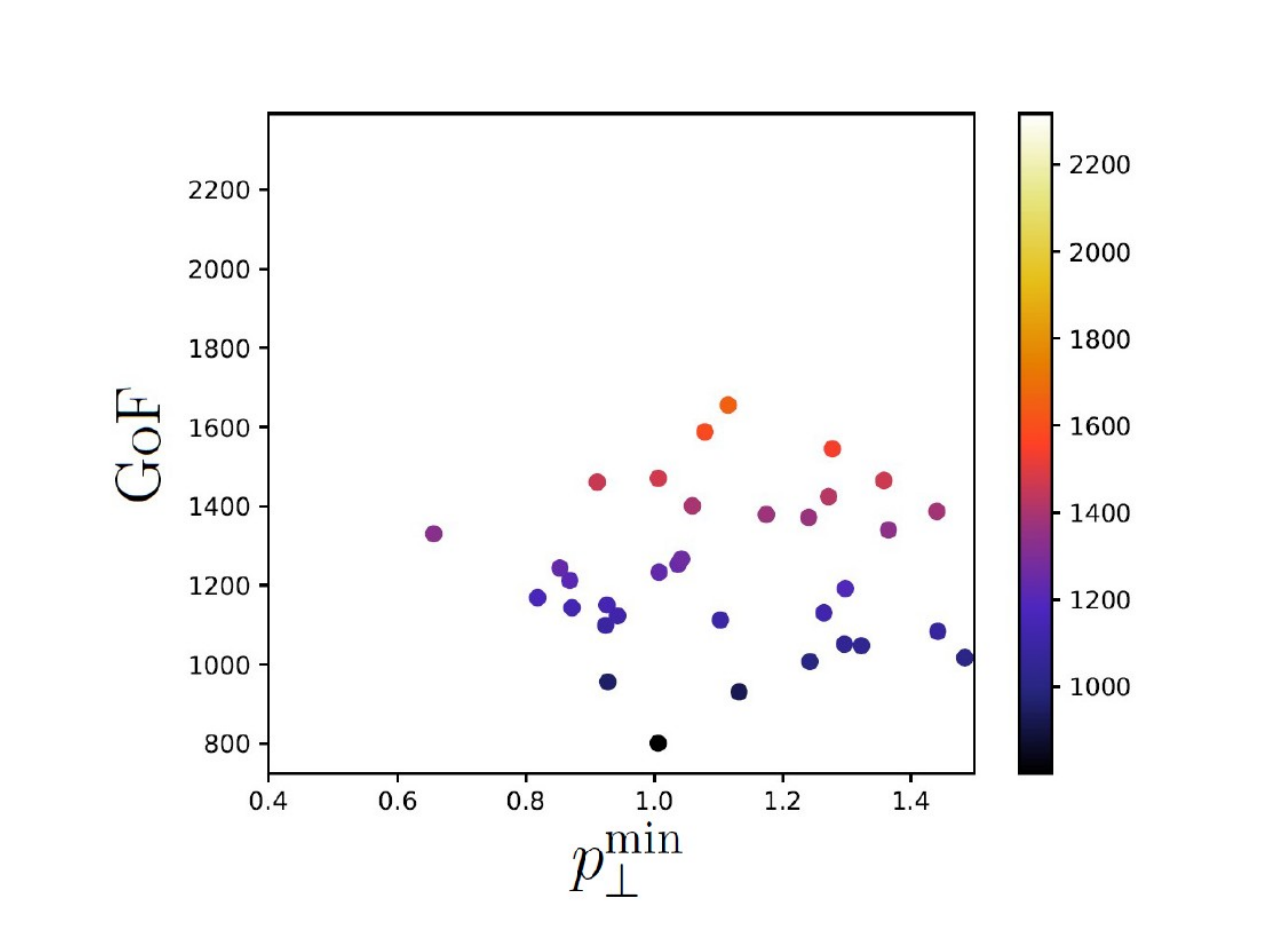}
        \caption{}
        \label{fig:chi2_LEP:b}
    \end{subfigure}
\caption{Final-state AOPS parameters $\alpha_{\mathrm{S}}^{\mathrm{FSR}}$ and $p_{\perp}^{\mathrm{min}}$ as examples of distribution of tuned values for these parameters before engaging the sequential parameter-fixing procedure.}
\label{fig:chi2_LEP}
\end{figure}

For the tuning procedure, we sample the 9-dimensional hypercube of parameters, from flat uniform distributions in the range of each parameter as listed in Table~\ref{tab:frag_params}.
We sample a total of 575 parameter vectors $\boldsymbol{p}$, and generate independent runs consisting of 1 million events for each of these.
From these, we construct 40 different interpolations using 400 generator runs for each, where the runs used for interpolation are chosen randomly from the full set of generator runs.\footnote{The generator runs used for the interpolations can overlap and hence the different interpolations are not independent.}
The minimal $\chi^2$ values for the tunes obtained from each interpolation are plotted against the optimal parameter values, as shown in Fig.~\ref{fig:chi2_LEP} for the \AOPS parameters $\alpha^{\mathrm{FSR}}_\mathrm{S}$ and $p_{\perp}^{\mathrm{min}}$.
Parameters whose optimised values exhibit low statistical uncertainty between tunes within their ranges, such as $\alpha_{\mathrm{S}}^{\mathrm{ISR}}$ in Fig.~\ref{fig:chi2_LEP:a}, are fixed to their lowest $\chi^2$ value (i.e. $0.126$ here) and the process of tuning is repeated (using the original interpolation). 
This is done until no more parameters can be visibly fixed. This step-by-step tuning can result in a well-constrained $\chi^2$ distribution even for parameters originally exhibiting larger statistical uncertainties in their ranges, e.g. the $p_{\perp}^{\mathrm{min}}$ parameter which is shown in Fig.~\ref{fig:chi2_LEP:b}. 
Such a distribution indicates that there is a wide space for finding the optimal values of other parameters.

Subsequently, in the second stage of tuning, eleven flavour parameters listed in Table~\ref{tab:flavor_params}, are tuned.
The first four parameters govern how a new flavour quantum number is selected in the hadronization process. 
The parameter \texttt{probStoUD} gives the relative strangeness suppression compared to the ordinary production of $u$ and $d$ quarks, \texttt{probQQtoQ} provides the suppression of diquark production relative to single quark production. \texttt{probSQtoQQ} controls the suppression of strange diquark production relative to light diquark production and \texttt{probQQ1toQQ0} governs the suppression of spin-1 diquark production to spin-0 one. 
The \texttt{popcornRate} parameter steers the intermediate meson production rate with respect to classical two-point string breaking, i.e. string breaking leads to production of baryon--meson--antibaryon instead of usual baryon--antibaryon case. 
All the remaining parameters contribute to standard meson production. The \texttt{etaSup} and \texttt{etaPrimeSup} parameters provide suppression of the $\eta$ and $\eta'$ mesons, and \texttt{mesonUDvector}, \texttt{mesonSvector}, \texttt{mesonCvector}, and \texttt{mesonBvector} parameters govern the relative production of vector particles to pseudoscalar ones for the lightest quarks, and separately for the $s$, $c$, and $b$ quarks~\cite{Bierlich:2022pfr}.

The observables used for tuning during this stage were the $\pi^{+}$ and $\pi^0$ multiplicities and the identified hadron yield with respect to the $\pi^{\pm}$ multiplicity measured at lepton colliders~\cite{ParticleDataGroup:2008zun}, the $b$-quark fragmentation function measured by the DELPHI Collaboration~\cite{Barker:2002iuq}, and again four different mean charged multiplicities as measured by the OPAL Collaboration~\cite{OPAL:1998arz} with reduced weights. 
These sets of observables along with their assigned weights are listed in Table~\ref{tab:LEPtuneweights2}, see \ref{appendix_tune}. 

For the flavour tuning, we proceed in the same way as with the fragmentation tuning. We sample 950 points from the 11-dimensional hypercube and generate runs consisting of 1 million events for each of these.
As before, we construct 40 different interpolations using 500 generator runs for each, from the full set of generator runs.
The final resultant tune corresponds to the optimal GoF value of all variations. The parameter values for the best tune (labelled Les Houches or in short \eetune\footnote{The name of the tune comes from the place where it was finished during the PhysTeV workshop at Les Houches.}) are listed in Table~\ref{tab:frag_params} and Table~\ref{tab:flavor_params}. 

\subsubsection{Hadron collisions}
Following the relatively clean environment of lepton collisions, the transition to hadron collisions introduces a significantly more complex and dynamic setting. Unlike elementary leptons, hadrons are composite particles, meaning that their collisions involve not only the hard scattering of individual partons but also a range of additional phenomena. 
\ISR can alter the partonic collision energy and together with the proton remnants can create a complex event structure. Furthermore, Multiple Parton Interactions lead to additional soft and semi-hard scatterings within a single event, enhancing overall activity. 
Colour reconnection further modifies the final-state structure by reshuffling colour connections between partons, influencing hadronization patterns. 
Together, these effects create a densely populated collision environment, requiring detailed comparisons between Monte Carlo predictions and experimental data to tune the \UE parameters, ensuring the most accurate description of hadronic interactions. 
Achieving a correct and universal modelling requires a multi-observable tuning strategy that balances different contributions without overfitting any single distribution.

\begin{figure}[h!]
\begin{subfigure}{0.38\textwidth}
    \includegraphics[width=\linewidth]{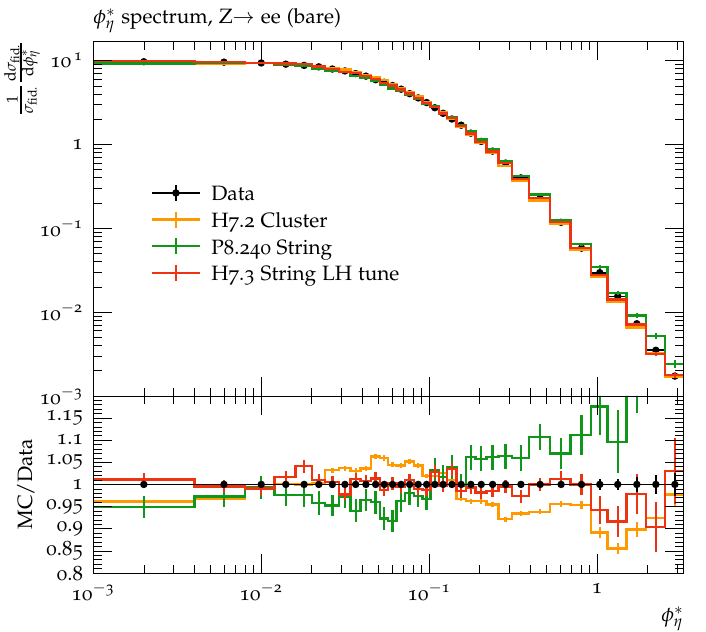}
    \caption{}
  \end{subfigure}\\
  \begin{subfigure}{0.38\textwidth}
    \includegraphics[width=\linewidth]{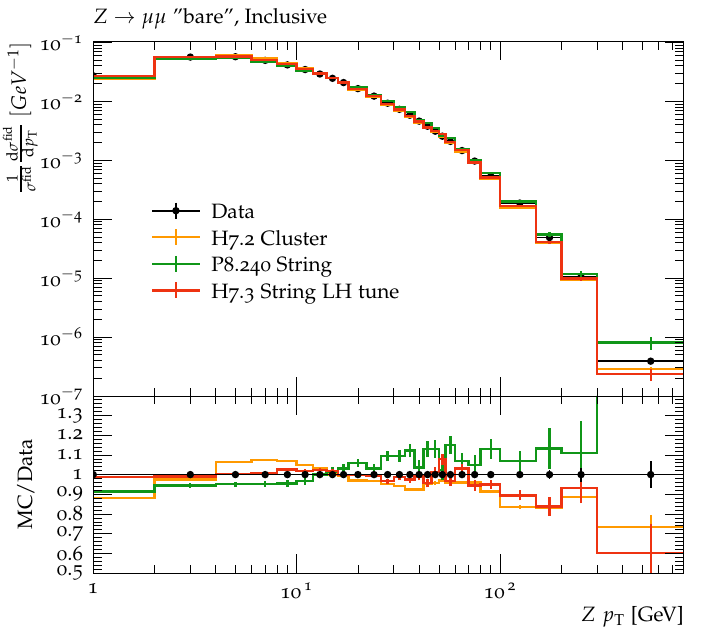}
    \caption{}
  \end{subfigure}
\caption{Normalized differential cross section of $Z$ boson production as a function of the (a) $\phi^*_{\eta}$ parameter of the $Z$ boson ($Z$ $\to$ $ee$) \cite{ATLAS:2012ewf} and (b) transverse momentum of the $Z$ boson ($Z$ $\to$ $\mu\mu$) \cite{ATLAS:2014alx}.}
\label{fig:Z_distributions}
\end{figure}

The next (third) stage of tuning involves tuning the primordial parton $k_{\mathrm{T}}$ and the \ISR strong coupling $\alpha_{\mathrm S}^{\mathrm{ISR}}(M_Z)$ with respect to the LHC $Z$-boson production data.
For event generation, we used \herwig{}'s default Parton Distribution Functions, CT14LO \cite{Dulat:2015mca} available via the LHAPDF library \cite{Buckley:2014ana}.
We perform the tune with respect to these analyses at 7 TeV \cite{ATLAS:2014alx,ATLAS:2012ewf,CMS:2011wyd} and find it sufficiently reliable for the \eetune. 
Following a previous \herwig{} tune \cite{Bewick:2021nhc}, which tried to avoid correlations among observables, we consider $Z$-boson transverse momentum distributions from the muon channel measurements and angular distributions from the electron channel data\footnote{Details on observable selection can be found in Table~\ref{tab:DYtuneweights}.}.
The distributions are shown in Fig.~\ref{fig:Z_distributions}, where it can be seen that the \eetune describes both $\phi^*$ and the transverse momentum of the $Z$ boson very well for small and intermediate values of these observables. For larger values of $p_\mathrm{T}$, this tune undershoots the data as expected, due to missing higher-order corrections in \herwig{} LO matrix element simulation. 
The resulting values of these parameters obtained for the \eetune are presented in Table~\ref{tab:Z_tune}.
Although $\alpha_{\mathrm S}^{\mathrm{ISR}}(M_{Z})$  and $\alpha_{\mathrm S}^{\mathrm{FSR}}(M_{Z})$ were tuned separately, their best tuned values are very similar and, in fact, could be unified to one common value.  
\begin{table}[h!]
    \centering
    \begin{tabular}{lcc}
        \centering
            Parameter & Range & \eetune \\
            \hline
            $\alpha_{\mathrm S}^{\mathrm{ISR}}$  & 0.100 - 0.134  & 0.124 \\
            $k_{\mathrm{T}}$ [GeV]     & 0.500 - 3.000    & 1.304 \\
    \end{tabular}
\caption{Resultant tunes obtained for $\alpha_{\mathrm{S}}^{\mathrm{ISR}}$ and $k_{\mathrm{T}}$ parameters subject to $Z$-boson production data from LHC.}
\label{tab:Z_tune}
\end{table}

\begin{table*}[]
    \centering
    \begin{tabular}{l c  c c  c  c  c }
        Parameter & Range & {\bf \eetune} & 0.9 TeV tune & 1.8 TeV tune & 7 TeV tune & 13 TeV tune \\
        \hline
        Power $c$                           & 0.0 - 2.0     & {\bf 0.23}  & 0.08   & 0.51   & 2.00   & 0.09  \\
        $p_{\perp,0}^{\mathrm{min}}$ [GeV]  & 2.0 - 5.0     & {\bf 3.13}  & 2.56   & 4.81   & 3.07   & 3.80  \\
        Offset $b$ [GeV]                    & 500.0 - 800.0 & {\bf 530.5} & 633.8  & 781.9  & 503.0  & 706.6  \\
        \hline
        $\mu^2$ [GeV$^{-2}$]                & 0.5 - 3.5     & {\bf 1.14}  & 1.19   & 0.69   & 0.93   & 1.18  \\
        ladderMult                          & 0.5 - 2.0     & {\bf 0.57}  & 0.74   & 1.10   & 0.79   & 0.65  \\
        ladderbFactor                       & 0.5 - 2.0     & {\bf 0.97}  & 0.92   & 0.95   & 1.33   & 0.90  \\
        \hline
        $R_{\mathrm{diff}}$                 & 0.0 - 0.5     & {\bf 0.21}  & 0.22   & 0.39   & 0.32   & 0.10  \\
        \texttt{m0} [GeV]                          & 0.1 - 3.0     & {\bf 2.87}  & 2.28   & 1.64   & 2.90   & 2.66  \\
        \texttt{junctionCorrection}                               & 0.05 - 5.0    & {\bf 4.55}  & 1.92   & 1.36   & 2.74   & 4.15  \\
    \end{tabular}
    \caption{Hadron collision tunes for \MB and \UE observables. \eetune compared to tunes for dedicated energy (0.9, 1.8, 7, 13 TeV) tunes.}
    \label{tab:UE_params}
\end{table*}

This is followed by the final (fourth) stage of tuning, which involves tuning \herwig's \UE model parameters using data from LHC and Tevatron at different energies (LHC at 0.9 TeV \cite{ALICE:2010syw,ATLAS:2011wqb,ATLAS:2010jvh,ATLAS:2010zmc,ATLAS:2010kmf}, Tevatron at 1.8 TeV \cite{CDF:2001hmt,Alexopoulos:1998bi,CDF:1989nkn,CDF:1988evs,CDF:2001onq}, LHC at 7 TeV \cite{ATLAS:2011wqb,ATLAS:2010jvh,ATLAS:2010kmf,ALICE:2010mty} and LHC at 13 TeV \cite{ATLAS:2016zba,ATLAS:2017blj}). 
The final results are summarised in Table~\ref{tab:UE_params}. 
The complete list of observables and tuning weights is provided in \ref{appendix_tune}, Tables~\ref{tab:MBUEtuneweights9} - \ref{tab:MBUEtuneweights13}. 
These parameters control MPI and a portion of diffraction events ($R_{\mathrm{diff}}$) within the Minimum-Bias event generation in \herwig7. 
This stage also includes tuning of two colour reconnection parameters within \pythiav{8} accessed via TheP8I. 
The so-called QCD based colour reconnection scheme \cite{Bierlich:2022pfr,Christiansen:2015yqa} is used in combination with the new beam remnant model \cite{Christiansen:2015yqa} and includes the formation of junction strings. We simplify the event generation by limiting the parameters to be tuned, thus disallowing several other functionalities within the scheme including double-junction remnants, time causality to string decays and special parity treatment for heavy quarks. 
This allows colour reconnection to be governed only by two parameters - \texttt{m0}, which is a lower bound on the mass of the colour reconnected system and \texttt{junctionCorrection}, which scales this mass for junction strings as a multiplicative factor controlling how easily junctions are formed.

We partially follow the tuning strategy for the latest MPI tune within \herwig7 \cite{Bellm:2019icn} and select four collision energies ($\sqrt{s}$ = 0.9, 1.8, 7, 13 TeV) and for each energy we obtained a separate tune.
Ultimately, we combined all the desired observables for each collision energy to create a single, energy-extrapolated \eetune. 
However, this combination of generator runs at multiple energies comes at the price of a reduced number of valid runs. From the original sample of 1000 generator runs, we have collected 265 runs for the \professor{} interpolation.

\begin{figure}[h!]
\centering
	\includegraphics[width=.5\textwidth]{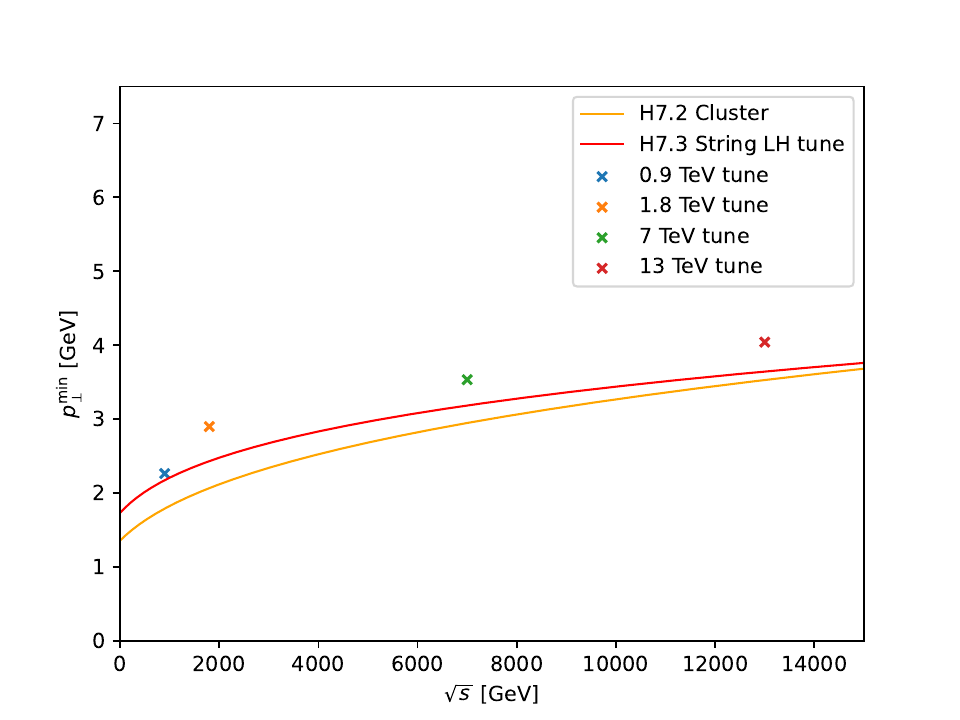}
	\caption{Evolution of perturbative threshold $p_{\perp}^{\mathrm{min}}$ with the centre-of-mass energies for individual tunes and the energy extrapolated \eetune along with a dedicated fit obtained from \cite{Bellm:2019icn}.}
        \label{fig:pTminevolution}
\end{figure}

With the combination of these two tuning strategies, we can also study the energy evolution of the minimal transverse momentum threshold $p_\perp^\mathrm{min}(s)$ separating semi-hard and soft parts of MPI modelling in Fig.~\ref{fig:pTminevolution}. 
This parameter has been removed from the list of free parameters in~\cite{Bellm:2019icn} and set to follow a power law governed by the three parameters. 
These are the initial transverse momentum scale $p_{\perp,0}^{\mathrm{min}}(s)$, the offset $b$ for the energy of the collision relative to the reference energy $E_{0}$, and the magnitude $c$ of the exponent in the power law
\begin{align}
\label{eqn:pTmin_s}
    \hspace{0.5in} p_{\perp}^{\mathrm{min}}(s) = p_{\perp,0}^{\mathrm{min}}(s) \left( \frac{b + \sqrt{s}}{E_0}   \right)^c ,
\end{align}
where the reference energy $E_0$ was fixed to 7 TeV for all points. 
Fig.~\ref{fig:pTminevolution} shows a red curve for the new \eetune along with the individual $p_\perp^{\mathrm{min}}(s)$ points obtained from fully independent tunes at four different energies. 

\begin{figure*}[]
\centering
\begin{subfigure}{.1435\textwidth}
    \includegraphics[width=\linewidth]{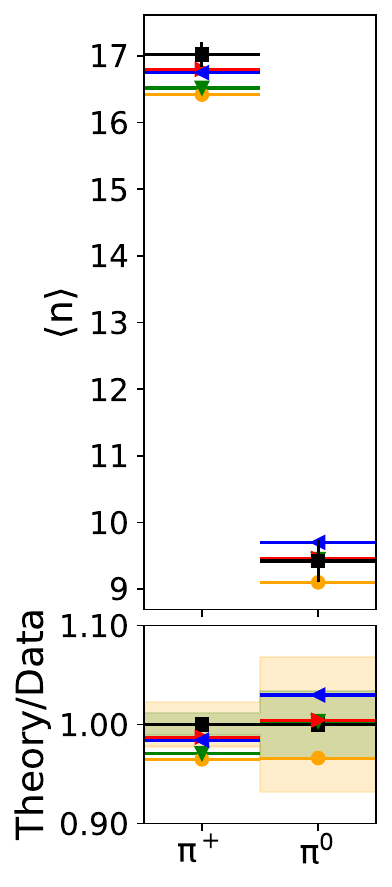}
    \caption{}
    \label{fig:pi_multiplicity}
\end{subfigure}
\begin{subfigure}{.4855\textwidth}
    \includegraphics[width=\linewidth]{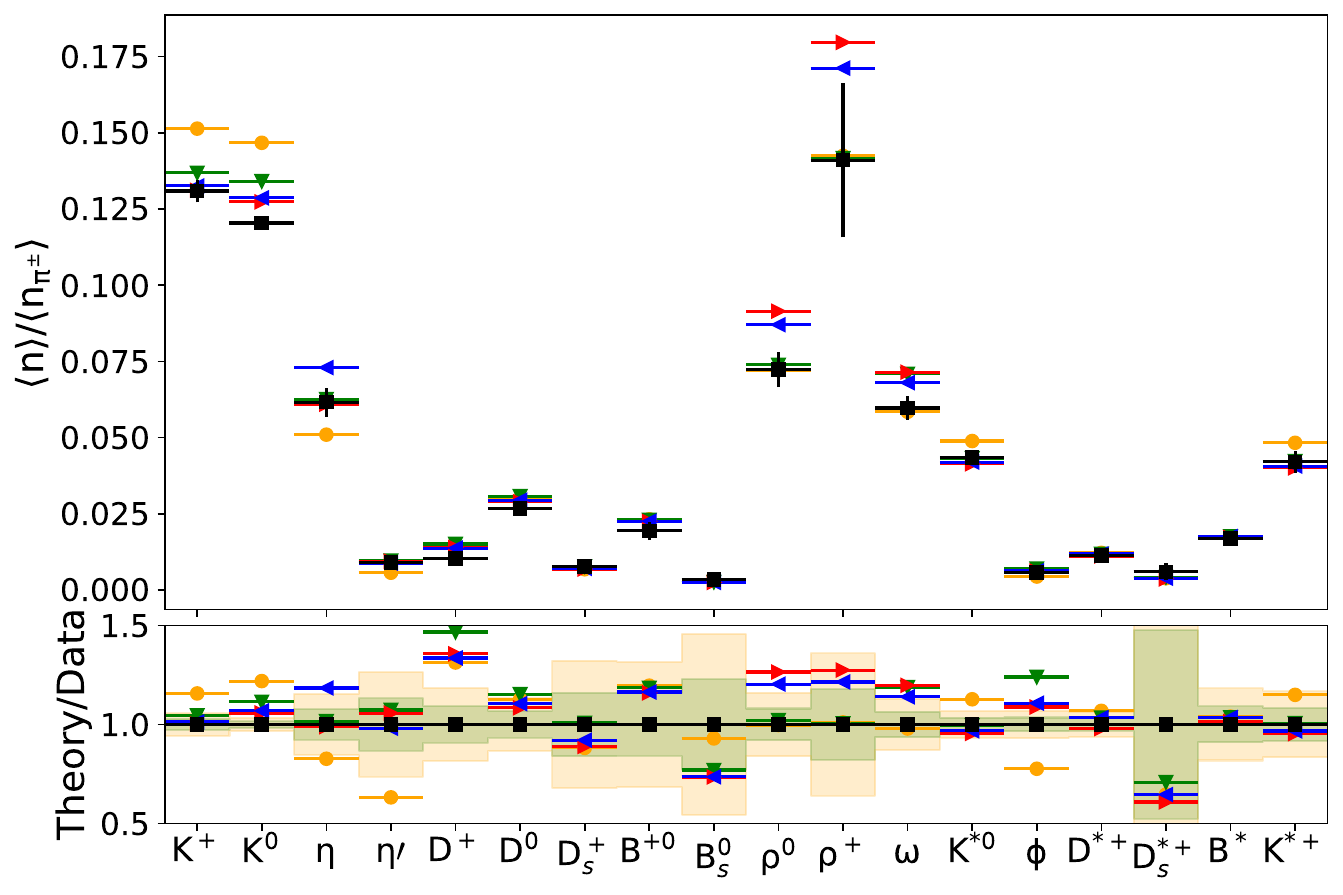}
    \caption{}
\label{fig:meson_mulitiplicity_ratios}
\end{subfigure}
\begin{subfigure}{.3144\textwidth}
    \includegraphics[width=\linewidth]{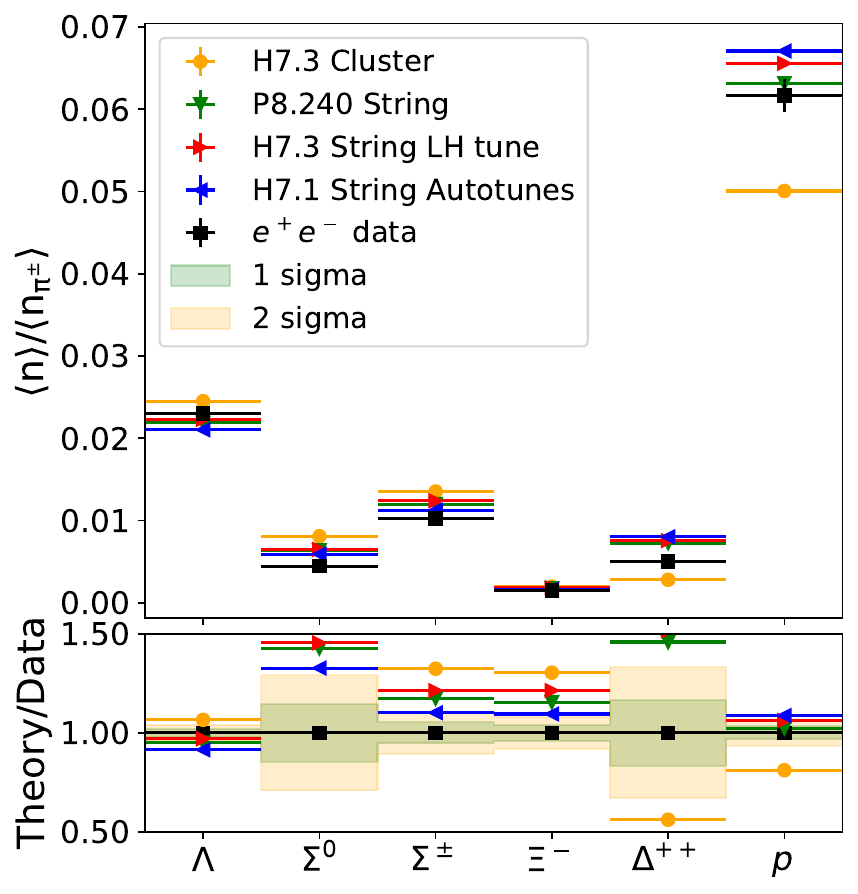}
    \caption{}
    \label{fig:baryon_multiplicity_ratios}
\end{subfigure}
\caption{Multiplicities of identified particles: (a) mean number of $\mathrm{\pi}^{+}$ and $\pi^{0}$, mean multiplicity ratios for identified (b) mesons (c) baryons with respect to $\pi^{\pm}$ multiplicity \cite{ParticleDataGroup:2008zun}.}
\label{fig:multiplicity_distributions}
\end{figure*}

It is important to note here that these individual tunes also depend on six other parameters (see Table~\ref{tab:UE_params}), thus potentially influencing the final particle outcome. 
However, it is interesting to observe that the individual points lie very close to the curve obtained from the \eetune, thus demonstrating the intrinsic power law that it obeys.  
Fig.~\ref{fig:pTminevolution} also shows an orange curve representing the result of a fit of the dedicated $p_{\perp}^{\mathrm{min}}$ cluster tune from~\cite{Bellm:2019icn}. We verified that the predictions of the individual energy tunes for the \UE and \MB observables are very similar to the \eetune, see Fig.~\ref{fig:EEvsIndep-UE} and \ref{fig:EEvsIndep-MB}.

%%%%%%%%%%%%%%%%%%%%%%%%%%%%%%%%%%%%%%%%%%%%%%%%%%%%%%%%%%%%%%%%%%%%%%%%%%%%%%%%%%%%%%%%%%%%%%%%%%%%%%%%%%%

\section{Comparison of the results}
\label{results}
In this section, we study the hadronization effects with the \AOPS in \herwig{} and compare them to the results obtained with the \pythia{} event generator. More precisely, we consider the following set of tunes for comparison:
\begin{itemize}
    \item \eetune (Les Houches Tune) using AOPS of \herwigv{7.3} with the string hadronization model,
    \item The \herwigv{7.3} default cluster tune~\cite{Bewick:2023tfi}, using AOPS with the cluster hadronization model%
    \footnote{As \herwigv{7.3} was not tuned to hadron collisions, an earlier version of \herwigv{7.2} was used to produce predictions for hadron colliders, as this version had been fitted to the hadronic data.},
    \item The \autotunes tune~\cite{Bellm:2019owc}, using AOPS of \herwigv{7.1} with the string hadronization model~\cite{Andersson:1983ia},
    \item The \pythiav{8.240} default string tune, Monash~\cite{Skands:2014pea}, using a $p_\mathrm{T}$-ordered shower with the string hadronization model.
\end{itemize}

We note here that the \autotunes tune was not tuned to LHC data; therefore, we only compare its predictions to LEP results. It also used an older version of AOPS than that used in \eetune.%
\footnote{There are notable differences in the shower between the two versions, i.e \herwigv{7.1} and \herwigv{7.3}}
Nevertheless, it is worth noting the differences in the predictions for both tunes in certain regions of phase space for various observables.

In the following, we discuss the phenomenological differences among the tunes listed above for the LEP observables in Section \ref{lep_results}, and for the hadron collision observables in Section \ref{hadron_results}, augmented with relevant definitions to give a comprehensive picture to the reader.

\subsection{Lepton collisions} 
\label{lep_results}
\begin{figure*}[]
\centering
\begin{subfigure}{.28\textwidth}
    \includegraphics[width=\linewidth]{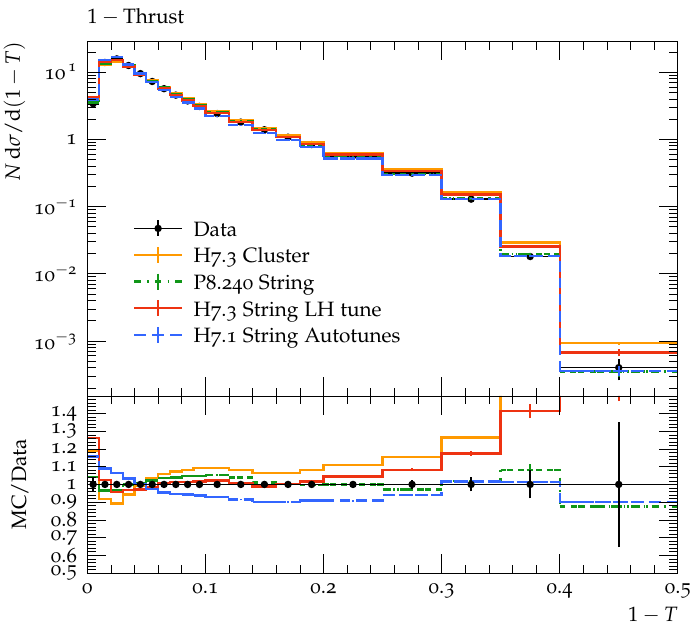}      %thrust  
    \caption{Thrust, $1 - T$}
    \label{fig:thrust_distribution}
\end{subfigure}
\begin{subfigure}{.28\textwidth}
    \includegraphics[width=\linewidth]{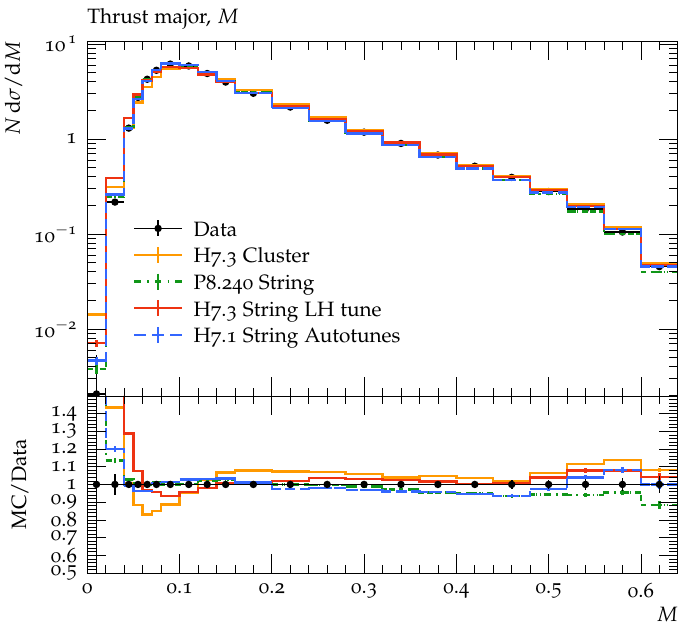}      %thrust major
    \caption{Thrust Major, $M$}
    \label{fig:major_distribution}
\end{subfigure}
\begin{subfigure}{.28\textwidth}
    \includegraphics[width=\linewidth]{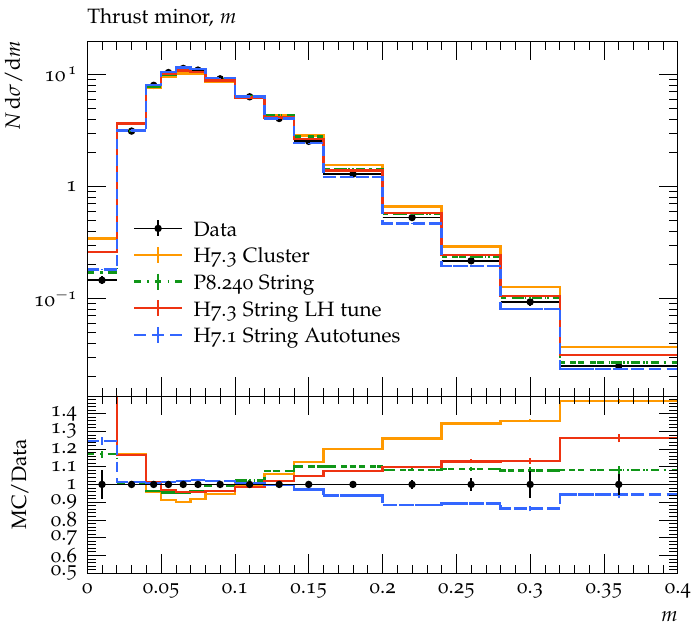}      %thrust minor
    \caption{Thrust Minor, $m$}
    \label{fig:minor_distribution}
\end{subfigure}
\begin{subfigure}{.28\textwidth}
    \includegraphics[width=\linewidth]{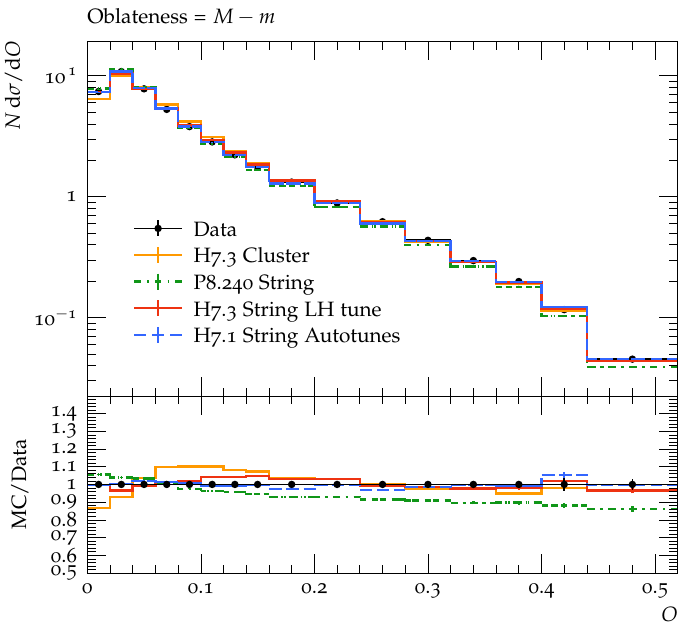}      %Oblateness
    \caption{Oblateness, $O$}
    \label{fig:oblateness_distribution}
\end{subfigure}
\begin{subfigure}{.28\textwidth}
    \includegraphics[width=\linewidth]{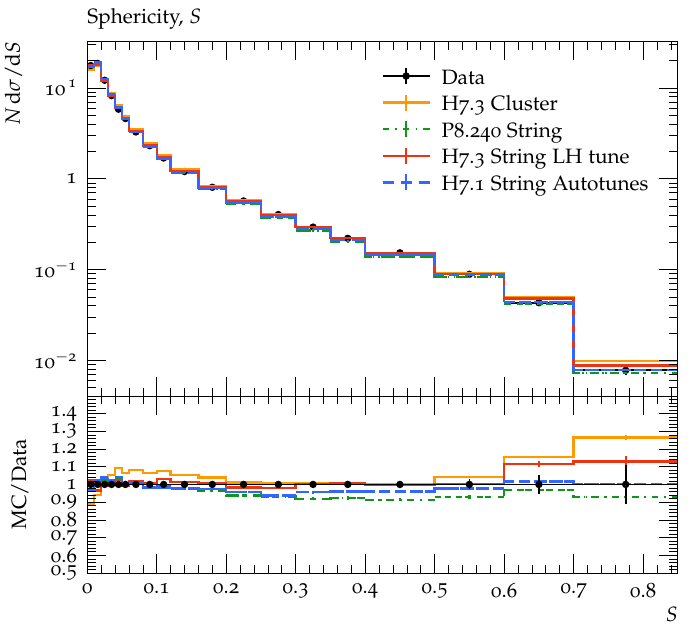}      %Sphericity
    \caption{Sphericity, $S$}
    \label{fig:sphericity_distribution}
\end{subfigure}
\begin{subfigure}{.28\textwidth}
    \includegraphics[width=\linewidth]{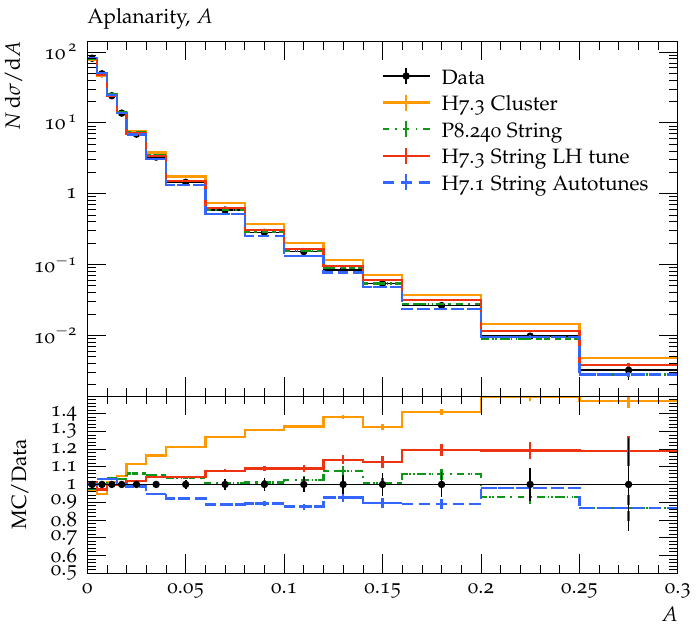}      %Aplaranity
    \caption{Aplanarity, $A$}
    \label{fig:aplanarity_distribution}
\end{subfigure}
\begin{subfigure}{.28\textwidth}
    \includegraphics[width=\linewidth]{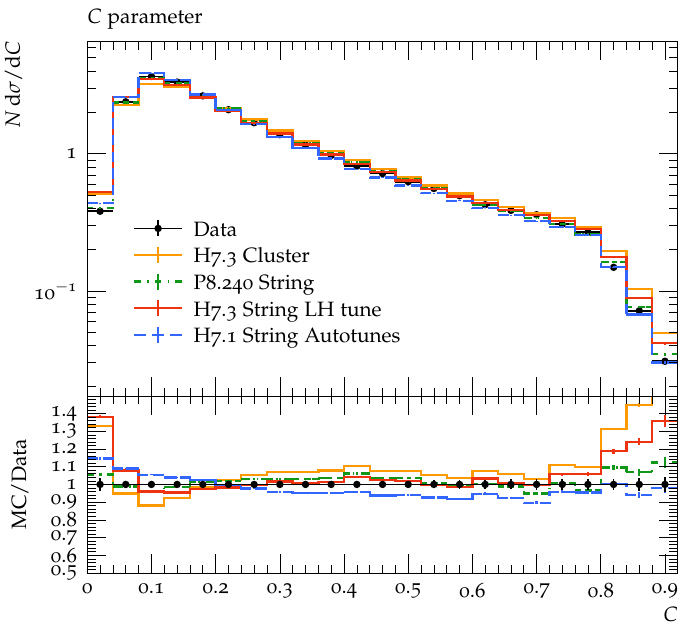}      %C parameter
    \caption{$C$ Parameter}
    \label{fig:C_distribution}
\end{subfigure}
\begin{subfigure}{.28\textwidth}
    \includegraphics[width=\linewidth]{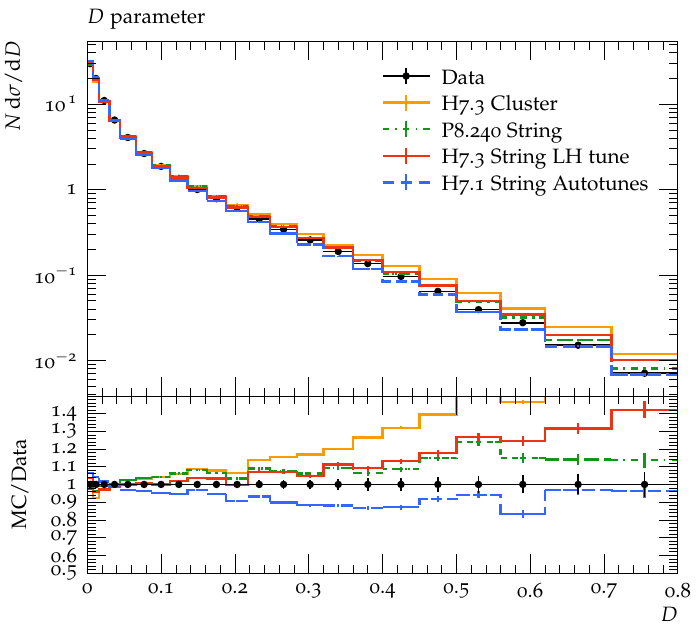}      %D parameter
    \caption{$D$ Parameter}
    \label{fig:D_distribution}
\end{subfigure}
\caption{Event Shape observables measured by DELPHI experiment at LEP \cite{DELPHI:1996sen}. These are assigned non-zero weights during the tuning procedure.}
\label{fig:LEPeventshapes}
\end{figure*}

\begin{figure*}[]
\centering
\begin{subfigure}{.3\textwidth}
    \includegraphics[width=\linewidth]{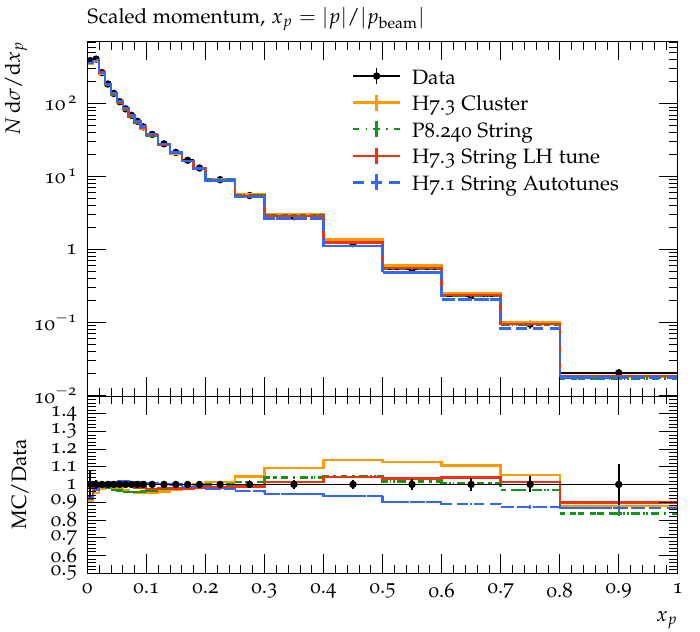}      %scaled momentum xp
    \caption{Scaled momentum, $x_p$}
    \label{fig:scaled_xp}
\end{subfigure}
\hspace{10pt}
\begin{subfigure}{.3\textwidth}
    \includegraphics[width=\linewidth]{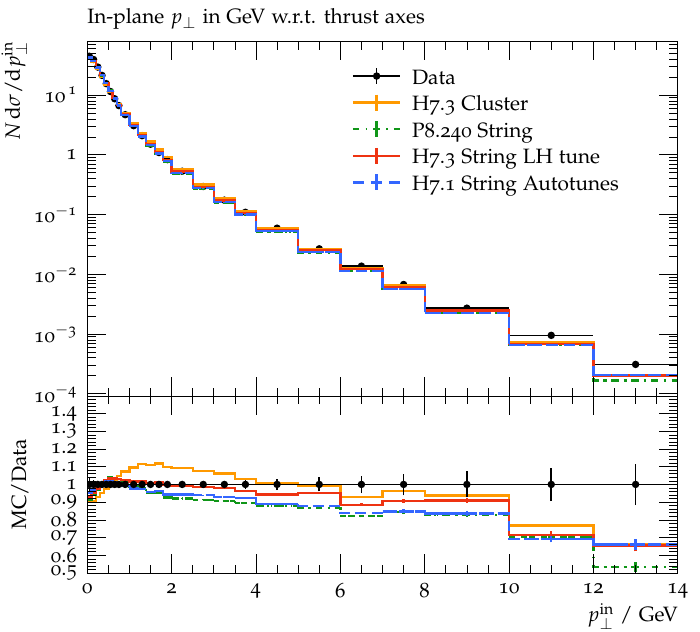}      %In-plane pT wrt thrust axes
    \caption{In-plane $p_{\perp}$ w.r.t thrust axis}
    \label{fig:pt_inplane}
\end{subfigure}
\hspace{10pt}
\begin{subfigure}{.3\textwidth}
    \includegraphics[width=\linewidth]{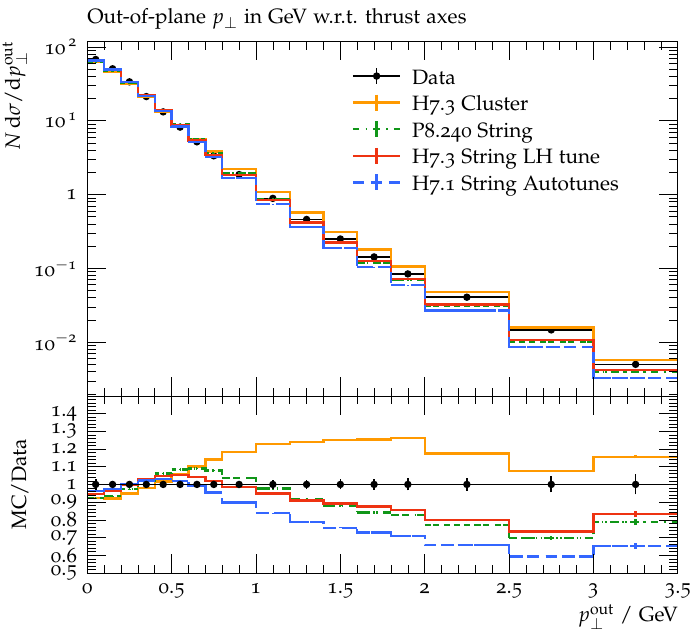}      %In-plane pT wrt thrust axes
    \caption{Out-of-plane $p_{\perp}$ w.r.t thrust axis}
    \label{fig:pt_outplane}
\end{subfigure}
\caption{Inclusive single particle momentum distributions measured by DELPHI experiment at LEP \cite{DELPHI:1996sen}. These are assigned non-zero weights during the tuning procedure.}
\label{fig:LEPmomentum}
\end{figure*}

\begin{figure*}[]
\centering
\begin{subfigure}{0.3\textwidth}
    \includegraphics[width=\linewidth]{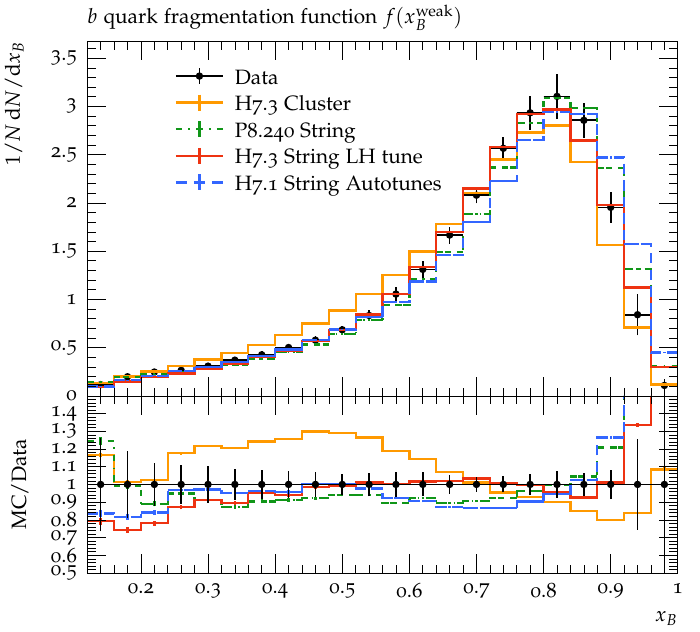}
    \caption{}
    \label{fig:eventshape_nottuned_a}
\end{subfigure}
\begin{subfigure}{0.3\textwidth}
    \includegraphics[width=\linewidth]{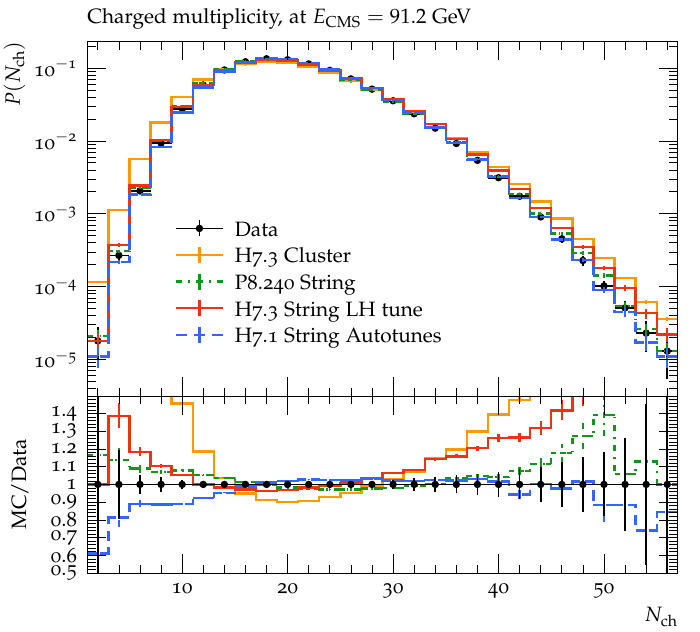}
    \caption{}
    \label{fig:charged_multiplicity_nottuned_b}
\end{subfigure}
\begin{subfigure}{0.3\textwidth}
    \includegraphics[width=\linewidth]{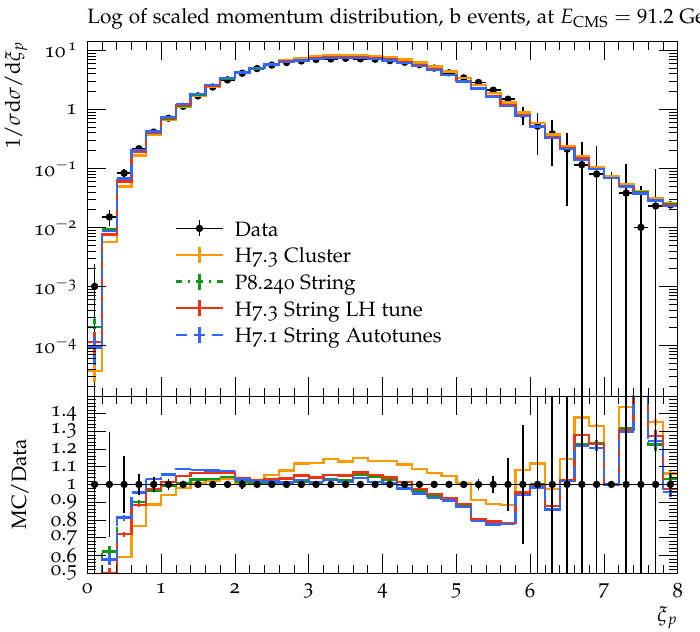}
    \caption{}
    \label{fig:scaled_mom_nottuned_c}
\end{subfigure}
\begin{subfigure}{0.3\textwidth}
    \includegraphics[width=\linewidth]{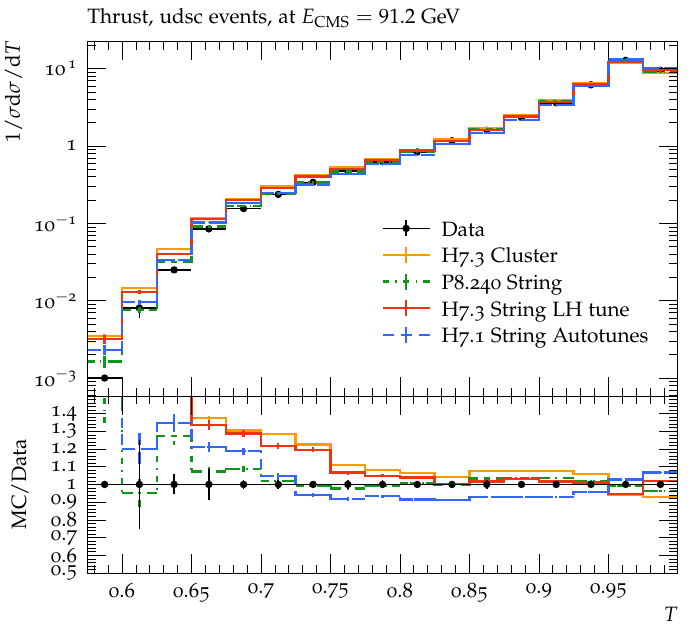}
    \caption{}
    \label{fig:thrust_nottuned_d}
\end{subfigure}
\begin{subfigure}{0.3\textwidth}
    \includegraphics[width=\linewidth]{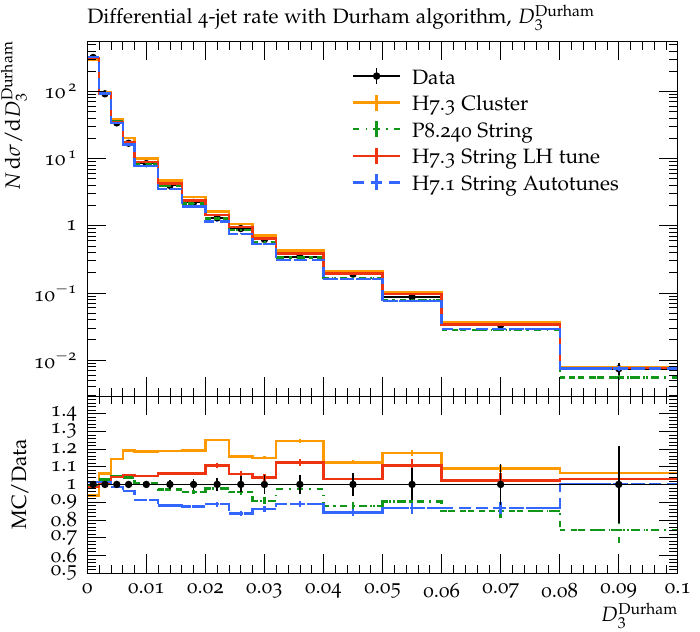}
    \caption{}
    \label{fig:diffjetrate3_nottuned_e}
\end{subfigure}
\begin{subfigure}{0.3\textwidth}
    \includegraphics[width=\linewidth]{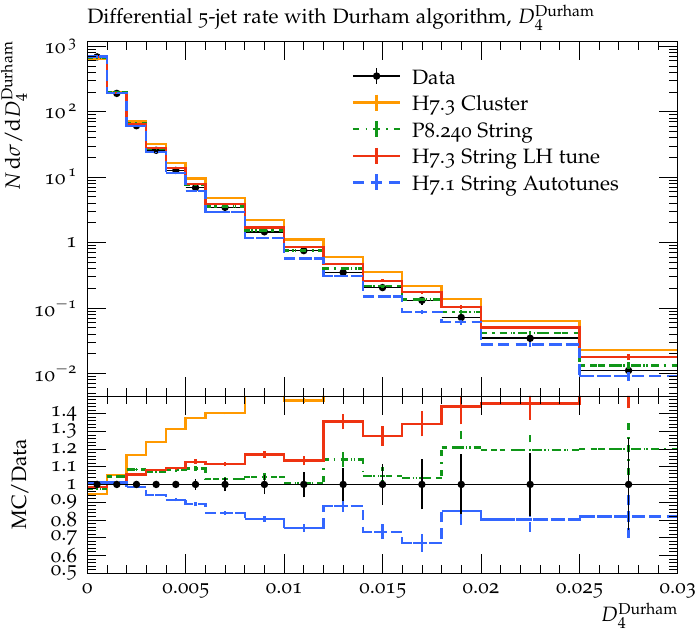}
    \caption{}
    \label{fig:diffjetrate4_nottuned_f}
\end{subfigure}
\begin{subfigure}{0.3\textwidth}
    \includegraphics[width=\linewidth]{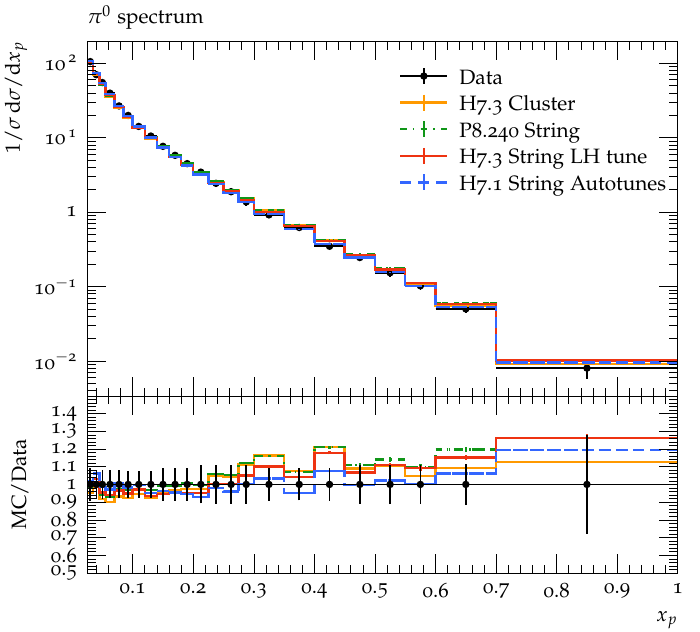}
    \caption{}
    \label{fig:pi0_scaled_mom_nottuned_g}
\end{subfigure}
\begin{subfigure}{0.3\textwidth}
    \includegraphics[width=\linewidth]{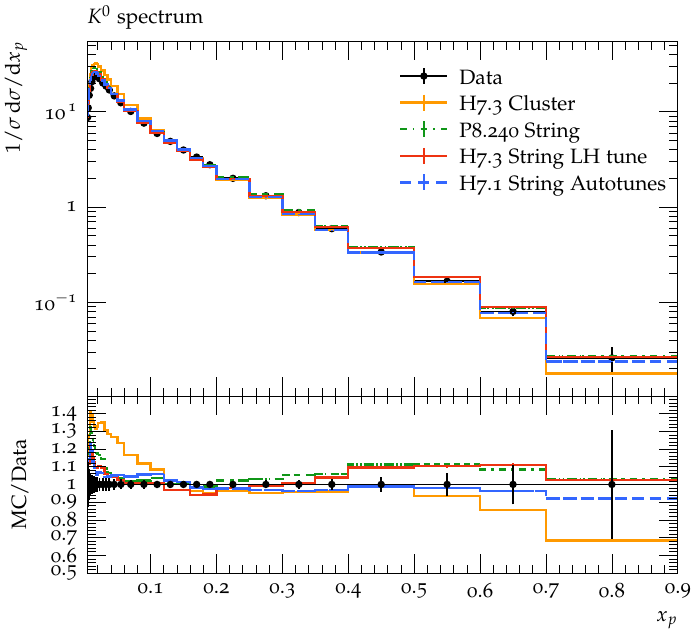}
    \caption{}
    \label{fig:K0_scaled_mom_nottuned_g}
\end{subfigure}
\begin{subfigure}{0.3\textwidth}
    \includegraphics[width=\linewidth]{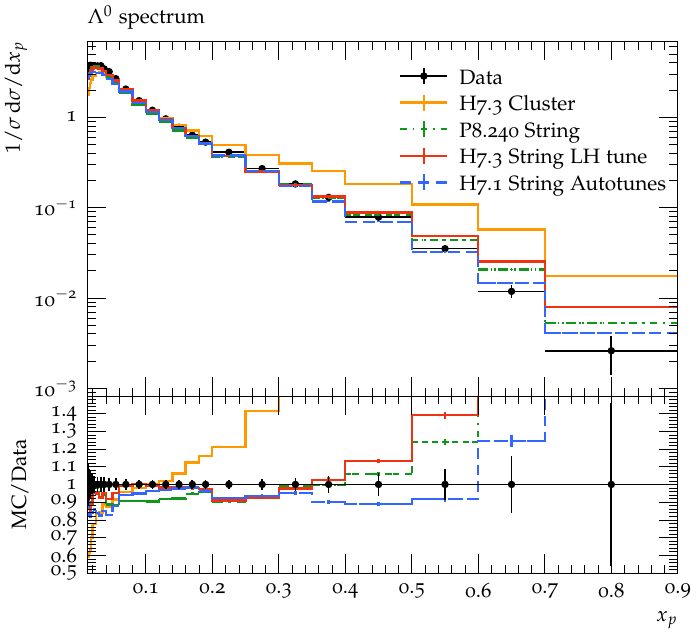}
    \caption{}
    \label{fig:lambda0_scaled_mom_nottuned_i}
\end{subfigure}
\caption{Observables measured by (a) SLD \cite{SLD:2002poq}, (b-d) L3 \cite{L3:2004cdh}, (e-f) DELPHI \cite{DELPHI:1996sen} and (g-i) ALEPH \cite{ALEPH:1996oqp} experiments at LEP and SLC. These observables are not assigned any weights in the tuning procedure, which do not bias the tunes directly. The \eetune (in red) lies within the bandwidth of other tunes and does reasonably well compared to data in these observables as well, as a test of robustness without overtuning.}
\label{fig:LEPnotunedplots}
\end{figure*}

The generator predictions for lepton collisions at $\sqrt{s} = 91.2\;\text{GeV}$ of \herwigv{7} \eetune compared to the data as well as other tunes are shown in Figs.~\ref{fig:multiplicity_distributions}-\ref{fig:LEPnotunedplots} for a number of observables across different regions of the phase space.
The mean multiplicities of $\pi^{+}$ and $\pi^{0}$ particles shown in Fig.~\ref{fig:pi_multiplicity} are well modelled, within the experimental error, by both \herwigv{7} tunes with the string model i.e. \autotunes and \eetune, while the default tunes of both \pythiav{8} with the string model and \herwigv{7} with the cluster model give slightly lower multiplicities of charged pions than the experimental data. 
This may have some impact on the results shown in Fig.~\ref{fig:meson_mulitiplicity_ratios} and Fig.~\ref{fig:baryon_multiplicity_ratios}, as they present the ratio of multiplicities of mesons and baryons to the charged pions multiplicity. In Fig.~\ref{fig:meson_mulitiplicity_ratios} we see that all tunes give a reasonable description of meson multiplicities, but all produce too many of the lightest particles containing charm quarks, namely $D^+$.
The predictions of the tunes using the string model show similar trends, especially the two string tunes in \herwig{}, while we see that the cluster model shows slightly different trends.
For example, the default \herwig{} tune with the cluster model produces a low multiplicity of $\rho$ and $\eta'$ mesons while producing too many kaons, all of which are bound states of the strange quark.
The multiplicity ratios for the baryons, see Fig. ~\ref{fig:baryon_multiplicity_ratios}, are slightly less well modelled compared to the mesons (with the exception of the $\Lambda$ baryon).
We see here as well that all string tunes give similar predictions, while the cluster tune shows a different trend especially for the $\Delta^{++}$ hadron, which consists of three up quarks.

An important class of observables are the event shapes, e.g. Thrust ($T$), Thrust major ($M$), Thrust minor ($m$), Sphericity ($S$), Aplanarity ($A$) etc. (see~\ref{appendix_b}) which provide a numerical description of the shape of the spread of particles in the final-state as the name suggests. 
More precisely, these give information about the flow of momenta in the final-state providing a good probe for studying QCD effects. 
Thus, they are particularly suited for tuning due to the existence of localised regions of phase space that are sensitive to phenomenological models. 
They typically vanish for planar events and are most sensitive to hadronization effects in those regions. These are shown in Fig.~\ref{fig:LEPeventshapes}. 

The thrust $T$ (or equivalently $1-T$) distribution has been one of the most important distributions for tuning efforts in the past. This provides a probe of the effects of hadronization in the region where $(1-T)$ has low values (approximately for $1-T \leq 0.3$). 
It can be seen that in this region (with the exception of the first bin), the \pythiav{8} tune and the \herwigv{7} \eetune are very close to the data and that the other tunes are within $\sim 10\%$ of the experimental data. 
Higher thrust values are poorly predicted by both the \herwigv{7} tunes with the cluster and the string hadronization models. However, this is to be expected, since higher-order perturbative corrections are important here, and were not included in both tunes.

We can see that some of the other event shapes such as Sphericity - event shape similar to Thrust but with different axes definitions, Thrust major, Oblateness ($O = M - m$) and $C$-Parameter are described by all tunes relatively well, about $\sim 10\%$ from experimental data for most bins. However, for some such as Aplanarity, $D$-Parameter, and Thrust minor, all the tunes differ more significantly, especially in the tails of the distributions.

In general, the \eetune predictions for this class of observables usually lie between the results of the default \herwigv{7} cluster tune and the \pythiav{8} string tune. 
In most regions of these distributions the \eetune follows the \pythiav{8} tune closely, however, in the regions more sensitive to the parton shower, their prediction can differ significantly (see the tail of $1-T$ distribution in Fig.~\ref{fig:thrust_distribution}).
This is not surprising since the \eetune and \pythiav{8} tune have very similar tuning strategies and therefore we can expect that in the region sensitive to hadronization they should be close to each other.
However, since both approaches use different shower algorithms, we can expect some differences.
  
A comparison between the \autotunes tune and the \eetune gives an idea of the spread of the predictions due to different tuning strategies of the string hadronization model in \herwig{}, reflected by a difference of $\sim 10\%$ between them.
This trend can also be seen in inclusive single particle momentum distributions in Fig.~\ref{fig:LEPmomentum}, where the \eetune describes the data slightly better than the \herwigv{7} cluster and \autotunes tunes, while closely following the \pythiav{8} tune.
While the \eetune does well in the in-plane $p_\perp$ distributions, it struggles to model the out-of-plane $p_\perp$ distribution which is expected as the inclusion of higher-order terms in the perturbative expansion of the strong coupling is required for better modelling.

In Fig.~\ref{fig:LEPnotunedplots} we show distributions that were not assigned any weights in the tuning procedure, hence, do not bias the GoF function directly. It can be seen that the \eetune generalises reasonably well for this set of observables. 
We observe that the b-quark fragmentation function in Fig.~\ref{fig:eventshape_nottuned_a} is generally well modelled by the string tunes.  
Other distributions including charged multiplicity, scaled momentum, and Thrust in Fig.~\ref{fig:charged_multiplicity_nottuned_b}-\ref{fig:thrust_nottuned_d} show reasonable agreement with relative differences within 10\% between the tunes in regions of interest, with the \eetune generally lying between the \herwigv{7} cluster and \pythiav{8} string tunes as seen before.
The differential jet rate distributions -- $D_3^{\text{Durham}}$ and $D_4^{\text{Durham}}$ in Fig.~\ref{fig:diffjetrate3_nottuned_e} and~\ref{fig:diffjetrate4_nottuned_f} are sensitive to perturbative modelling, and hence we observe larger uncertainty between the tunes here.
These uncertainties rise with the number of jets in the final-state.

The scaled momentum distributions for specific flavours are shown in Figs.~\ref{fig:pi0_scaled_mom_nottuned_g}-\ref{fig:lambda0_scaled_mom_nottuned_i}.
For the $\pi^0$ spectrum, all the tunes exhibit good agreement with the data, accompanied by relatively smaller uncertainties. 
This behaviour can be largely attributed to the tuned distribution shown in Fig.~\ref{fig:scaled_xp}, which shows the scaled momentum of all final-state particles. 
Since the final-state is predominantly composed of pions, we observe that the scaled momentum distribution $x_p$ of the $\pi^0$ follows a similar trend. 
However, this changes as we look at the distributions of flavours of particles that are less abundant in the final-state such as $K^0$ and $\Lambda^0$, where we observe larger relative uncertainties between the tunes, as can be seen in Figs.~\ref{fig:K0_scaled_mom_nottuned_g} and \ref{fig:lambda0_scaled_mom_nottuned_i}.

In general, the variations observed in the predictions across the different tunes serve as an indication of the level of uncertainty that arises when selecting different parton showers along with different hadronization models for a wide range of observables.

\subsection{Hadron collisions} 
\label{hadron_results}
\begin{figure*}[]
\centering
\begin{subfigure}{0.32\textwidth}
    \captionsetup{labelformat=empty}
    \caption[]{\bf $\mathbf{\sqrt{s}}$ = 0.9 TeV}
    \includegraphics[width=1\textwidth]{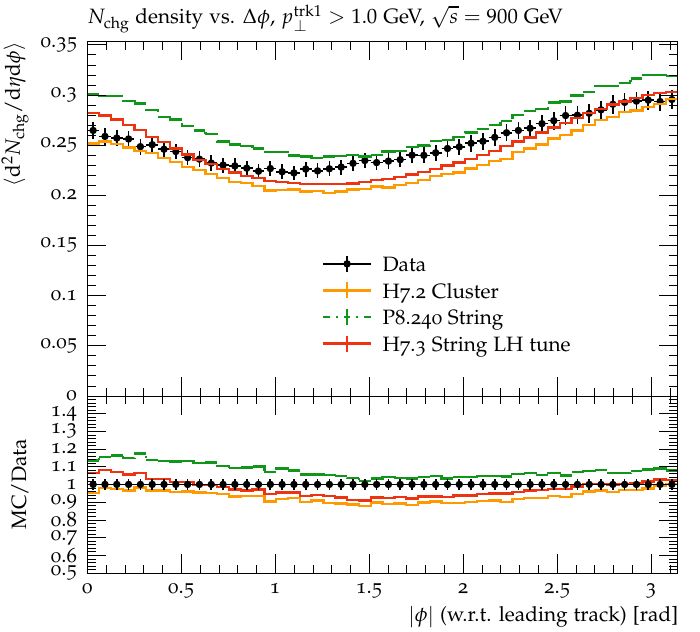}
\end{subfigure}
\begin{subfigure}{0.32\textwidth}
    \captionsetup{labelformat=empty}
    \caption[]{\bf $\mathbf{\sqrt{s}}$ = 7 TeV}
    \includegraphics[width=1\textwidth]{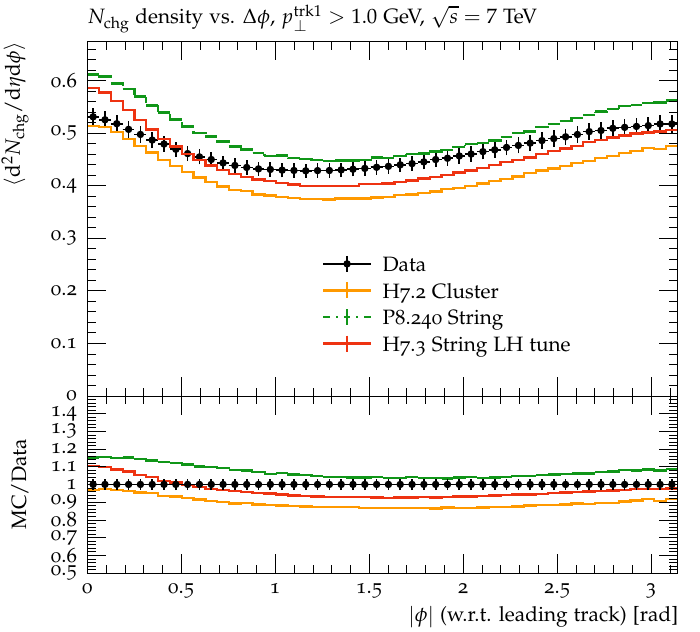}
\end{subfigure}
\begin{subfigure}{0.32\textwidth}
    \captionsetup{labelformat=empty}
    \caption[]{\bf $\mathbf{\sqrt{s}}$ = 13 TeV}
    \includegraphics[width=1\textwidth]{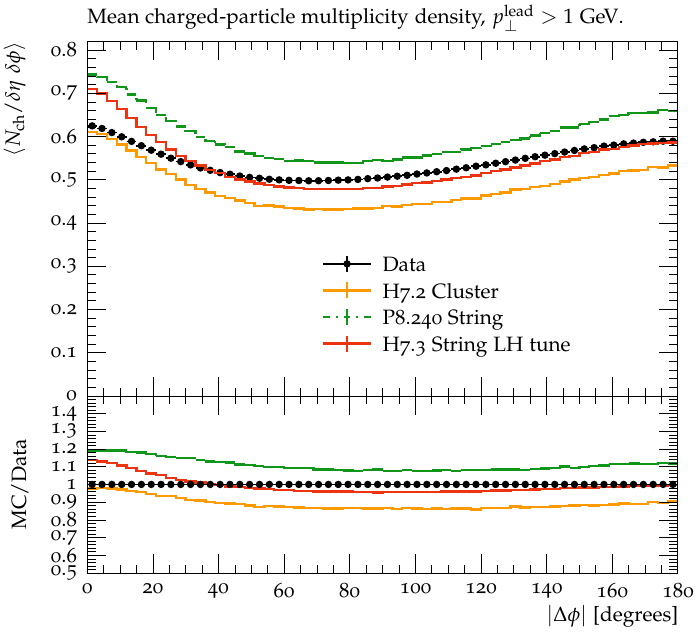}
\end{subfigure}\\
\vspace{0.5cm}
\begin{subfigure}{0.32\textwidth}
    \includegraphics[width=1\textwidth]{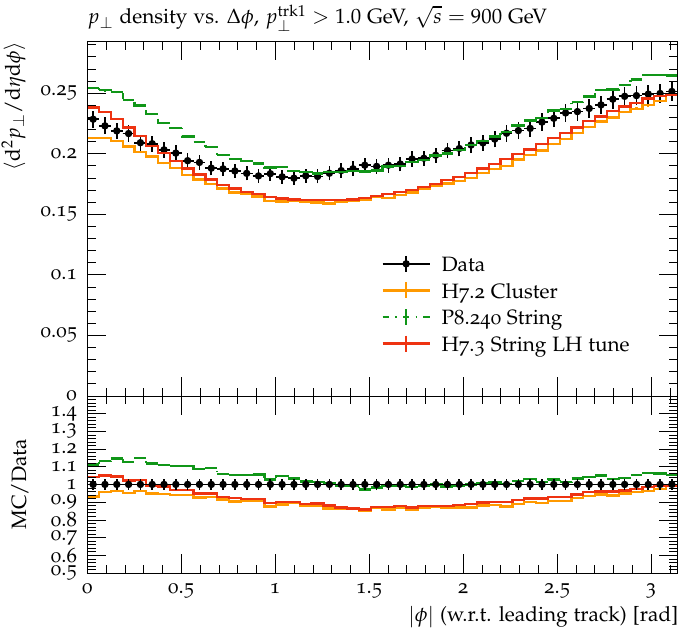}
\end{subfigure}
\begin{subfigure}{0.32\textwidth}
    \includegraphics[width=1\textwidth]{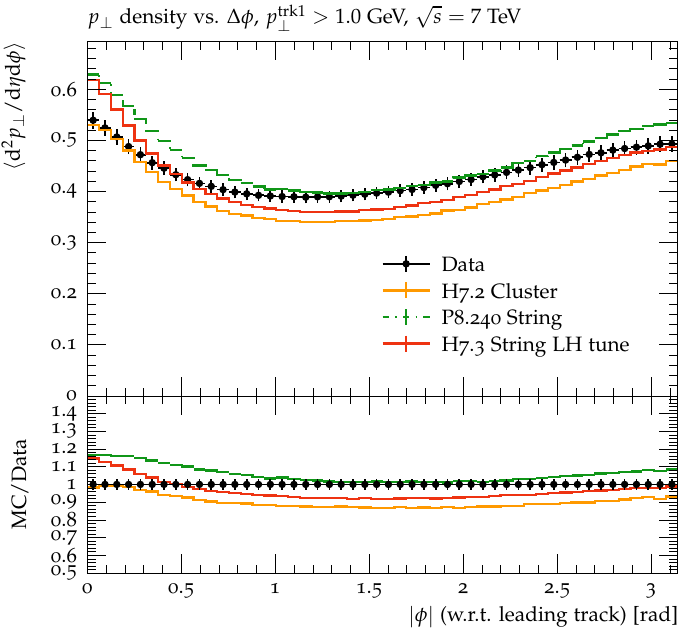}
\end{subfigure}
\begin{subfigure}{0.32\textwidth}
    \includegraphics[width=1\textwidth]{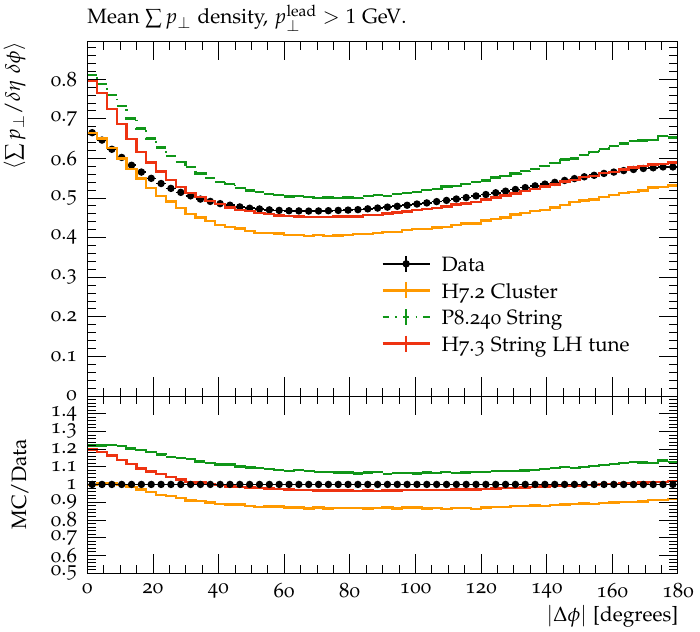}
\end{subfigure}
\caption{\UE particle multiplicity distributions in azimuthal plane relative to the leading object direction measured by the ATLAS experiment at different energies. The three columns relate to collision energies (from the left): 0.9 TeV, 7 TeV~\cite{ATLAS:2010kmf}, and 13 TeV~\cite{ATLAS:2017blj}. All shown observables were included in the tuning interpolation.}
\label{fig:UE_1}
\end{figure*}

\begin{figure*}[]
\centering
\begin{subfigure}{0.32\textwidth}
    \captionsetup{labelformat=empty}
    \caption[]{\bf Transverse region}
    \includegraphics[width=1\textwidth]{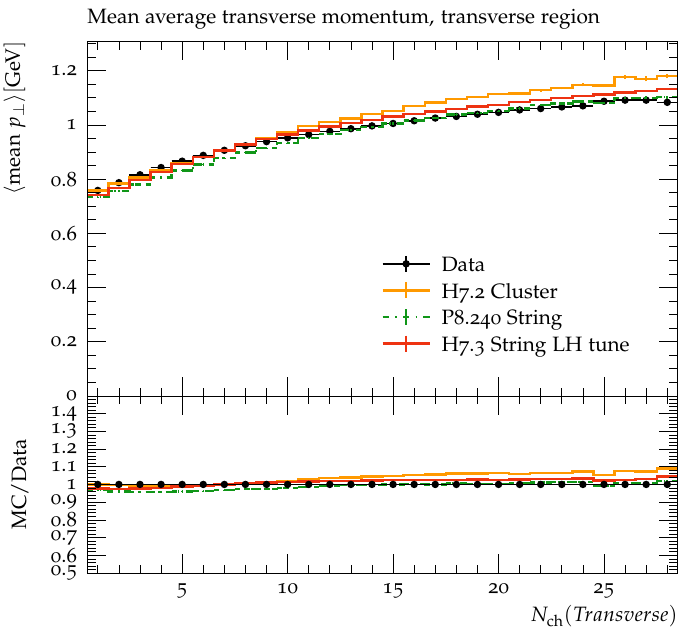}
\end{subfigure}
\begin{subfigure}{0.32\textwidth}
    \captionsetup{labelformat=empty}
    \caption[]{\bf Towards region}
    \includegraphics[width=1\textwidth]{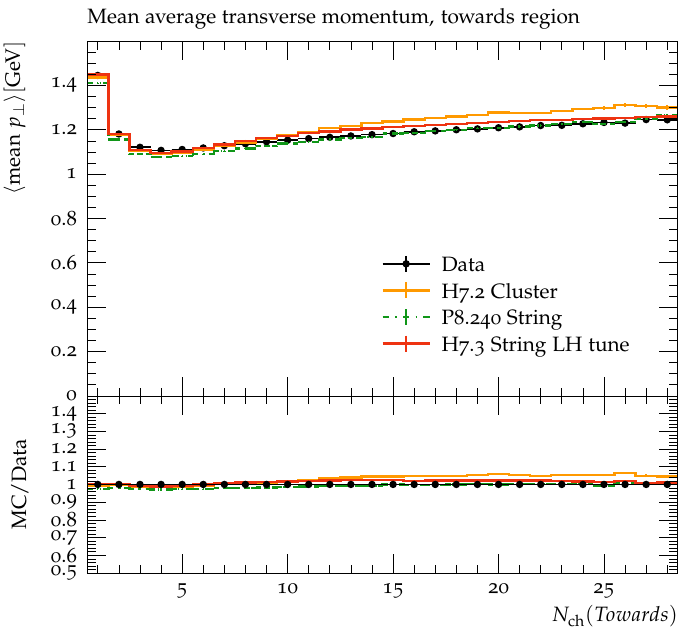}
\end{subfigure}
\begin{subfigure}{0.32\textwidth}
    \captionsetup{labelformat=empty}
    \caption[]{\bf Away region}
    \includegraphics[width=1\textwidth]{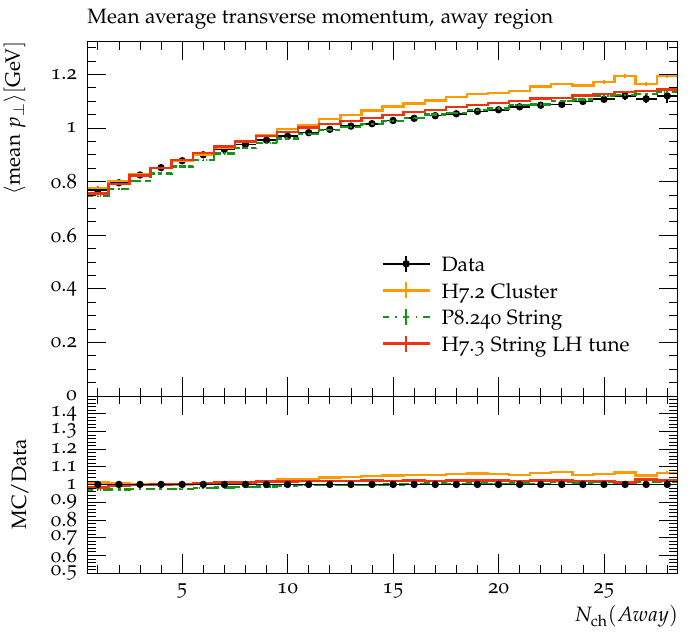}
\end{subfigure}\\
\begin{subfigure}{0.32\textwidth}
    \includegraphics[width=1\textwidth]{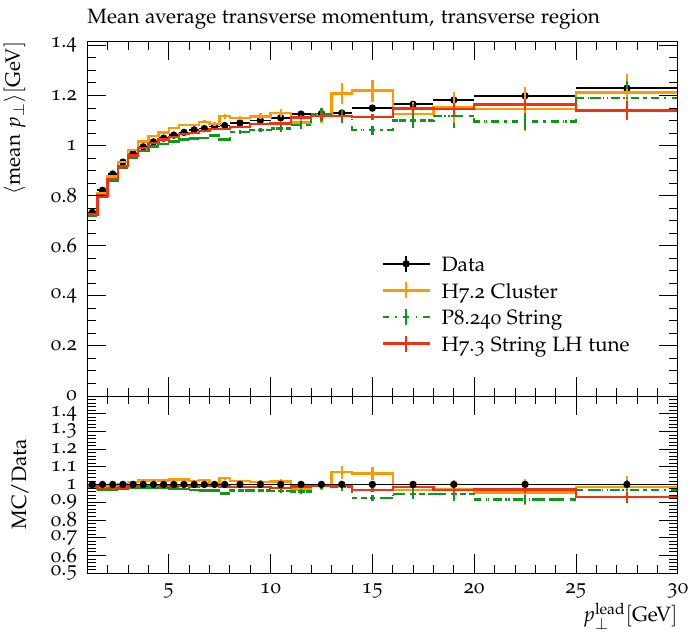}
    %\caption{\bf $\mathbf{\sqrt{s}}$ = 13 TeV}
\end{subfigure}
\begin{subfigure}{0.32\textwidth}
    \includegraphics[width=1\textwidth]{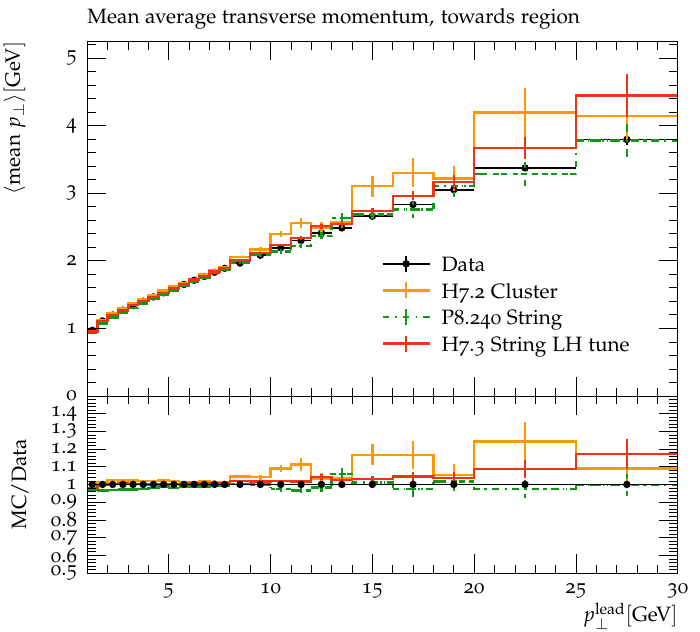}
    %\caption{\bf $\mathbf{\sqrt{s}}$ = 13 TeV}
\end{subfigure}
\begin{subfigure}{0.32\textwidth}
    \includegraphics[width=1\textwidth]{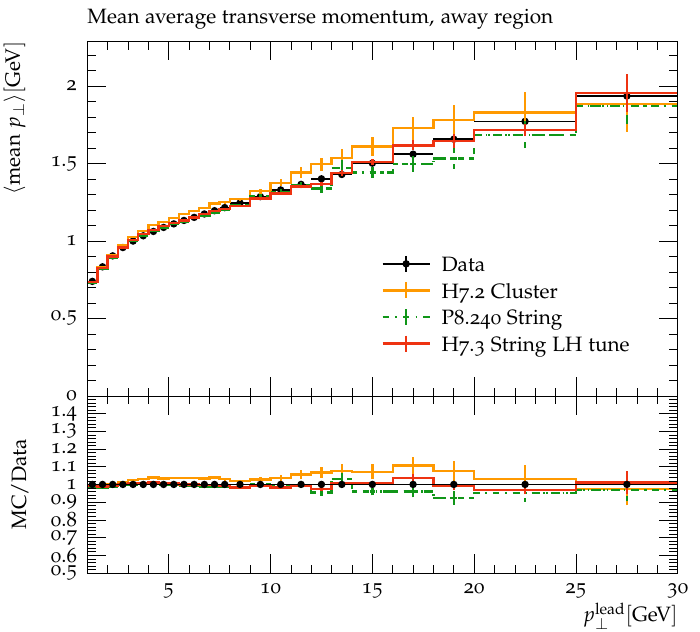}
    %\caption{\bf $\mathbf{\sqrt{s}}$ = 13 TeV}
\end{subfigure}\\
\begin{subfigure}{0.32\textwidth}
    \includegraphics[width=1\textwidth]{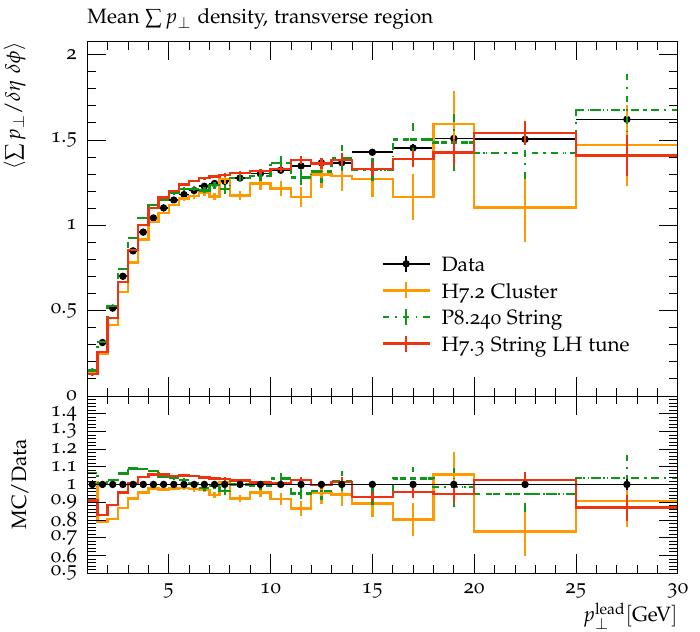}
    %\caption{\bf $\mathbf{\sqrt{s}}$ = 13 TeV}
\end{subfigure}
\begin{subfigure}{0.32\textwidth}
    \includegraphics[width=1\textwidth]{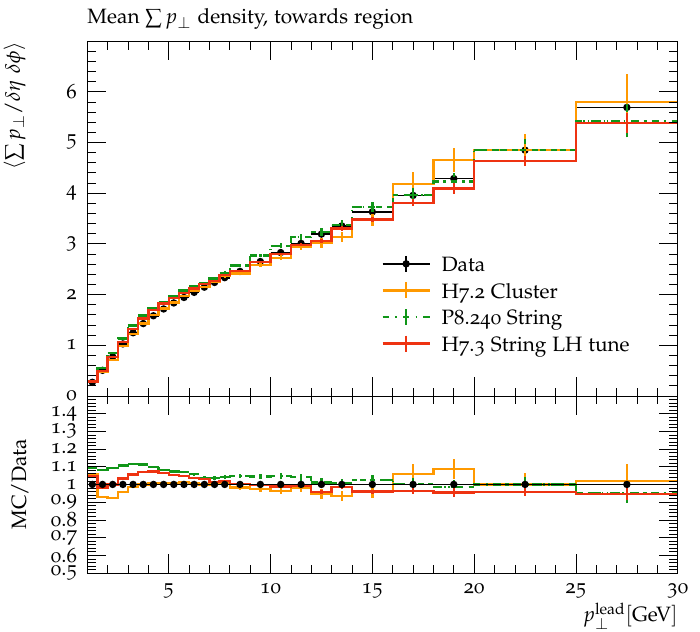}
    %\caption{\bf $\mathbf{\sqrt{s}}$ = 13 TeV}
\end{subfigure}
\begin{subfigure}{0.32\textwidth}
    \includegraphics[width=1\textwidth]{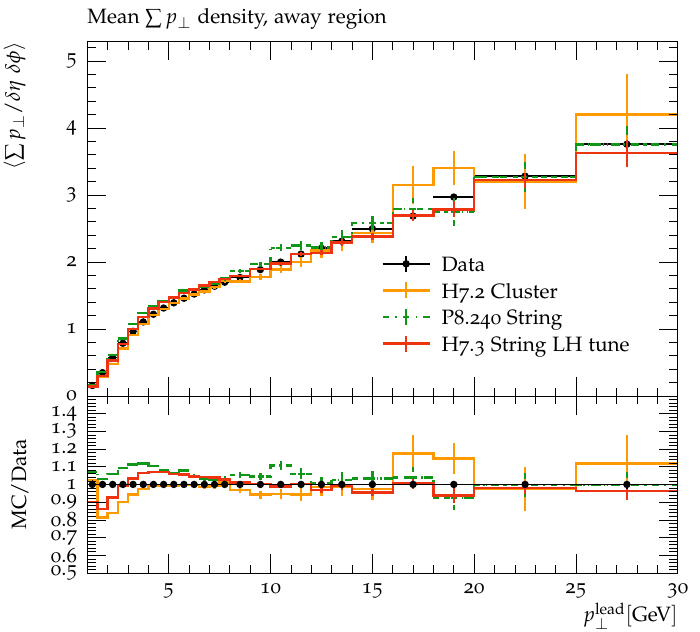}
    %\caption{\bf $\mathbf{\sqrt{s}}$ = 13 TeV}
\end{subfigure}
\caption{\UE observables measured by the ATLAS experiment at 13 TeV proton collisions~\cite{ATLAS:2017blj} which were included in the tuning interpolation. Plots are organised in the columns (from the left): particles in the transverse region, particles in the towards region, and particles in the away region; and in the rows (from the top): the mean average particle $p_{\perp}$ as a function of the multiplicity, the mean average $p_{\perp}$ as a function of the leading object $p_{\perp}^{\mathrm{lead}}$, and the mean scalar sum of particle $p_{\perp}$ per solid angle bin as a function of the leading object $p_{\perp}^{\mathrm{lead}}$. }
\label{fig:UE_2}
\end{figure*}

\begin{figure*}[]
\centering
\begin{subfigure}{.3\textwidth}
    \includegraphics[width=\linewidth]{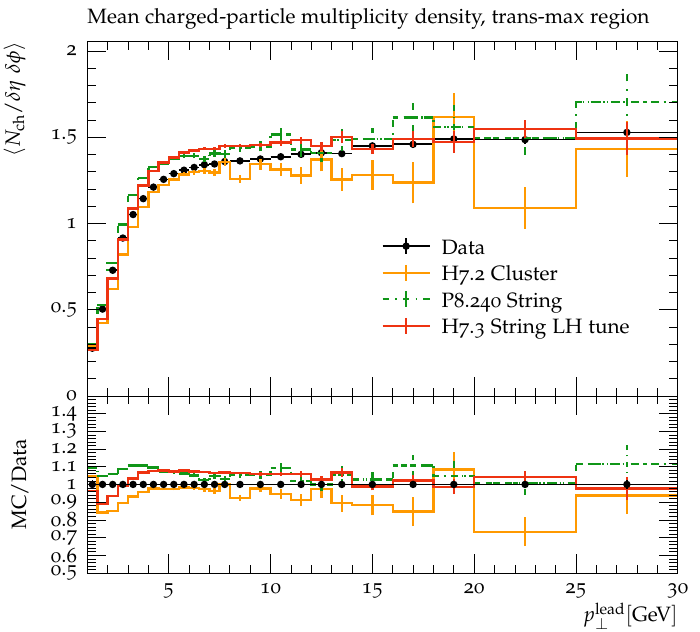}      
    \caption{}
    \label{fig:UE_3:a}
\end{subfigure}
\begin{subfigure}{.3\textwidth}
    \includegraphics[width=\linewidth]{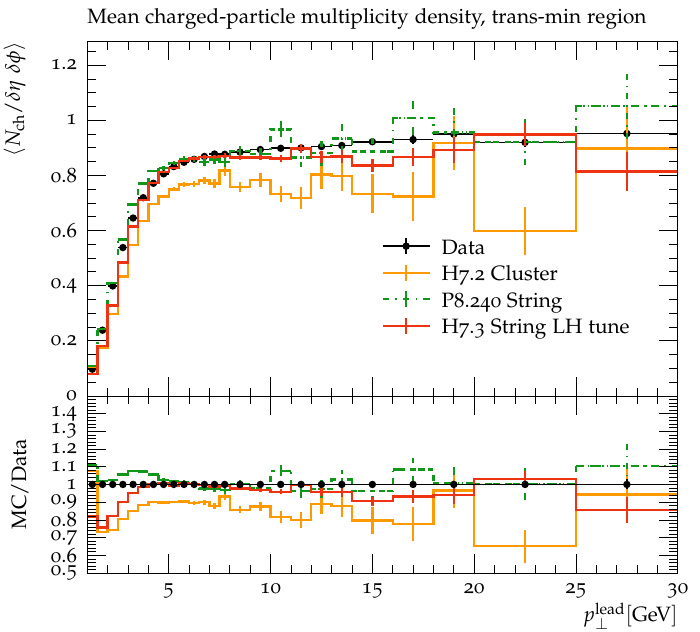}      
    \caption{}
    \label{fig:UE_3:b}
\end{subfigure}
\begin{subfigure}{.3\textwidth}
    \includegraphics[width=\linewidth]{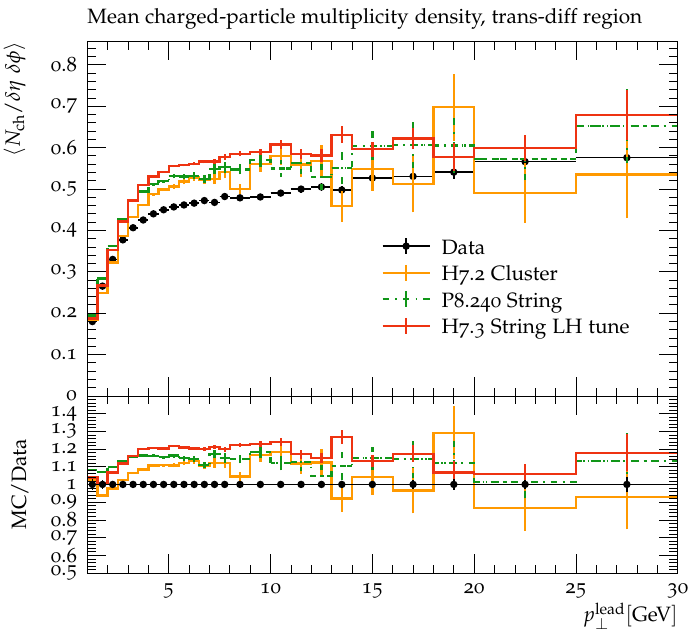}
    \caption{}
    \label{fig:UE_3:c}
\end{subfigure}
        \caption{\UE observables measured by the ATLAS experiment at 13 TeV proton collisions~\cite{ATLAS:2017blj} which were not included in the tuning interpolation. The particle multiplicity per solid angle bin evaluated in two azimuthal plane regions: trans-min and trans-max, as well as after their subtraction, trans-diff.}
\label{fig:UE_3}
\end{figure*}

We compare the \eetune with the other tunes%
\footnote{We consider the \herwigv{7.2} cluster tune here, which is the latest version tuned to hadron data.} 
for hadron collisions at multiple energies compared to LHC (0.9, 7 and 13 TeV) and Tevatron (1.8 TeV) data as shown in Figs.~\ref{fig:UE_1}-\ref{fig:MB_2}.
We are primarily interested in two types of measurements of the final-state at hadron colliders -- \UE and \MB, as they provide a good probe for the sensitivity to relevant phenomenological parameters steering the global event properties.

\subsubsection{\UE}
First, we focus on observables that describe particle distributions in the azimuthal angle plane $\phi$ calculated with respect to the leading object, which is set at $\phi = 0$. 
In these measurements, the leading object is a charged particle (i.e. a track) with transverse momentum greater than $1$ GeV and pseudorapidity $|\eta| < 2.5$. 
\UE analyses \cite{ATLAS:2010kmf} (0.9 TeV and 7 TeV) and \cite{ATLAS:2017blj} (13 TeV) provide sufficiently similar conditions of the experimental trigger and comparable final-state phase-spaces allowing us to perform a reliable tune of the UE parameters. 

Fig.~\ref{fig:UE_1} shows the charged particle multiplicity $N_\mathrm{ch}$ and the transverse momentum $p_\perp$ (the mean sum of transverse momenta of particles) as a function of $\phi$ measured at the LHC. 
We can see that the predictions of the \eetune lie within the default \herwigv{7} cluster and \pythiav{8} string tunes. 
It follows a shape similar to the \herwigv{7} cluster tune but shows slightly better agreement with the data except in the $\phi$ $\to$ 0 region (i.e. in the direction of the leading particle) where it approaches the \pythiav{8} tune. 
In this region, the difference between the \eetune and the default \herwigv{7} cluster tune increases up to 20\%, while in most of the bins (roughly for $\phi$ > 0.3 rad) their ratio is within 10\%.
This is true for all three energies, and the width of the envelope for the different predictions remains constant.

Next, we take a closer look at more UE distributions (Figs.~\ref{fig:UE_2},~\ref{fig:UE_3},~\ref{fig:UE_aux_09},~\ref{fig:UE_aux_18}, and~\ref{fig:UE_aux_7}), where we use the standard definition of three kinematic regions in the azimuthal angle plane used by \UE analyses. 
The {\it towards} region is around the leading object up to |$\phi$| < $\pi$/3, the {\it away} region is |$\phi$| > 2$\pi$/3 (usually containing a recoiling hard object), and the remaining {\it transverse} region, where we expect the highest sensitivity to \UE modelling. 

Fig.~\ref{fig:UE_2} shows the mean average transverse momentum spectra as a function of either the charged multiplicity
(the first row) or $p_\perp$ of the leading particle and the mean scalar sum of $p_\perp$ of all tracks as a function of $p_\perp$ of the leading particle (second and third rows). In the three columns, we compare the measurements in {\it transverse}, {\it towards} and {\it away} regions respectively.
We observe that all tunes give good predictions for the observables at 13 TeV, however, we notice that \eetune slightly improves the predicted spectra of the \herwigv{7} cluster tune for higher values of multiplicities $N_\mathrm{ch}$, where we expect higher sensitivity to colour reconnection and hadronization. 
More UE observables for $pp$ collisions at 0.9, 7 and 13 TeV which we used to obtain the \eetune can be seen in Fig.~\ref{fig:UE_aux_09},~\ref{fig:UE_aux_18}, and~\ref{fig:UE_aux_7}, however, the conclusions stay the same. 

We validate the \eetune with the distributions shown in Fig.~\ref{fig:UE_3}, where we present observables that were not taken into account in the \professor{} fit.
This allows us to independently assess the extrapolation capabilities of the tune.
%and the robustness of the simulation parameters.
The mean charged particle multiplicity as a function of the leading particle $p_\mathrm{\perp}^{\mathrm{lead}}$ exhibits a different behaviour depending on the choice of the part of the transverse region in the azimuthal plane.
The region with the larger scalar sum of charged particle $p_\perp$ is labelled {\it trans-max}, the other region is {\it trans-min}. 
The motivation for this division is to isolate the regions most sensitive to \UE as possible. 
While the {\it trans-max} region may contain recoiled particles from the leading (usually hard) object, the {\it trans-min} is more likely to contain ``Underlying Event'' particles. 
Fig.~\ref{fig:UE_3:a} shows that all tunes give good predictions for the relatively harder part of the event generation with a deviation mostly less than 10 \% from the data. 
\herwigv{7} cluster tune slightly underestimates the ATLAS data in the distribution in Fig.~\ref{fig:UE_3:b}, while the \eetune and the \pythiav{8} string tune agree well with the data.
The difference between these observables, i.e. $\text{\it trans-diff} = \text{\it trans-max}~ - ~\text{\it trans-min},$ is constructed to emphasise the harder part of the event, usually formed during the parton shower evolution, by subtracting the softer event component, which stems mostly from MPI populating the entire phase space in no preferred direction. This $\text{\it trans-diff}$ distribution is shown in Fig.~\ref{fig:UE_3:c}. 
We can see that all the tunes overestimate the data by $\sim 10-20\%$ with the \herwigv{7} cluster tune being slightly better among the rest.

\subsubsection{\MB}

\begin{figure*}[]
\captionsetup[subfigure]{labelformat=empty}
\begin{subfigure}{0.32\textwidth}
    \captionsetup{labelformat=empty}
    \caption[]{\bf $\mathbf{\sqrt{s}}$ = 0.9 TeV}        \includegraphics[width=1\textwidth]{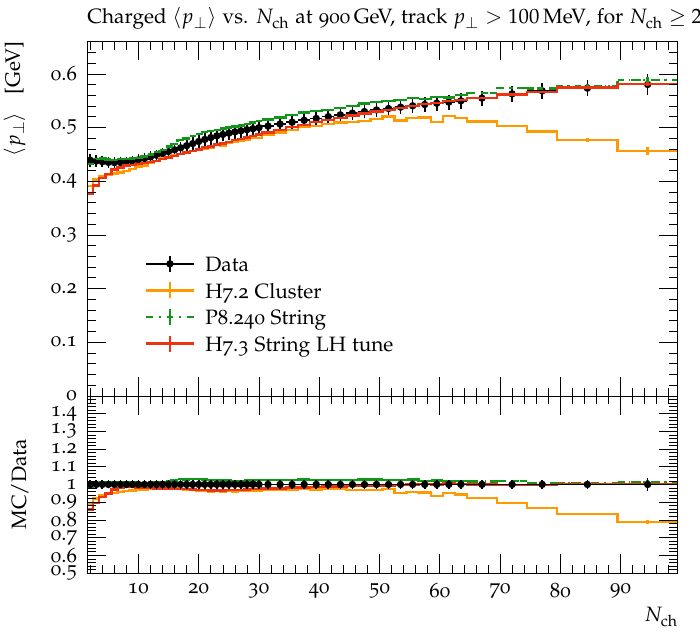}
\end{subfigure}
\begin{subfigure}{0.32\textwidth}        
    \captionsetup{labelformat=empty}
    \caption[]{\bf $\mathbf{\sqrt{s}}$ = 7 TeV}        \includegraphics[width=1\textwidth]{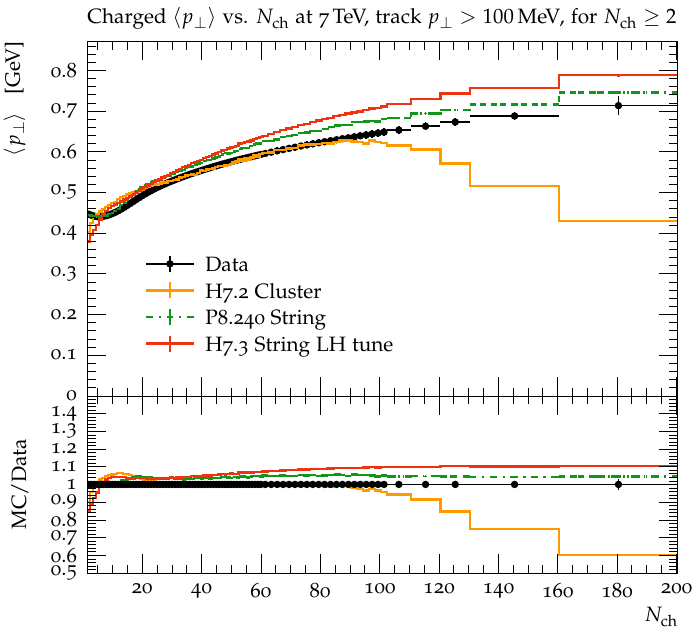}
\end{subfigure}
\begin{subfigure}{0.32\textwidth}        
    \captionsetup{labelformat=empty}
    \caption[]{\bf $\mathbf{\sqrt{s}}$ = 13 TeV}
        \includegraphics[width=1\textwidth]{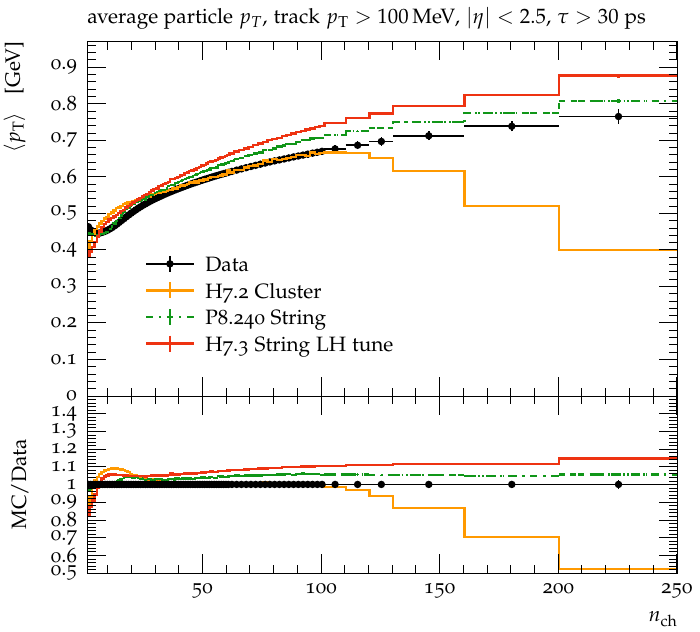}
\end{subfigure}\\
\begin{subfigure}{0.32\textwidth}
        \includegraphics[width=1\textwidth]{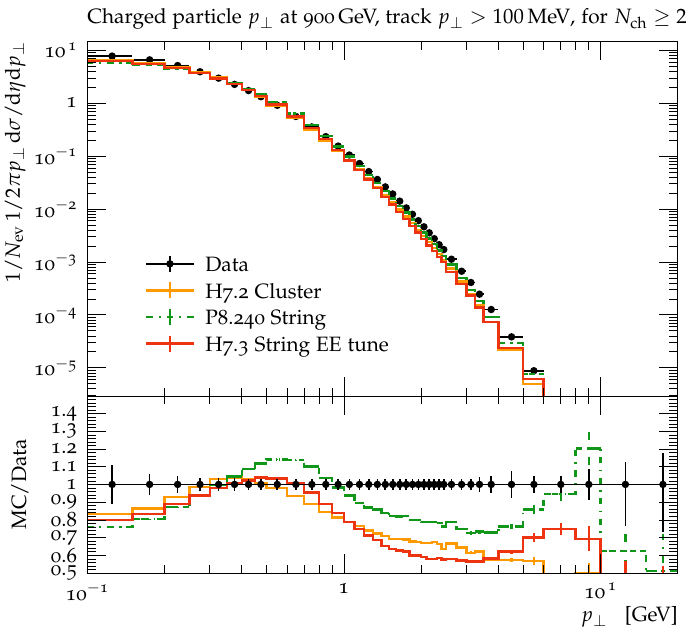} 
\end{subfigure}
\begin{subfigure}{0.32\textwidth} 
        \includegraphics[width=1\textwidth]{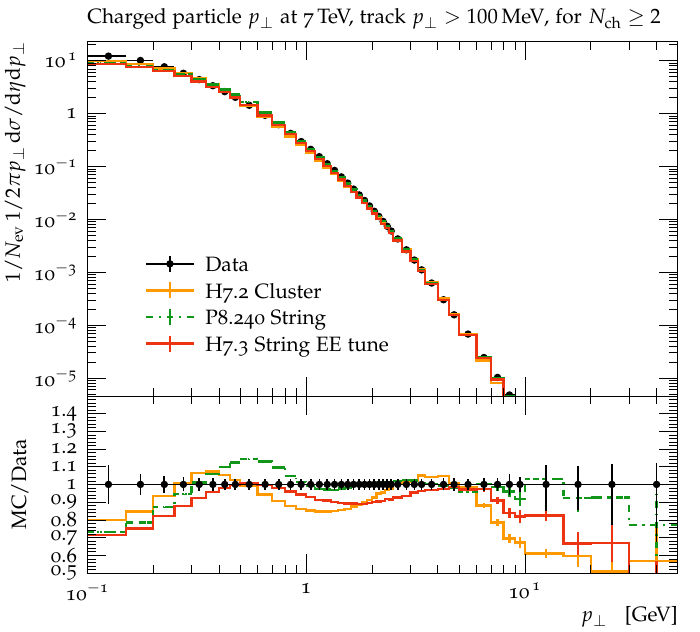}
\end{subfigure}
\begin{subfigure}{0.32\textwidth}
        \includegraphics[width=1\textwidth]{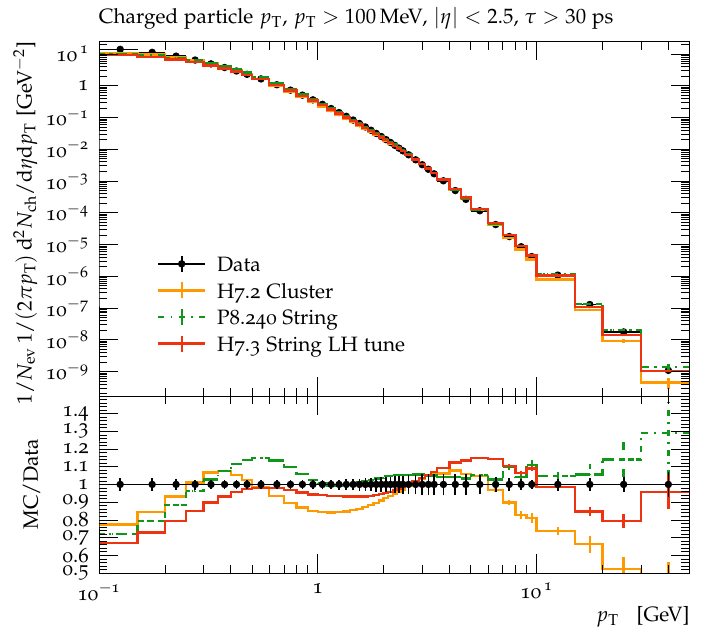}
\end{subfigure}\\
\begin{subfigure}{0.32\textwidth}
        \includegraphics[width=1\textwidth]{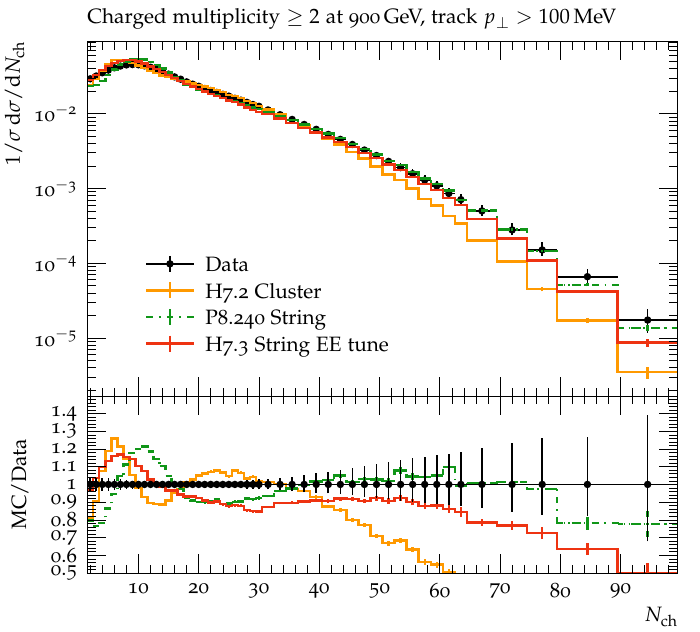} 
\end{subfigure}
\begin{subfigure}{0.32\textwidth}
        \includegraphics[width=1\textwidth]{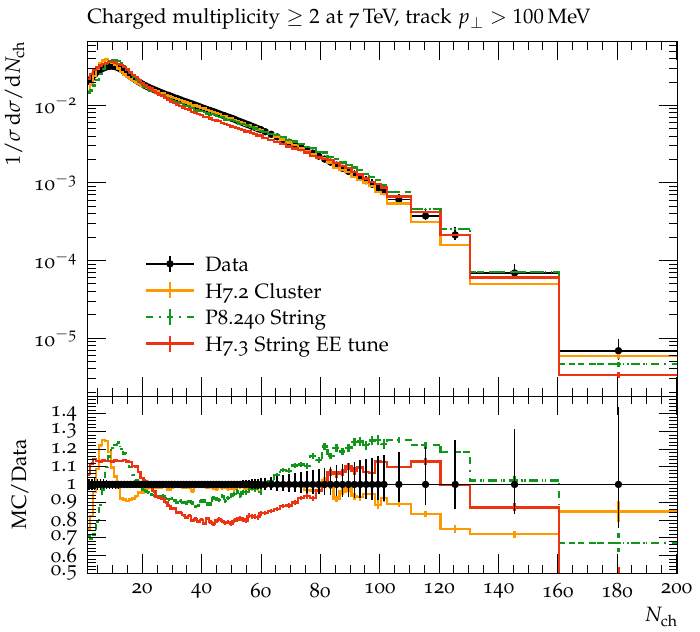}
\end{subfigure}
\begin{subfigure}{0.32\textwidth}
        \includegraphics[width=1\textwidth]{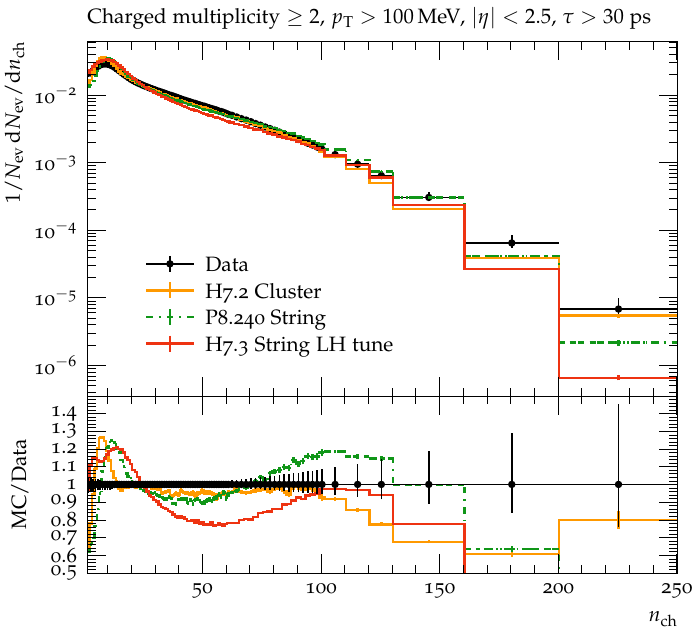}
\end{subfigure}    \\
\begin{subfigure}{0.32\textwidth}
        \includegraphics[width=1\textwidth]{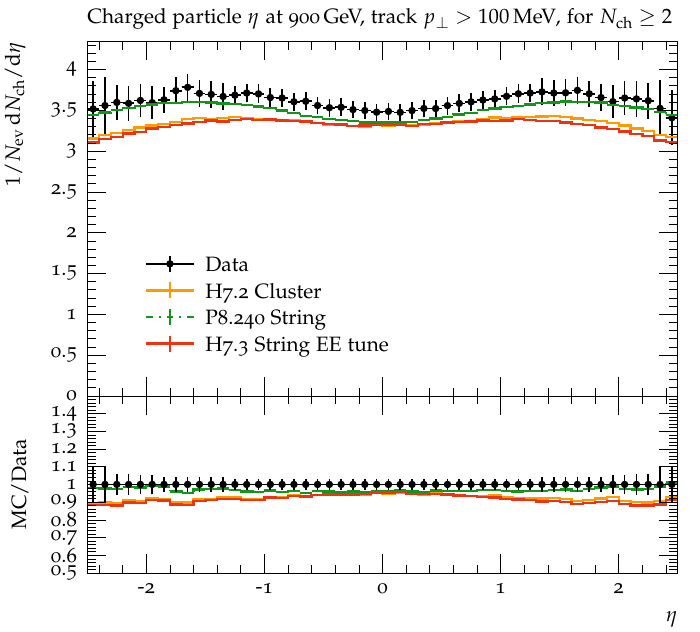} 
\end{subfigure}
\begin{subfigure}{0.32\textwidth}        
        \includegraphics[width=1\textwidth]{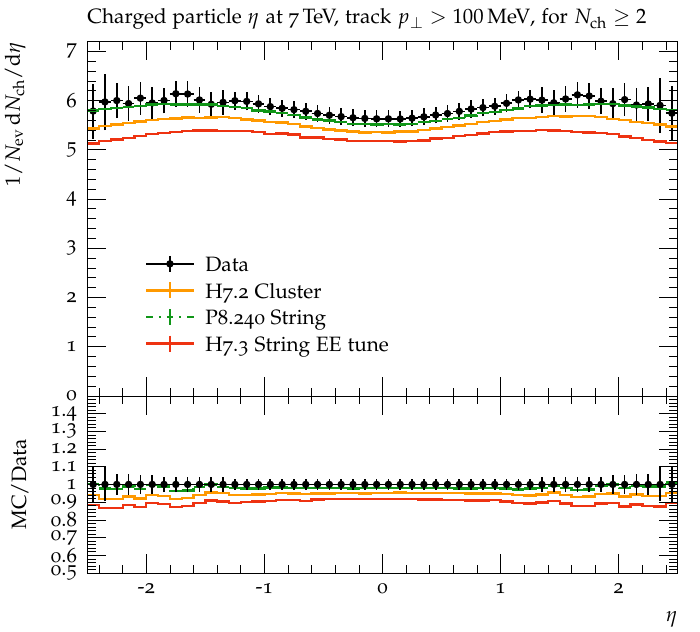}
\end{subfigure}
\begin{subfigure}{0.32\textwidth}        
        \includegraphics[width=1\textwidth]{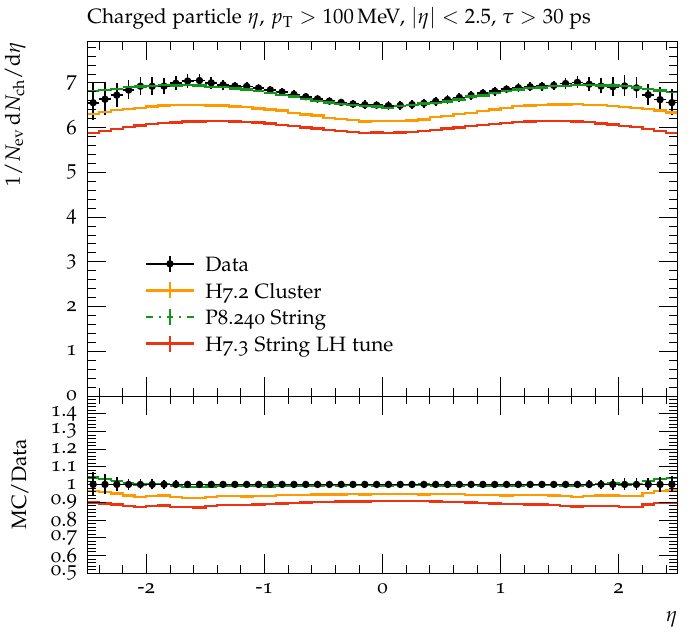}
\end{subfigure}
\caption{\MB observables measured by the ATLAS experiment at 0.9 and 7 TeV~\cite{ATLAS:2010jvh}, and at 13 TeV~\cite{ATLAS:2016zba} proton collisions which were included in the tuning interpolation. Plots are organised in columns according to collision energies (from the left): 0.9 TeV, 7 TeV, and 13 TeV; and in the rows (from the top): mean particle $p_{\perp}$ as a function the particle multiplicity, particle $p_{\perp}$ distribution, particle multiplicity distribution, and particle pseudorapidity distribution. All three analyses consider tracks with $p_{\perp}$ > 100 MeV.}
\label{fig:MB_1}
\end{figure*}     

\begin{figure*}[]
        \includegraphics[width=0.32\textwidth]
        {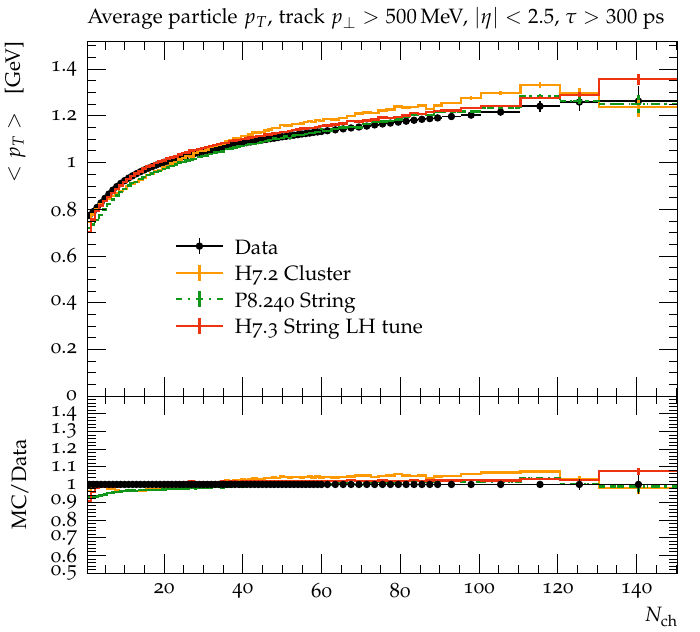}
        \includegraphics[width=0.32\textwidth]
        {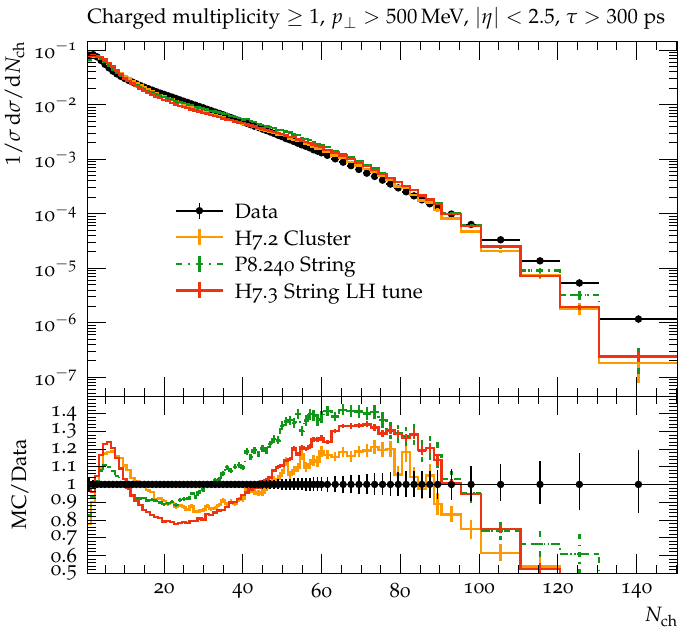}
        \includegraphics[width=0.32\textwidth]{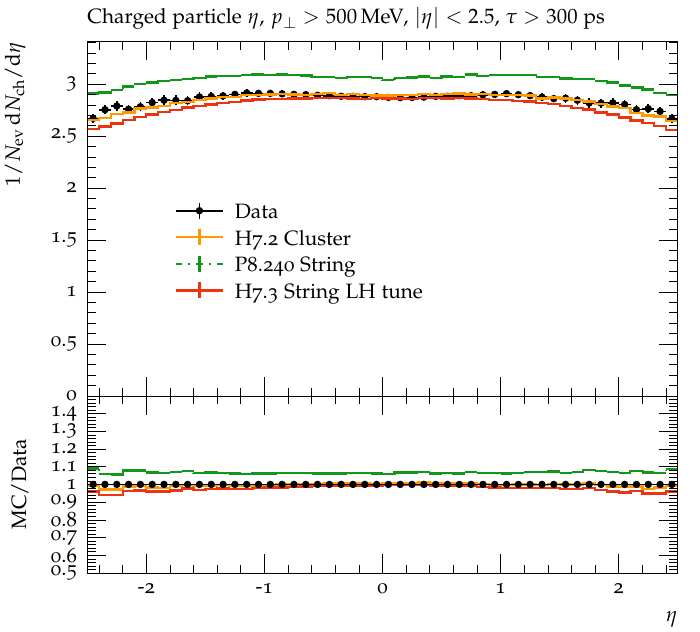}
        \caption{\MB observables measured by the ATLAS experiment at 13 TeV~\cite{ATLAS:2016zkp} proton collisions which were not included in the tuning interpolation. From the left: mean $p_{\perp}$ as a function of particle multiplicity, particle multiplicity distribution, and particle pseudorapidity distribution. Here, particles with $p_{\perp}$ > 500 MeV were considered.}
\label{fig:MB_2}
\end{figure*}

The charged particle distributions used in the \professor{} interpolation for \MB events at the LHC are shown in Figs.~\ref{fig:MB_1}, \ref{fig:MB_aux_09}, \ref{fig:MB_aux_7} and at the Tevatron in Fig.~\ref{fig:MB_aux_18}. 
We can see bigger differences among the predictions of different tunes, even up to $\sim 40\%$ for some kinematic regions, which is in contrast to the predictions for \UE which were within $\sim 10-20\%$ of the data.
This is to be expected since the \UE measurements trigger on a hard object, which is not the case in the \MB data. Therefore, it is more sensitive to soft physics.
 
In general, it can be seen here that the \eetune is competitive with the \herwigv{7} cluster and the \pythiav{8} string tunes. Some notable differences arise in certain regions of these distributions as discussed below, which also highlights the systematic uncertainties of different hadronization models along with their combinations with different parton shower algorithms.

Fig.~\ref{fig:MB_1} displays four MB observables (in rows) for three collision energies (in columns). 
It starts with the average charged particle $p_\perp$ as a function of the charged multiplicity (the first row). 
These distributions show that the \eetune in the low $N_\mathrm{ch}$ region ($N_\mathrm{ch}$ $<$ 8) follows the \herwigv{7} cluster tune which slightly underestimates the data, most likely due to the same diffractive model. 
On the other hand, in the high $N_\mathrm{ch}$ region, the \eetune follows the \pythiav{8} string tune prediction and performs significantly better than the \herwigv{7} cluster tune.
This is most likely due to the different colour reconnection models used in the \herwigv{7} string and cluster tunes. 
We observe a similar behaviour for all LHC energies considered in this study, with the disagreement with the data for the high $N_\mathrm{ch}$ events increasing with higher collision energies. 
An overall disagreement of about 10-30\% among all generator tunes and the data can be seen in Tevatron collisions at 1.8 TeV in Fig.~\ref{fig:MB_aux_18}.

The Monte Carlo distributions for particle $p_{\perp}$ and $N_{\mathrm{ch}}$ (second and third rows in Fig.~\ref{fig:MB_1}) exhibit significant variations, with deviations from the data up to 30-40\%.
On the other hand, the charged particle pseudorapidity distribution (last row in Fig.~\ref{fig:MB_1}) shows no improvement for \herwigv{7} with the \eetune, which is farthest from the data. 
However, the relative difference between the tunes remains small, around 10$\%$. 

To demonstrate the variations in Monte Carlo predictions we also present Fig.~\ref{fig:MB_2} with three observables (the average transverse momentum of charged particles, the charged multiplicity, and the charged particle pseudorapidity) that were not included in the \eetune and which are equivalent to those in the 13 TeV column in Fig.~\ref{fig:MB_1}. 
The difference lies in the track selection criteria. Observables shown in Fig.~\ref{fig:MB_2} counts tracks with $p_{\perp}$ > 500 GeV and lifetime $\tau$ > 300 ps~\cite{ATLAS:2016zkp}, while the distributions shown for 13 TeV collisions in Fig.~\ref{fig:MB_1} were obtained for a lower threshold of $p_{\perp}$ > 100 GeV and lifetime $\tau$ > 30 ps.
This comparison thus provides insight on the effect of the different kinematic and lifetime constraints on the predictions. 
For instance, the \herwigv{7} cluster tune in Fig.~\ref{fig:MB_2} overestimates the data for the average $p_{\perp}$ versus multiplicity of charged particles, contrary to Fig.~\ref{fig:MB_1} which shows the opposite, likely due to the higher $p_{\perp}$ cut.
The charged particle multiplicity exhibits comparable disagreement along the data points. The only difference is that the \eetune has a smooth single peak at low $N_{\mathrm{ch}}$ in Fig.~\ref{fig:MB_2}, while a double peak can be seen in Fig.~\ref{fig:MB_1}.  
The last observable in Fig.~\ref{fig:MB_2} is the pseudorapidity distribution. 
Here, the \eetune and the \herwigv{7} cluster tune describe the data very closely while the \pythiav{8} tune overestimates the data by a constant 10\%. 
It exhibits another counter-trend with respect to the 100 MeV track pseudorapidity distributions in Fig.~\ref{fig:MB_1}, where the \eetune deviates most from the data.

Another set of distributions comparing the different tunes at 0.9 and 7 TeV collision energies are shown in Fig.~\ref{fig:UEMB_aux_09} and Fig.~\ref{fig:UEMB_aux_7} in the Appendix. 
Qualitative conclusions drawn from the 13 TeV observables also hold at lower energies.
The charged particle $p_{\perp}$ distributions are well modelled by all the tunes in intermediate bins of the distributions; however, fall off at very low and high $p_{\perp}$ regions. 
Charged multiplicity distributions show large differences among the three tunes, demonstrating the systematic uncertainties associated with hadronization models in different regions. 

\begin{figure}[]
\centering
\includegraphics[width=0.4\textwidth]{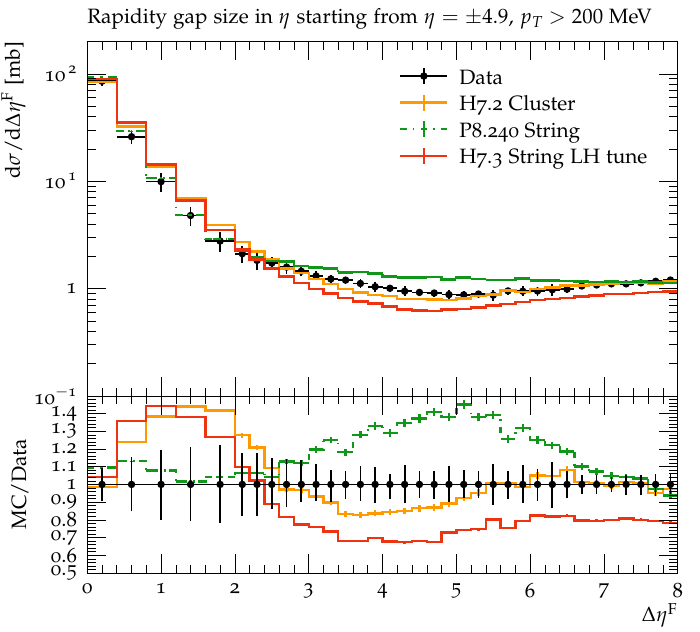}
\caption{Forward rapidity gaps as measured at ALTAS for $\sqrt{s}$ = 7 TeV \cite{ATLAS:2012djz}.}
\label{fig:MB_diff}
\end{figure}

We also compare the \eetune to the default tunes of \herwigv{7} and \pythiav{8} for diffraction processes as well, to assess the modelling capabilities of the diffraction phenomena. 
Despite the fact that we did not use dedicated diffraction-sensitive observables for our \eetune, we obtain reasonable agreement with the \herwigv{7} cluster tune up to $\sim 20\%$ as shown in Fig.~\ref{fig:MB_diff}.
The distribution shown here is the forward rapidity gap as measured at the LHC~\cite{ATLAS:2012djz}.
This provides us another insight into the uncertainties associated with different non-perturbative models. 
We observe here that the low $\mathrm{\Delta}\eta^\mathrm{F}$ region, i.e. $\mathrm{\Delta}\eta^\mathrm{F} < 2.5$ displays non-diffractive events, which are better modelled by the \pythiav{8} tune and overestimated by the \herwigv{7} cluster and the \herwigv{7} string tunes. 
In regions of larger $\mathrm{\Delta}\eta^\mathrm{F}$ i.e. $\mathrm{\Delta}\eta^\mathrm{F} > 2.5$,
where diffraction plays the dominant role, the trend is reversed, i.e. the \pythiav{8} tune overestimates the data and the \herwigv{7} with the string model underestimates the data, while the \herwigv{7} with the cluster model agrees relatively well with the data.

\begin{figure*}[hbt]
\begin{subfigure}{0.32\textwidth}
        \includegraphics[width=1.0\textwidth]
        {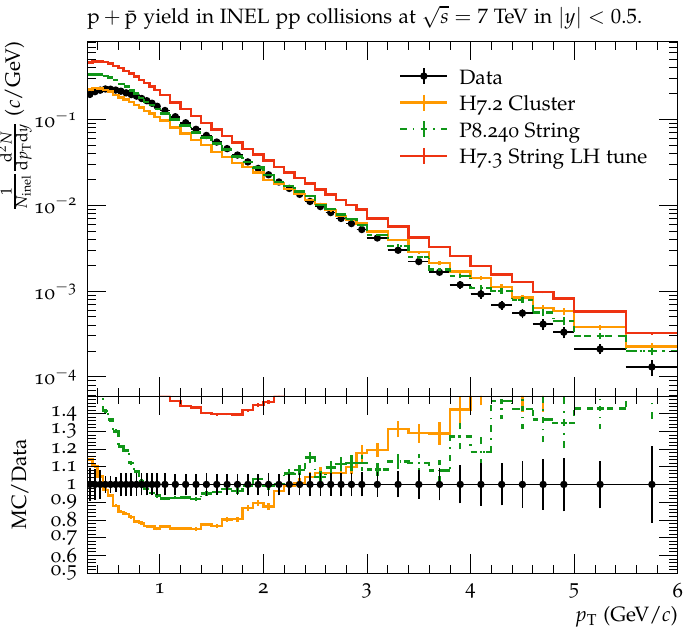}
        \caption{}
\end{subfigure}
\begin{subfigure}{0.32\textwidth}
        \includegraphics[width=1.0\textwidth]{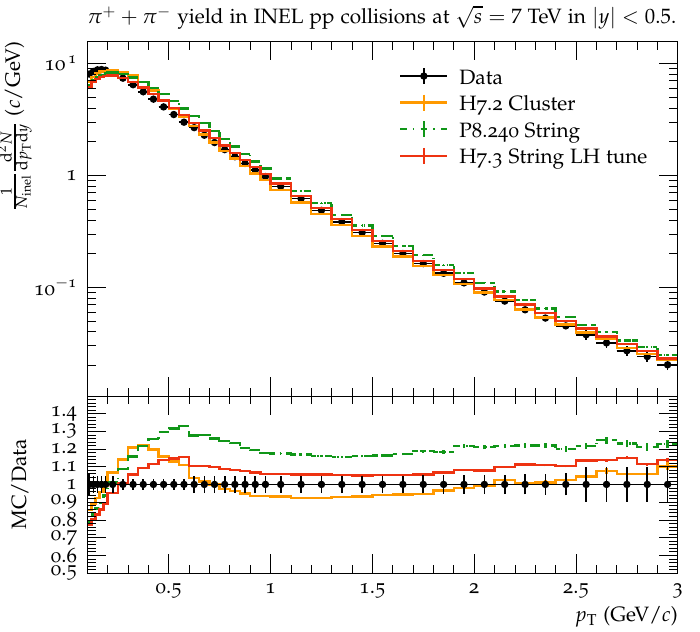}
        \caption{}
\end{subfigure}        
\begin{subfigure}{0.32\textwidth}
        \includegraphics[width=1.0\textwidth]{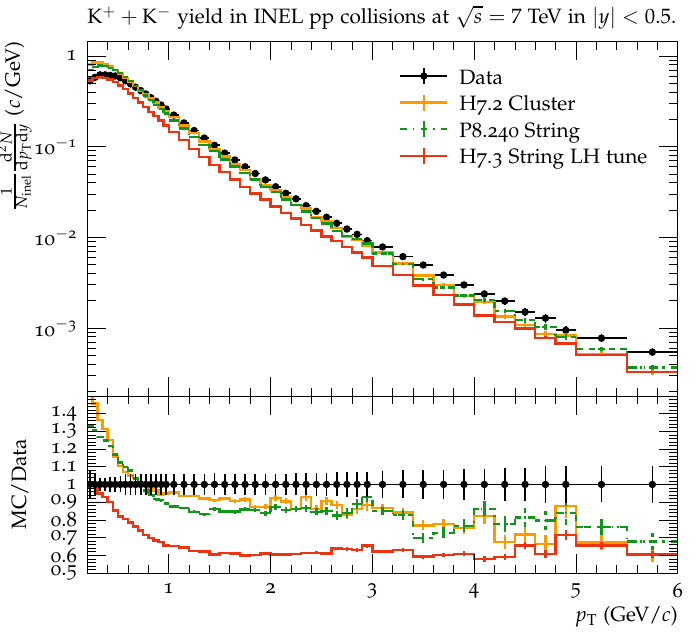}
        \caption{}
\end{subfigure}\\
\begin{subfigure}{0.245\textwidth}        
        \includegraphics[width=1.0\textwidth]{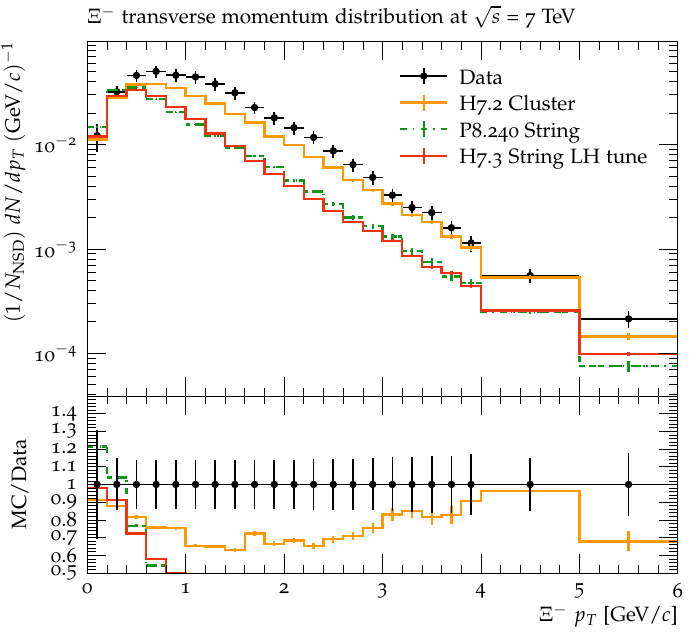}
        \caption{}
\end{subfigure}      
\begin{subfigure}{0.245\textwidth}
        \includegraphics[width=1.0\textwidth]{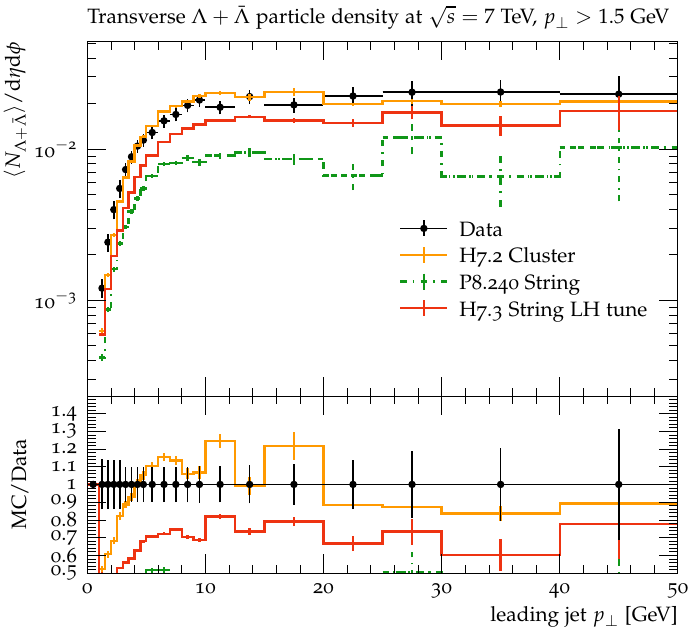}
        \caption{}
\end{subfigure}        
\begin{subfigure}{0.245\textwidth}
        \includegraphics[width=1.0\textwidth]{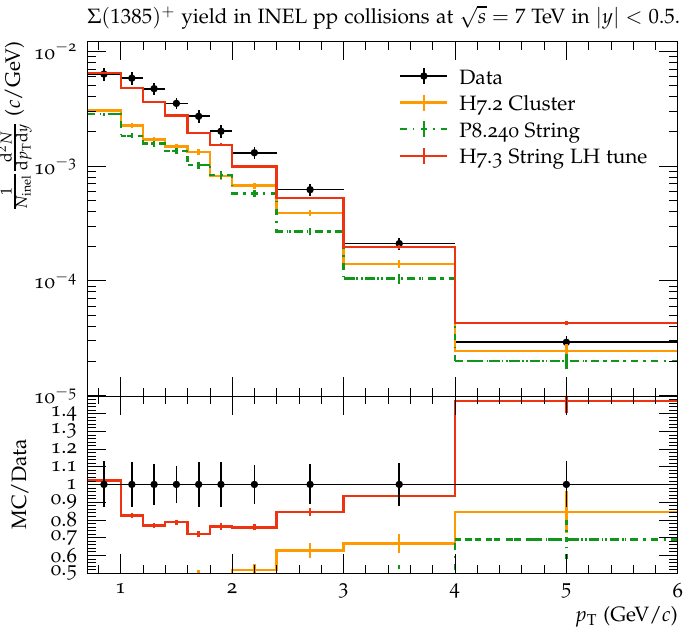}
        \caption{}
\end{subfigure}        
\begin{subfigure}{0.245\textwidth}
        \includegraphics[width=1.0\textwidth]{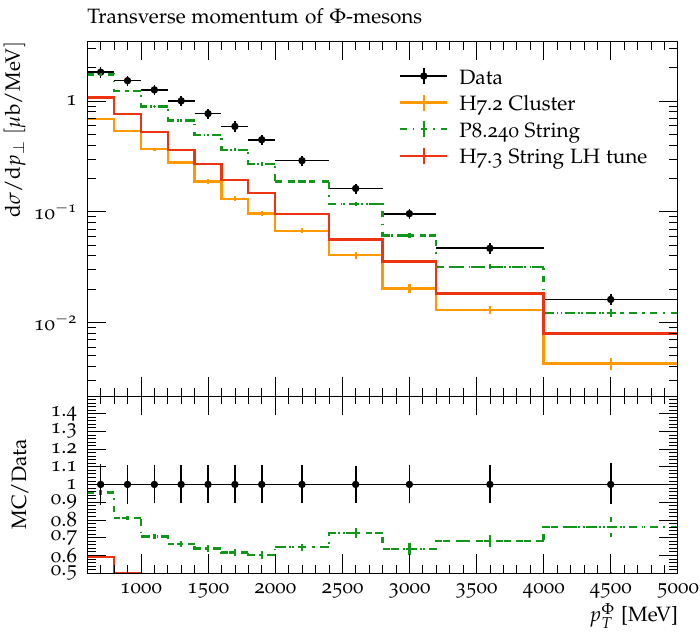}
        \caption{}
\end{subfigure}        
\caption{Identified particle transverse momentum spectra for proton-proton collisions at 7 TeV. Individual plots are: relative yields of (a) protons plus anti-protons,  (b) pions plus anti-pions, and (c) kaons plus anti-kaons in the central rapidity bin as measured by the ALICE experiment \cite{ALICE:2015ial}; (d) a relative yield of $\Xi^-$ as measured by the CMS experiment \cite{CMS:2011jlm}; (e) the number of $\Lambda$ and $\bar{\Lambda}$ mesons with respect to the leading jet $p_{\perp}$ measured by the CMS experiment \cite{CMS:2011fsn}; (f) a relative yield of $\Sigma(1385)^+$ mesons measured by the ALICE experiment \cite{ALICE:2014zxz}; (g) a differential cross section for the $\Phi$ meson production as measured by the LHCb experiment \cite{LHCb:2011ijs}.}
\label{fig:MB_3}
\end{figure*}

\subsubsection{Flavour}
So far, we have tuned the fragmentation parameters that govern the flavour composition of partons using data exclusively from LEP, without incorporating constraints from hadron collisions. 
Thus, we assess and validate the accuracy of the specific-flavour particle composition, using identified particle spectra from the LHC. 
To this end, we focus on comparing the spectra of protons, pions, kaons, and other mesons against experimental data. 
This analysis helps to determine whether the flavour-dependent aspects of fragmentation, constrained at LEP, remain consistent when applied to the hadronic environment of the LHC, where additional QCD effects come into play. 
Fig.~\ref{fig:MB_3} provides a closer look at seven observables selected from four dedicated measurements of proton collisions at 7 TeV: ALICE \cite{ALICE:2015ial,ALICE:2014zxz}, CMS \cite{CMS:2011jlm}, and LHCb \cite{LHCb:2011ijs}. 
The selection covers the absolute or relative yields of protons and mesons ($\pi$, $K$, $\Xi$, $\Lambda$, $\Sigma$, $\Phi$) as a function of the particle transverse momentum. 
One exception is the spectrum of the $\Lambda$ meson, which is related to the transverse momentum of the leading jet.

It can be concluded from the results in Fig.~\ref{fig:MB_3} that none of the tunes can describe the flavour data well, and the non-perturbative systematic uncertainties associated with these should be taken into consideration in future phenomenological and experimental studies. 
The disagreement among the different models spans almost constantly over most of the $p_{\perp}$ range, and its overall magnitude reaches up to 50\%. 
This level of uncertainty can not be easily overcome by additional tuning of flavour-related fragmentation parameters, as the improvement in one spectrum undoubtedly leads to the deterioration of another spectrum.
Instead, further improvements in the underlying phenomenological models would be required to bridge these large discrepancies and provide more reliable predictions.

\begin{figure*}[htp]
\begin{subfigure}{.32\textwidth}
        \includegraphics[width=\textwidth]{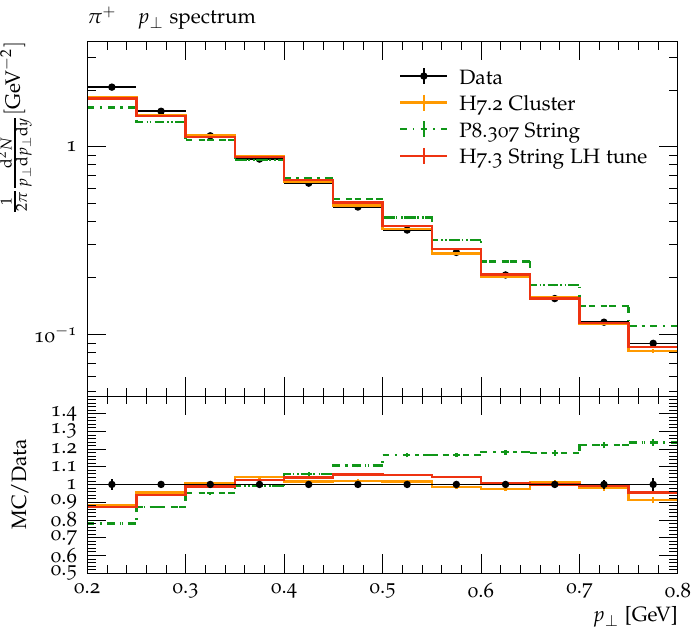}
        \caption{}
        \label{fig:15a}
\end{subfigure}
\begin{subfigure}{.32\textwidth}
        \includegraphics[width=\textwidth]{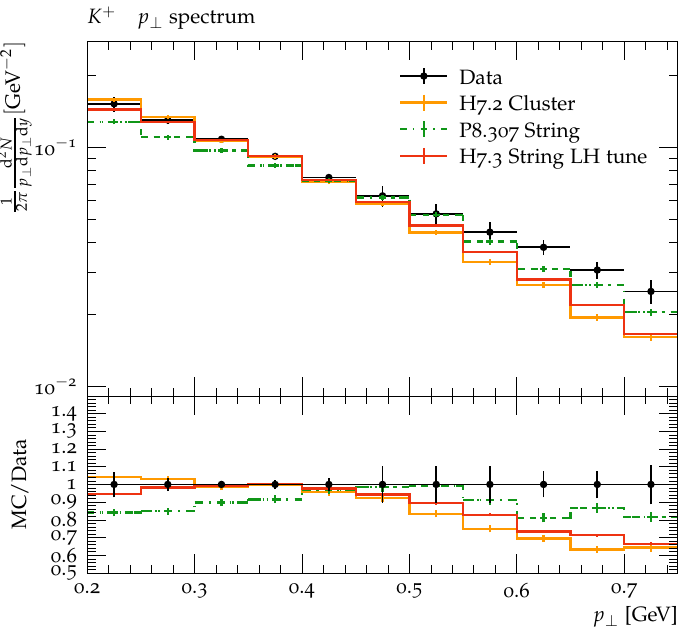}
        \caption{}
        \label{fig:15b}
\end{subfigure}
\begin{subfigure}{.32\textwidth}
        \includegraphics[width=\textwidth]{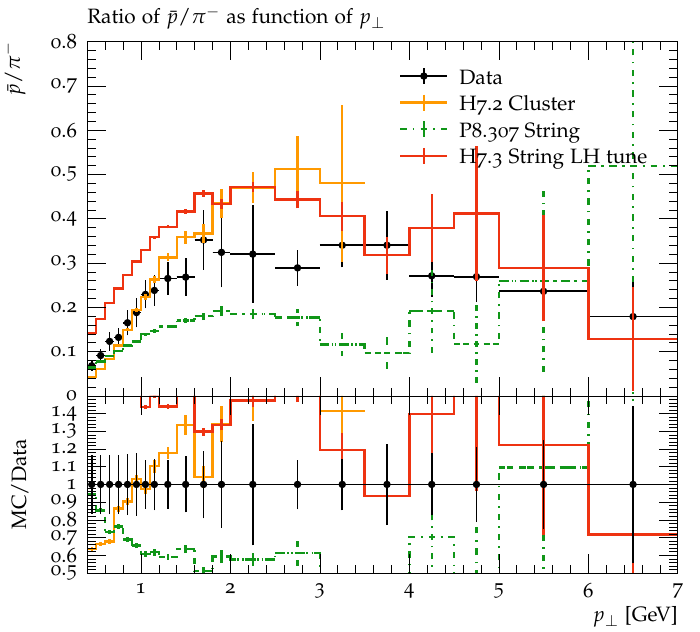}
        \caption{}
        \label{fig:15c}
\end{subfigure} \\
\begin{subfigure}{.32\textwidth}
        \includegraphics[width=\textwidth]{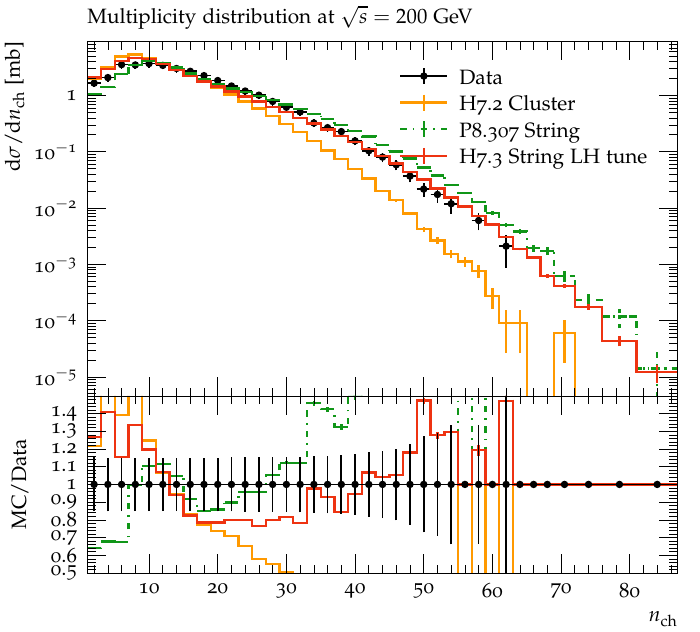}
        \caption{}
        \label{fig:15d}
\end{subfigure}
\begin{subfigure}{.32\textwidth}
        \includegraphics[width=\textwidth]{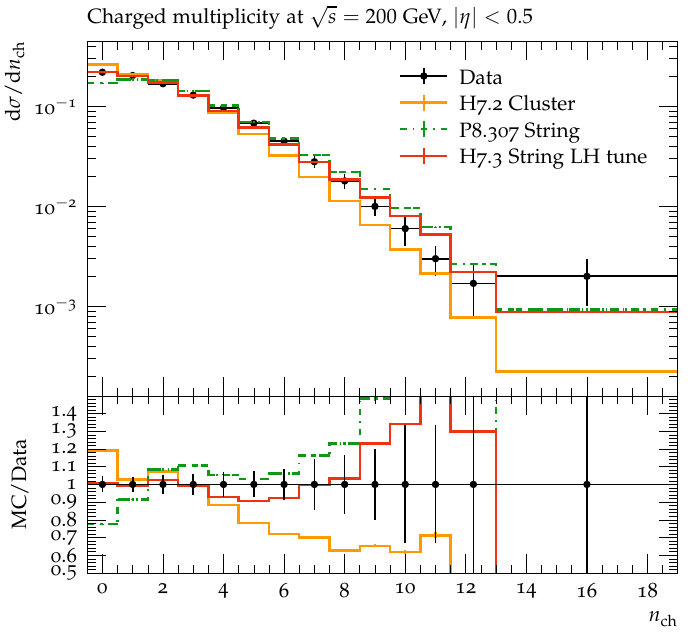}
        \caption{}
        \label{fig:15e}
\end{subfigure}
\begin{subfigure}{.32\textwidth}
        \includegraphics[width=\textwidth]{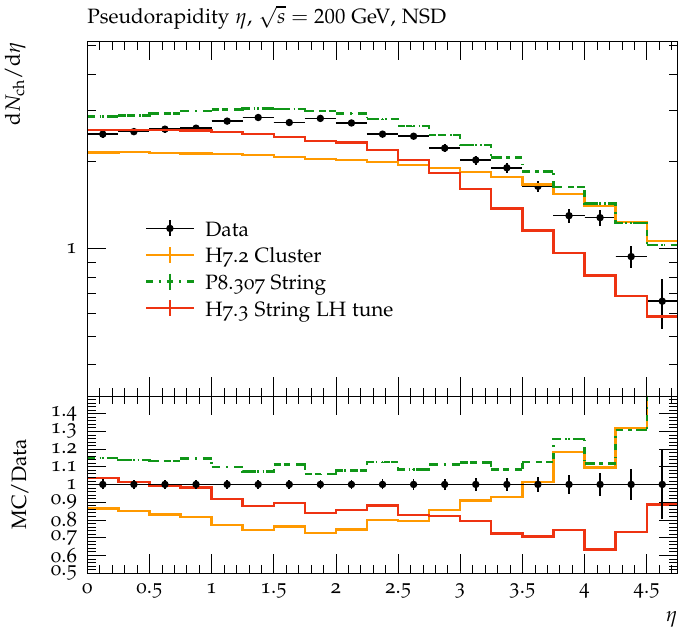}
        \caption{}
        \label{fig:15f}
\end{subfigure}
\caption{Observables measured at STAR \cite{STAR:2006xud,STAR:2008med} and SppS (using UA1-UA5 detectors) \cite{UA5:1986yef, UA5:1988gup, UA1:1989bou} experiments at $\sqrt{s}=$ 200 GeV. Individual plots are: Transverse momentum spectrum of (a) $\pi^{+}$ pions, (b) $K^{+}$ kaons, (c) Ratio of anti-protons ($\bar{p}$) to pions ($\pi^{-}$) as a function of $p_{\perp}$, (d) multiplicity distribution, (e) charged multiplicity distribution and (f) pseudorapidity distribution for NSD.}
\label{fig:STAR_SppS_200}
\end{figure*}

As a final check, we also explore the functionality of the \eetune settings at lower energies%
\footnote{We check internally that the \eetune can be used to generate events for lower energies down to 50 GeV.}. 
For instance, we compare the \eetune predictions for observables measured at the STAR (proton-proton collisions) and SppS (proton-antiproton collisions) experiments for $\sqrt{s}$ = 200 GeV, as shown in Fig.~\ref{fig:STAR_SppS_200}. 
The \eetune shows reasonable agreement with the data for this set of observables. Generally, The \eetune follows the \herwigv{7} cluster tune closely in the particle $p_{\perp}$ spectrum distributions in Fig.~\ref{fig:15a} and \ref{fig:15b}, while it is closer to the \pythiav{8} string tune for multiplicity distributions in Fig.~\ref{fig:15d} and \ref{fig:15e}.

%%%%%%%%%%%%%%%%%%%%%%%%%%%%%%%%%%%%%%%%%%%%%%%%%%%%%%%%%%%%%%%%%%%%%%%%%%%%%%%%%%%%%%%%%%%%%%%%%%%%%%%%%%%

\section{Conclusion and outlook}
\label{summary}
In this article, we studied the effects of hadronization in combination with the \AOPS in both lepton-lepton and hadron-hadron collisions. 
Two different hadronization models –- cluster and string, were used in combination with the latest version of AOPS implemented in \herwigv{7.3}.
Since \herwig{} uses only cluster hadronization by default, we utilised \ThePI interface to integrate the modern string hadronization model from \pythia{} within \herwig{}.
In order to study hadron collisions, we have expanded the \ThePI interface to access the colour reconnection model in \pythia{}, which is important for describing LHC data.
These modifications will be available in the \texttt{herwig-bootstrap} installation of the \herwigv{7.4} release.

To compare the predictions of the string model with AOPS, we first performed a dedicated tuning procedure.
This was done in four stages, focused on lepton and hadron collision observables separately. 
For lepton collisions at LEP, we achieved an accurate description of fragmentation and flavour observables. 
We then extended our tuning strategy to include hadron collisions at multiple centre-of-mass energies, including 0.9, 1.8, 7 and 13 TeV. 
This resulted in a tune that was extrapolated across the range of energies, which we call `Les Houches' tune (in short \eetune).

By comparing the new \eetune with the existing \herwig cluster and \pythia string tunes, we were able to see the relative differences between the models and identify regions where the predictions are very close and where they diverge significantly (for example, in identified particle transverse momentum spectra for proton-proton collision, Fig.~\ref{fig:MB_3}).
Therefore, the use of a ``mixture'' of different models to describe perturbative and non-perturbative quantum chromodynamics allows us to roughly assess the uncertainties associated with these effects.
In particular, we can study the uncertainties related to non-perturbative hadronization effects by fixing all the other components of the simulation in \herwig{} and switching the hadronization models.
The \eetune proposed in this paper has already been used for this purpose in a recent research on a comparative study on flavoured jet algorithms~\cite{Behring:2025ilo}.  
Moreover, the \eetune generalises well beyond distributions that have been explicitly tuned and can be used, for example, for research in the STAR experiment, for which a dedicated cluster tune has recently been created \cite{Qureshi:2024eqz}.

A possible continuation of this study would be to perform a similar analysis using the dipole cascade available in \herwigv{7}. 
This would enable the investigation of different systematic effects in \herwigv{7} associated with the use of two distinct hadronization models and two different parton showers.
Such comparisons of different model combinations are a step in the right direction when estimating uncertainties associated with predictions from Monte Carlo generators (it could be also extended to use different colour reconnection models~\cite{Bellm:2019wrh,Kiebacher:2025fye} which were shown to be important for example for $t\bar{t}$ final-states, see~\cite{Argyropoulos:2014zoa}).
Another possible extension of the work would also be to provide a set of systematic variations for a given tune, for example, based on the eigentune method, as in the case of \pythia A14 tune~\cite{ATL-PHYS-PUB-2014-021}.
Finally, many interesting measurements sensitive to hadronization effects have been made at the LHC, for example Lund plane, LHC event shapes etc. that are not yet used in tuning efforts.
This is primarily due to the computing constraints on simultaneously tuning dozens of parameters using data from both the LEP and the LHC. 
However, efforts towards such a tune would be beneficial and is worth attempting in the future, also as new data becomes available at collider experiments.

%%%%%%%%%%%%%%%%%%%%%%%%%%%%%%%%%%%%%%%%%%%%%%%%%%%%%%%%%%%%%%%%%%%%%%%%%%%%%%%%%%%%%%%%%%%%%%%%%%%%%%%%%%%

\section{Acknowledgement}
We are grateful to Andreas Papaefstathiou, Simon Pl\"{a}tzer and James Whitehead for their thoughtful comments on the manuscript.
This work was supported by grant 2019/34/E/ST2/00457 of the National Science Centre, Poland.
AS and PS are also supported by the Priority Research Area Digiworld under the program 
`Excellence Initiative -- Research University'
at the Jagiellonian University in Krakow. MM acknowledges the financial support from Fundamental constituents of matter through frontier technologies (FORTE) project No. CZ.02.01.01/00/22\_008/0004632.
We gratefully acknowledge Polish high-performance computing infrastructure PLGrid (HPC Center: ACK Cyfronet AGH) for providing computer facilities and support within computational grant no.\@ PLG/2024/016990.
AS and PS would like to thank the organisers of the Les Houches PhysTeV 2025 workshop, where the work on the \eetune was finalised and acknowledges CERN TH Department for hospitality while this research was being carried out.

%%%%%%%%%%%%%%%%%%%%%%%%%%%%%%%%%%%%%%%%%%%%%%%%%%%%%%%%%%%%%%%%%%%%%%%%%%%%%%%%%%%%%%%%%%%%%%%%%%%%%%%%%%%
\onecolumn

\appendix
\label{appendix}

\section{Observable weights for tuning interpolation}
\label{appendix_tune}

\begin{table}[h!]
\hspace{28mm}
    \begin{tabular}{c c l}
    Rivet Analysis Identifier & Weight & Observable \\
    \hline
    \multirow{16}{*}{DELPHI\_1996\_S3430090} & 1.0 & In-plane $p_\perp$ in GeV w.r.t. thrust axes (charged) \\
    & 1.0 & Out-of-plane $p_\perp$ in GeV w.r.t. thrust axes (charged) \\        
    & 1.0 & In-plane $p_\perp$ in GeV w.r.t. sphericity axes (charged) \\
    & 1.0 & Out-of-plane $p_\perp$ in GeV w.r.t. sphericity axes (charged) \\
    & 1.0 & Scaled momentum, $x_{\mathrm p} = |p|/|p_{\mathrm{beam}}|$ (charged) \\
    & 1.0 & Log of scaled momentum, $\log(1/x_p)$ (charged) \\
    & 1.0 & 1-Thrust, $1-T$ (charged) \\
    & 1.0 & Thrust major, $M$ (charged) \\
    & 1.0 & Thrust minor, $m$ (charged) \\
    & 1.0 & Oblateness = $M - m$ (charged) \\
    & 1.0 & Sphericity, $S$ (charged) \\
    & 1.0 & Aplanarity, $A$ (charged) \\
    & 1.0 &  Planarity, $P$ (charged)    \\
    & 1.0 &  $C$ parameter (charged)    \\
    & 1.0 &  $D$ parameter (charged)    \\
    & 160.0 &  Mean charged multiplicity    \\
    \hline
    \multirow{4}{*}{OPAL\_1998\_S3780481} &   20.0 & $uds$ events mean charged multiplicity     \\
    & 20.0 & $c$ events mean charged multiplicity     \\
    & 20.0 & $b$ events mean charged multiplicity    \\
    & 60.0 & all events mean charged multiplicity
    \end{tabular}
    \caption{Weights for LEP observables during the first stage of tuning to optimise the fragmentation parameters of \eetune.}
    \label{tab:LEPtuneweights1}
\end{table}

\begin{table}[h!]
   \begin{tabular}{r c l}
    Rivet Analysis Identifier & Weight & Observable \\
    \hline
        \multirow{2}{*}{DELPHI\_2002\_069\_CONF\_603} & 
        1.0 & $b$ quark fragmentation function $f(x_{B}^{\mathrm{weak}})$ \\
        & 1.0 & Mean of $b$ quark fragmentation function $f(x_{B}^{\mathrm{weak}})$ \\
        \hline
        \multirow{2}{*}{PDG\_HADRON\_MULTIPLICITIES} &
        1.0 & $\pi^+$ multiplicity \\
        & 1.0 & $\pi^0$ multiplicity \\
        \hline
        \multirow{26}{*}{PDG\_HADRON\_MULTIPLICITIES\_RATIOS}  &
        6.0 & $\pi^0$ multiplicity \\
        & 6.0 & $K^+$ multiplicity\\
        & 6.0 & $K^0$ multiplicity \\
        & 2.0 & $\eta$ multiplicity    \\
        & 1.0 & $\eta'(958)$ multiplicity     \\
        & 1.0 & $D^+$ multiplicity     \\
        & 1.0 & $D^0$ multiplicity    \\
        & 2.0 & $D_s^+$ multiplicity     \\
        & 1.0 & $B^+$, $B_d^0$ multiplicity     \\
        & 2.0 & $B_s^0$ multiplicity    \\
        & 9.0 & $\rho(770)^0$ multiplicity    \\
        & 9.0 & $\rho(770)^+$ multiplicity     \\
        & 9.0 & $\omega(782)$ multiplicity     \\
        & 2.0 & $K^*(892)^+$ multiplicity     \\
        & 2.0 & $K^*(892)^0$ multiplicity     \\
        & 1.0 & $\phi(1020)$ multiplicity     \\
        & 1.0 & $D^*(2010)^+$ multiplicity     \\
        & 1.0 & $D_s^*(2112)^+$ multiplicity     \\
        & 1.0 & $B^*$ multiplicity     \\
        & 3.0 & $p$ multiplicity     \\
        & 4.0 & $\Lambda$ multiplicity     \\
        & 2.0 & $\Sigma^0$ multiplicity     \\
        & 2.0 & $\Sigma^\pm$ multiplicity     \\
        & 1.0 & $\Xi^-$ multiplicity     \\
        & 1.0 & $\Delta(1232)^++$ multiplicity     \\
        & 1.0 & $\Sigma(1385)^\pm$ multiplicity    \\
        \hline
        \multirow{4}{*}{OPAL\_1998\_S3780481} &
        1.0 & $uds$ events mean charged multiplicity \\
        & 1.0 & $c$ events mean charged multiplicity \\        
        & 1.0 & $b$ events mean charged multiplicity \\
        & 1.0 & all events mean charged multiplicity 
    \end{tabular}
    \caption{Weights for LEP observables during the second stage of tuning to optimise the flavour parameters of \eetune.}
    \label{tab:LEPtuneweights2}
\end{table}

\begin{table}[]
\vspace{1cm}
\hspace{30.6mm}    
    \begin{tabular}{c c l}
    Rivet Analysis Identifier & Weight & Observable \\
    \hline
    \multirow{8}{*}{ATLAS\_2012\_I1204784} &
        1.0 & $\phi_\eta^*$ spectrum, $\mathrm{Z} \to e^+e^-$ (bare)  \\
        & 1.0 & $\phi_\eta^*$ spectrum, $\mathrm{Z} \to e^+e^-$ (bare), $\vert\eta_\mathrm{Z}\vert < 0.8$ \\
        & 1.0 & $\phi_\eta^*$ spectrum, $\mathrm{Z} \to e^+e^-$ (bare), $\vert\eta_\mathrm{Z}\vert < 1.6$ \\
        & 1.0 & $\phi_\eta^*$ spectrum, $\mathrm{Z} \to e^+e^-$ (bare), $\vert\eta_\mathrm{Z}\vert > 1.6$ \\
        & 1.0 & $\phi_\eta^*$ spectrum, $\mathrm{Z} \to e^+e^-$ (dressed) \\
        & 1.0 & $\phi_\eta^*$ spectrum, $\mathrm{Z} \to e^+e^-$ (dressed), $\vert\eta_\mathrm{Z}\vert < 0.8$ \\
        & 1.0 & $\phi_\eta^*$ spectrum, $\mathrm{Z} \to e^+e^-$ (dressed), $\vert\eta_\mathrm{Z}\vert < 1.6$ \\
        & 1.0 & $\phi_\eta^*$ spectrum, $\mathrm{Z} \to e^+e^-$ (dressed), $\vert\eta_\mathrm{Z}\vert > 1.6$ \\
        \hline
        \multirow{2}{*}{ATLAS\_2014\_I1300647} &
        1.0 & $\mathrm{Z} \; p_{\mathrm{T}}$, $\mathrm{Z} \to \mu^+ \mu^-$ (bare), Inclusive   \\
        & 1.0 & $\mathrm{Z} \; p_{\mathrm{T}}$, $\mathrm{Z} \to \mu^+ \mu^-$ (dressed), Inclusive \\
        \hline
        \multirow{2}{*}{CMS\_2012\_I941555}  &
        1.0 & Z boson $p_{\mathrm{T}}$ with dressed muons \\
        & 1.0 & Z boson $p_{\mathrm{T}}$ with dressed muons (low $p_{\mathrm{T}}$) 
    \end{tabular}
    \caption{Weights for DY observables during the third stage of tuning to optimise the intrinsic $k_\perp$ and $\alpha_{\mathrm{S}}^{\mathrm{ISR}}$ of \eetune.}
    \label{tab:DYtuneweights}
\end{table}

\begin{table}[]
\vspace{1cm}
\hspace{30mm}       
    \begin{tabular}{c c l}
    Rivet Analysis Identifier & Weight & Observable \\
        \hline
        \multirow{2}{*}{ALICE\_2010\_S8706239} &
        1.0 & Invariant Yield  \\
        & 1.0 & Avg. transv. momentum vs. $N_{\mathrm{ch}} (0.15 < p_{\perp} < 4 GeV)$ \\
        \hline
        \multirow{2}{*}{ATLAS\_2011\_S8994773} &
        1.0 & Transverse $N$ density vs $p_{\perp}^{\mathrm{clus1}}$ \\
        & 1.0 & Transverse sum $p_{\perp}$ density vs $p_{\perp}^{\mathrm{clus1}}$ \\
        \hline
        \multirow{18}{*}{ATLAS\_2010\_S8918562}  &
        10.0 & Charged particle $\eta$, track $p_{\perp} > 100 \, \mathrm{MeV}, N_{\mathrm{ch}} \geq 2$ \\
        & 1.0 & Charged particle $\eta$, track $p_{\perp} > 100 \, \mathrm{MeV}, N_{\mathrm{ch}} \geq 20$ \\
        & 10.0 & Charged particle $\eta$, track $p_{\perp} > 500 \, \mathrm{MeV}, N_{\mathrm{ch}} \geq 1 $ \\
        & 10.0 & Charged particle $\eta$, track $p_{\perp} > 500 \, \mathrm{MeV}, N_{\mathrm{ch}} \geq 6 $ \\
        & 1.0 & Charged particle $\eta$, track $p_{\perp} > 2500 \, \mathrm{MeV}, N_{\mathrm{ch}} \geq 1 $ \\
        & 1.0 & Charged particle $p_{\perp}$, track $p_{\perp} > 100 \, \mathrm{MeV}, N_{\mathrm{ch}} \geq 2 $ \\
        & 1.0 & Charged particle $p_{\perp}$, track $p_{\perp} > 100 \, \mathrm{MeV}, N_{\mathrm{ch}} \geq 20 $ \\
        & 1.0 & Charged particle $p_{\perp}$, track $p_{\perp} > 500 \, \mathrm{MeV}, N_{\mathrm{ch}} \geq 1 $ \\
        & 1.0 & Charged particle $p_{\perp}$, track $p_{\perp} > 500 \, \mathrm{MeV}, N_{\mathrm{ch}} \geq 6 $ \\
        & 1.0 & Charged particle $p_{\perp}$, track $p_{\perp} > 2500 \, \mathrm{MeV}, N_{\mathrm{ch}} \geq 1 $ \\
        & 1.0 & Charged multiplicity, track $p_{\perp} > 500 \, \mathrm{MeV} $ \\
        & 1.0 & Charged multiplicity $\geq$ 2, track $p_{\perp} > 100 \, \mathrm{MeV} $ \\
        & 1.0 & Charged multiplicity $\geq$ 20, track $p_{\perp} > 100 \, \mathrm{MeV} $ \\
        & 1.0 & Charged multiplicity $\geq$ 6, track $p_{\perp} > 500 \, \mathrm{MeV} $ \\
        & 1.0 & Charged multiplicity $\geq$ 1, track $p_{\perp} > 2500 \, \mathrm{MeV} $ \\
        & 1.0 & Charged mean $p_{\perp}$ vs $N_{\mathrm{ch}}$, $p_{\perp} > 100 \, \mathrm{MeV}, N_{\mathrm{ch}} \geq 2 $ \\
        & 1.0 & Charged mean $p_{\perp}$ vs $N_{\mathrm{ch}}$, $p_{\perp} > 500 \, \mathrm{MeV}, N_{\mathrm{ch}} \geq 1 $ \\
        & 1.0 & Charged mean $p_{\perp}$ vs $N_{\mathrm{ch}}$, $p_{\perp} > 2500 \, \mathrm{MeV}, N_{\mathrm{ch}} \geq 1 $ \\
        \hline
        \multirow{4}{*}{ATLAS\_2010\_S8591806} &
        10.0 &  Charged particle multiplicity as function of $\eta$ \\
        & 1.0 &  Charged particle multiplicity as function of $p_\perp$ \\
        & 1.0 &  Charged particle density \\
        & 1.0 &  Average transverse momentum as function of $N_{\mathrm{ch}}$ \\
        \hline
        \multirow{13}{*}{ATLAS\_2010\_S8894728} &
        1.0 &  Transverse $N_{\mathrm{chg}}$ density vs $p_{\perp}^{\mathrm{trk1}}$ \\
        & 1.0 &  Toward $N_{\mathrm{chg}}$ density vs $p_{\perp}^{\mathrm{trk1}}$ \\
        & 1.0 &  Away $N_{\mathrm{chg}}$ density vs $p_{\perp}^{\mathrm{trk1}}$ \\
        & 1.0 &  Transverse sum $p_\perp$ density vs $p_{\perp}^{\mathrm{trk1}}$ \\
        & 1.0 &  Toward sum $p_\perp$ density vs $p_{\perp}^{\mathrm{trk1}}$ \\
        & 1.0 &  Away sum $p_\perp$ density vs $p_{\perp}^{\mathrm{trk1}}$ \\
        & 1.0 &  Transverse mean $p_\perp$ vs $p_{\perp}^{\mathrm{trk1}}$ \\
        & 1.0 &  Toward mean $p_\perp$ vs $p_{\perp}^{\mathrm{trk1}}$ \\
        & 1.0 &  Away mean $p_\perp$ vs $p_{\perp}^{\mathrm{trk1}}$ \\
        & 1.0 &  Transverse mean $p_\perp$ vs $N_{\mathrm{chg}}$ \\
        & 1.0 &  Toward mean $p_\perp$ vs $N_{\mathrm{chg}}$ \\
        & 1.0 &  $N_{\mathrm{chg}}$ density vs $\Delta \phi$, $p_{\perp}^{\mathrm{trk1}} > 1 \, \mathrm{GeV}$ \\
        & 1.0 &  $p_\perp$ density vs $\Delta \phi$, $p_{\perp}^{\mathrm{trk1}} > 1 \, \mathrm{GeV}$ 
    \end{tabular}
    \caption{Weights assigned to LHC observables for the fourth stage of tuning of \UE parameters for proton-proton collisions at $\sqrt{s}$ = 0.9 TeV.}
    \label{tab:MBUEtuneweights9}
\end{table}
        
\begin{table}[]
\vspace{1cm}
\hspace{33.5mm} 
    \begin{tabular}{c c l}
    Rivet Analysis Identifier & Weight & Observable \\
    \hline
        CDF\_2002\_S4796047  & 1.0 & Mean $p_\perp$ vs multiplicity, $\vert\eta\vert < 1, \; p_\perp > 0.4 \, \mathrm{GeV}$  \\ 
        \hline
        E735\_1998\_S3905616 & 1.0 & Charged multiplicity  \\
        \hline
        CDF\_1990\_S2089246  & 1.0 & Pseudorapidity distribution \\
        \hline
        CDF\_1988\_S1865951 & 1.0 &  $p_\perp$ distribution   \\
        \hline
        \multirow{9}{*}{CDF\_2001\_S4751469} &
        1.0 & Mean $N_{\mathrm{ch}}$ vs $\Delta \phi$ from leading jet $(p_\perp^{\mathrm{lead}} > 2 \, \mathrm{GeV})$ \\
        & 1.0 & Mean $p_\perp^{\mathrm{sum}}$ vs $\Delta \phi$ from leading jet $(p_\perp^{\mathrm{lead}} > 2 \, \mathrm{GeV})$   \\
        & 1.0 & $N_{\mathrm{ch}}$ (toward) for min-bias   \\
        & 1.0 & $N_{\mathrm{ch}}$ (transverse) for min-bias   \\
        & 1.0 & $N_{\mathrm{ch}}$ (away) for min-bias   \\
        & 1.0 & $p_\perp^{\mathrm{sum}}$ (toward) for min-bias   \\
        & 1.0 & $p_\perp^{\mathrm{sum}}$ (transverse) for min-bias   \\
        & 1.0 & $p_\perp^{\mathrm{sum}}$ (away) for min-bias   \\
        & 1.0 & $p_\perp$ distribution (transverse, $p_\perp^{\mathrm{lead}} > 2 \, \mathrm{GeV}$)
    \end{tabular}
    \caption{Weights assigned to Tevatron observables during the fourth stage of tuning for \UE parameters for proton-antiproton collisions at $\sqrt{s}$ = 1.8 TeV.}
    \label{tab:MBUEtuneweights18}
\end{table}        

\begin{table}[]
\vspace{1cm}
\hspace{30mm} 
\begin{tabular}{c c l}
    Rivet Analysis Identifier & Weight & Observable \\
    \hline
        ALICE\_2010\_S8625980 & 1.0 & Charged multiplicity   \\
        \hline
        \multirow{2}{*}{ATLAS\_2011\_S8994773} &
        1.0 & Transverse $N$ density vs $p_\perp^{\mathrm{clus}1}$ \\
        & 1.0 & Transverse sum $p_\perp$ density vs $p_\perp^{\mathrm{clus}1}$ \\
        \hline
        \multirow{18}{*}{ATLAS\_2010\_S8918562}  &
        10.0 & Charged particle $\eta$, track $p_\perp > 100 \, \mathrm{MeV}, N_\mathrm{ch} \geq 2$ \\
        & 1.0 & Charged particle $\eta$, track $p_\perp > 100 \, \mathrm{MeV}, N_\mathrm{ch} \geq 20$ \\
        & 10.0 & Charged particle $\eta$, track $p_\perp > 500 \, \mathrm{MeV}, N_\mathrm{ch} \geq 1 $ \\
        & 10.0 & Charged particle $\eta$, track $p_\perp > 500 \, \mathrm{MeV}, N_\mathrm{ch} \geq 6 $ \\
        & 1.0 & Charged particle $\eta$, track $p_\perp > 2500 \, \mathrm{MeV}, N_\mathrm{ch} \geq 1 $ \\
        & 1.0 & Charged particle $p_\perp$, track $p_\perp > 100 \, \mathrm{MeV}, N_\mathrm{ch} \geq 2 $ \\
        & 1.0 & Charged particle $p_\perp$, track $p_\perp > 100 \, \mathrm{MeV}, N_\mathrm{ch} \geq 20 $ \\
        & 1.0 & Charged particle $p_\perp$, track $p_\perp > 500 \, \mathrm{MeV}, N_\mathrm{ch} \geq 1 $ \\
        & 1.0 & Charged particle $p_\perp$, track $p_\perp > 500 \, \mathrm{MeV}, N_\mathrm{ch} \geq 6 $ \\
        & 1.0 & Charged particle $p_\perp$, track $p_\perp > 2500 \, \mathrm{MeV}, N_\mathrm{ch} \geq 1 $ \\
        & 1.0 & Charged multiplicity $\geq$ 2, track $p_\perp > 100 \, \mathrm{MeV} $ \\
        & 1.0 & Charged multiplicity $\geq$ 20, track $p_\perp > 100 \, \mathrm{MeV} $ \\
        & 1.0 & Charged multiplicity $\geq$ 1, track $p_\perp > 500 \, \mathrm{MeV} $ \\
        & 1.0 & Charged multiplicity $\geq$ 6, track $p_\perp > 500 \, \mathrm{MeV} $ \\
        & 1.0 & Charged multiplicity $\geq$ 1, track $p_\perp > 2500 \, \mathrm{MeV} $ \\
        & 1.0 & Charged mean $p_\perp$ vs $N_\mathrm{ch}$, $p_\perp > 100 \, \mathrm{MeV}, N_\mathrm{ch} \geq 2 $ \\
        & 1.0 & Charged mean $p_\perp$ vs $N_\mathrm{ch}$, $p_\perp > 500 \, \mathrm{MeV}, N_\mathrm{ch} \geq 1 $ \\
        & 1.0 & Charged mean $p_\perp$ vs $N_\mathrm{ch}$, $p_\perp > 2500 \, \mathrm{MeV}, N_\mathrm{ch} \geq 1 $ \\
        \hline
        \multirow{14}{*}{ATLAS\_2010\_S8894728} &
        1.0 &  Transverse $N_{\mathrm{chg}}$ density vs $p_\perp^{\mathrm{trk}1}$ \\
        & 1.0 &  Toward $N_{\mathrm{chg}}$ density vs $p_\perp^{\mathrm{trk}1}$ \\
        & 1.0 &  Away $N_{\mathrm{chg}}$ density vs $p_\perp^{\mathrm{trk}1}$ \\
        & 1.0 &  Transverse sum $p_\perp$ density vs $p_\perp^{\mathrm{trk}1}$ \\
        & 1.0 &  Toward sum $p_\perp$ density vs $p_\perp^{\mathrm{trk}1}$ \\
        & 1.0 &  Away sum $p_\perp$ density vs $p_\perp^{\mathrm{trk}1}$ \\
        & 1.0 &  Transverse mean $p_\perp$ vs $p_\perp^{\mathrm{trk}1}$ \\
        & 1.0 &  Toward mean $p_\perp$ vs $p_\perp^{\mathrm{trk}1}$ \\
        & 1.0 &  Away mean $p_\perp$ vs $p_\perp^{\mathrm{trk}1}$ \\
        & 1.0 &  Transverse mean $p_\perp$ vs $N_{\mathrm{chg}}$ \\
        & 1.0 &  Toward mean $p_\perp$ vs $N_{\mathrm{chg}}$ \\
        & 1.0 &  Away mean $p_\perp$ vs $N_{\mathrm{chg}}$ \\
        & 1.0 &  $N_{\mathrm{chg}}$ density vs $\Delta \phi$, $p_\perp^{\mathrm{trk}1}$ > 1 GeV    \\
        & 1.0 &  $p_\perp$ density vs $\Delta \phi$, $p_\perp^{\mathrm{trk}1}$ > 1 GeV
    \end{tabular}
    \caption{Weights assigned to LHC observables during the fourth stage of tuning for \UE parameters for proton-proton collisions at $\sqrt{s}$ = 7 TeV.}
    \label{tab:MBUEtuneweights7}
\end{table}

\begin{table}[]
\vspace{1cm}
\hspace{30.5mm} 
    \begin{tabular}{c c l}
    Rivet Analysis Identifier & Weight & Observable \\
    \hline
        \multirow{12}{*}{ATLAS\_2017\_I1509919} &
        1.0 & Normalized distribution of $p_\perp^{\mathrm{lead}}$ ($p_\perp^{\mathrm{lead}}$ > 1 GeV)  \\ 
        & 1.0 & Mean average $p_\perp$ vs $N_\mathrm{ch}$, transverse region  \\ 
        & 1.0 & Mean average $p_\perp$ vs $N_\mathrm{ch}$, toward region  \\ 
        & 1.0 & Mean average $p_\perp$ vs $N_\mathrm{ch}$, away region  \\ 
        & 1.0 & Mean average $p_\perp$ vs $p_\perp^{\mathrm{lead}}$, transverse region  \\ 
        & 1.0 & Mean average $p_\perp$ vs $p_\perp^{\mathrm{lead}}$, toward region  \\
        & 1.0 & Mean average $p_\perp$ vs $p_\perp^{\mathrm{lead}}$, away region  \\
        & 1.0 & Mean charged multiplicity density, $p_\perp^{\mathrm{lead}}$ > 1 GeV  \\ 
        & 1.0 & Mean sum $p_\perp$ density, $p_\perp^{\mathrm{lead}}$ > 1 GeV  \\ 
        & 1.0 & Mean sum $p_\perp$ density, transverse region  \\
        & 1.0 & Mean sum $p_\perp$ density, towards region  \\
        & 1.0 & Mean sum $p_\perp$ density, away region  \\
        \hline
        \multirow{4}{*}{ATLAS\_2016\_I1467230} &
        5.0 & $dN_{\mathrm{ch}}/d\eta \geq 2$, $p_\perp$ > 100 MeV, |$\eta$| < 2.5, $\tau$ > 30 ps \\
        & 5.0 & Charged particle $p_\perp$, $p_\perp$ > 100 MeV, |$\eta$| < 2.5, $\tau$ > 30 ps \\
        & 5.0 & Charged particle $\eta$, $p_\perp$ > 100 MeV, |$\eta$| < 2.5, $\tau$ > 30 ps \\
        & 5.0 & Average particle $p_\perp$, $p_\perp$ > 100 MeV, |$\eta$| < 2.5, $\tau$ > 30 ps
    \end{tabular}
    \caption{Weights assigned to LHC observables during the fourth stage of tuning for \UE parameters for proton-proton collisions at $\sqrt{s}$ = 13 TeV.}
    \label{tab:MBUEtuneweights13}
\end{table}

\twocolumn

\section{Definitions of Observables}
\label{appendix_b}

\subsection{Inclusive Single Particle Observables}
\label{appendix_singleparticle}

\textbf{Transverse momenta, $p^{\mathrm{in}}_{\perp}$, $p^{\mathrm{out}}_{\perp}$}:
In-plane transverse momentum with respect to the thrust axis is defined as $$p^{\mathrm{out}}_{\perp} = \overrightarrow{p} \cdot \overrightarrow{n}_{\mathrm{Major}}$$
Similarly, the out-of-plane transverse momentum is defined as $$p^{\mathrm{in}}_{\perp} = \overrightarrow{p} \cdot \overrightarrow{n}_{\mathrm{minor}}$$
where $\overrightarrow{n}_{\mathrm{Major}}$ and $\overrightarrow{n}_{\mathrm{minor}}$ are thrust major and minor axes respectively.
These can be thought of as components of transverse momenta in and out of the event plane, defined by the thrust and beam axes.

\subsection{Event Shape Observables}
\label{appendix_eventshapes}
Event shape observables describe the shape of the spread of momenta in the final-state. 
This is usually done using different variables, some of which are described below. More observable definitions can be found in \cite{Dasgupta:2003iq,ATLAS:2012tch} in detail.

\textbf{Thrust, Major and Minor}:
Thrust is defined as
$$T = \underset{\overrightarrow{n}_T}{\text{max}} \frac{\sum_{i} |\overrightarrow{p_i}.\overrightarrow{n}_T|}{\sum_{i}|\overrightarrow{p_i}|}$$
where $\overrightarrow{p_i}$ are the momenta of the final-state particles and $\overrightarrow{n_T}$ is the thrust axis. As a common practice, the distributions are binned as $\mathrm{1 -T}$ to preserve the common shape of event shape observables which peaks at lower values and higher values indicating a departure from a two-body system.

Two other orthogonal axes can be defined with respect to the thrust axes and the thrust quantities measured with respect to those axes are called Thrust major (M) and Thrust minor (m) such that
$$M = \underset{\overrightarrow{n}_M}{\text{max}} \frac{\sum_{i} |\overrightarrow{p_i}.\overrightarrow{n}_M|}{\sum_{i}|\overrightarrow{p_i}|}, 
\;\;\;
m = \underset{\overrightarrow{n}_m}{\text{max}} \frac{\sum_{i} |\overrightarrow{p_i}.\overrightarrow{n}_m|}{\sum_{i}|\overrightarrow{p_i}|}$$
The $\overrightarrow{n_M}$ and $\overrightarrow{n_m}$ are major and minor axes, respectively, with the additional constraints $\overrightarrow{n_M} \cdot \overrightarrow{n_T} = 0$ and $\overrightarrow{n_m} = \overrightarrow{n_T} \times \overrightarrow{n_M}$.

\textbf{Sphericity, Aplanarity}:
The Sphericity, $S$ and Aplanarity, $A$ are defined with respect to the eigenvalues $\lambda_1$, $\lambda_2$ and $\lambda_3$ of the global momentum tensor of the event, $M_{xyz}$.
$$M_{xyz} = 
    \begin{pmatrix}
        p_{xi}^2 & p_{xi}p_{yi} & p_{xi}p_{zi}\\
        p_{yi}p_{xi} & p_{yi}^2 & p_{yi}p_{zi}\\
        p_{zi}p_{xi} & p_{zi}p_{yi} & p_{zi}^2 
    \end{pmatrix}$$
The eigenvalues are normalized and ordered such that $\lambda_1 > \lambda_2 > \lambda_3$ and $\sum_{i} \lambda_i = 1$.
The observables are expressed as: $$S = \frac{3}{2} (\lambda_2 + \lambda_3), \;\;\;
A = \frac{3}{2} \lambda_3$$

\textbf{$C$, $D$-Parameters}:
The $C$ and $D$-Parameters are defined with respect to the eigenvalues of the linear momentum tensor $\Theta_{i.j}$.
$$\Theta_{i,j} = \frac{1}{\sum_{k=1}^{N} \vert \overrightarrow{p}_{k} \vert} \cdot \sum_{k=1}^{N} \frac{p_k^i p_k^j}{\vert \overrightarrow{p}_k \vert}$$
where $p_k^i$ and $p_k^j$ are the $i^{th}$ and $j^{th}$ components of the 3-vector $\overrightarrow{p_k}$ of the $k^{th}$ particle. The observables are expressed in terms of the eigenvalues as: $$C = 3 \; (\lambda_1 \lambda_2 + \lambda_2 \lambda_3 + \lambda_3 \lambda_1 ), \;\;\; D = 27 \; \lambda_1 \lambda_2 \lambda_3$$

\section{Modifications to \ThePI}
\label{appendix_thep8i}
\ThePI C\texttt{++} package \cite{TheP8I} provides an interface to \pythiav{8} classes enabling the use of the string hadronization model with showered events from \herwigv{7}. 
This gives us the framework to systematically study non-perturbative effects of the combined setup using lepton and hadron collision data. 
While the standalone version of \ThePI is sufficient for generating lepton collision events which can be tuned to LEP data, modifications are needed to account for colour reconnections for comparison to hadron collision data in order to generate events with realistic final-states. 
These modifications are highlighted here.

\ThePI interface primarily uses the \texttt{StringFragmentation} and \texttt{Pythia8Interface} classes to hadronise the system of particles once these are converted to a \pythiav{8} readable format from input \herwigv{7} events. 
Instances of \texttt{ColourReconnection} and \texttt{JunctionSplitting} classes are used to perform colour reconnection before hadronization of showered events within \pythiav{8}. 
The set of colour reconnection parameters is added to the interface via the \texttt{StringFragmentation} class and interfaced to \herwigv{7}, which gives us control over their values so that they can be tuned.
These parameters are listed in Table~\ref{tab:CR_params}. 
\texttt{CRreconnect} is used to switch on and off colour reconnection. 
These modifications are implemented as a patch, to the \ThePI version in \cite{TheP8I}, in the \texttt{herwig-bootstrap} script which will be available with the \herwigv{7.4} release.

\onecolumn

\begin{table}[t]
    \centering
    \begin{tabular}{l r c}
          & Parameters & Allowed values \\
         \hline
         \multirow{4}{*}{Steering parameters} & \texttt{CRreconnect} & 0 (\texttt{off}), 1 (\texttt{on})\\
         & \texttt{CRmode} & 0, 1, 2, 3, 4 \\
         & \texttt{CRforceHadronLevel} & 0 (\texttt{off}), 1 (\texttt{on})\\
         & \texttt{BRremnantMode} & 0 (\texttt{off}), 1 (\texttt{on})\\
         \hline
         \multirow{8}{*}{QCD-based scheme} & \texttt{CRlambdaForm} & 0, 1\\
         \multirow{8}{*}{(\texttt{CRmode} = 1, \texttt{BRremnantMode} = 1)} & \texttt{CRm0} & 0.1 - 5.0 \\
         & \texttt{CRjunctionCorrection} & 0.01 - 10.0\\
         & \texttt{CRnColours} & 1 - 30\\
         & \texttt{CRsameNeighbourColours} & 0 (\texttt{off}), 1 (\texttt{on}) \\
         & \texttt{CRallowJunctions} & 0 (\texttt{off}), 1 (\texttt{on}) \\
         & \texttt{CRallowDoubleJunRem} & 0 (\texttt{off}), 1 (\texttt{on}) \\
         & \texttt{CRtimeDilationMode} & 0, 1, 2, 3, 4, 5 \\
         & \texttt{CRtimeDilationPar} & 0.0 - 100.0 \\
         \hline
         \multirow{3}{*}{Gluon-move scheme} & \texttt{CRm2Lambda} & 0.25 - 16.0 \\
         \multirow{3}{*}{(\texttt{CRmode} = 2)} & \texttt{CRfracGluon} & 0.0 - 1.0 \\
         & \texttt{CRdLambdaCut} & 0.0 - 10.0 \\
         & \texttt{CRflipMode} & 0, 1, 2, 3, 4 
    \end{tabular}
    \caption{Parameters (and their allowed ranges) steering Colour Reconnection as implemented in \pythiav{8}, which can be accessed via the \ThePI interface and tweaked as input to \herwigv{7} for event generation.}
    \label{tab:CR_params}
\end{table}

\section{Comparison of results for \eetune and tunes for individual energies}
\label{appendix_EEvsIndep}

\begin{figure}[h!]
\centering
\begin{subfigure}{0.3\textwidth}
    \captionsetup{labelformat=empty}
    \caption[]{\bf $\mathbf{\sqrt{s}}$ = 0.9 TeV}
    \includegraphics[width=1\textwidth]{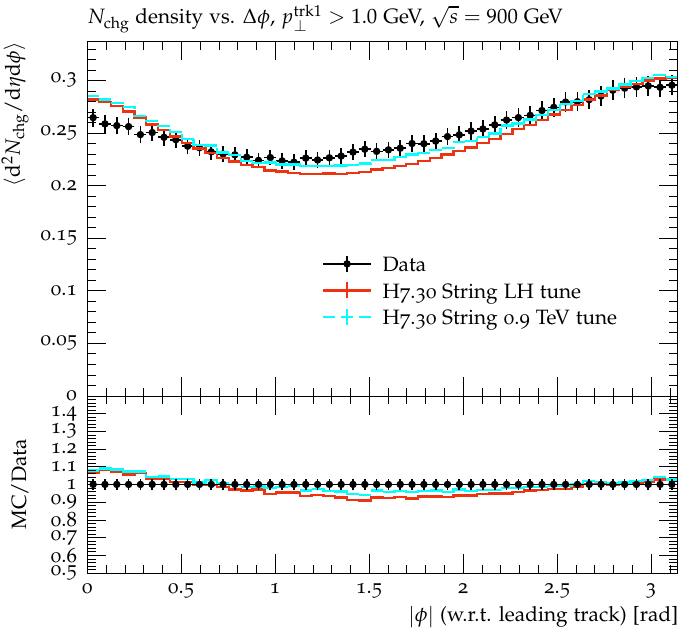}
\end{subfigure}
\begin{subfigure}{0.3\textwidth}
    \captionsetup{labelformat=empty}
    \caption[]{\bf $\mathbf{\sqrt{s}}$ = 7 TeV}
    \includegraphics[width=1\textwidth]{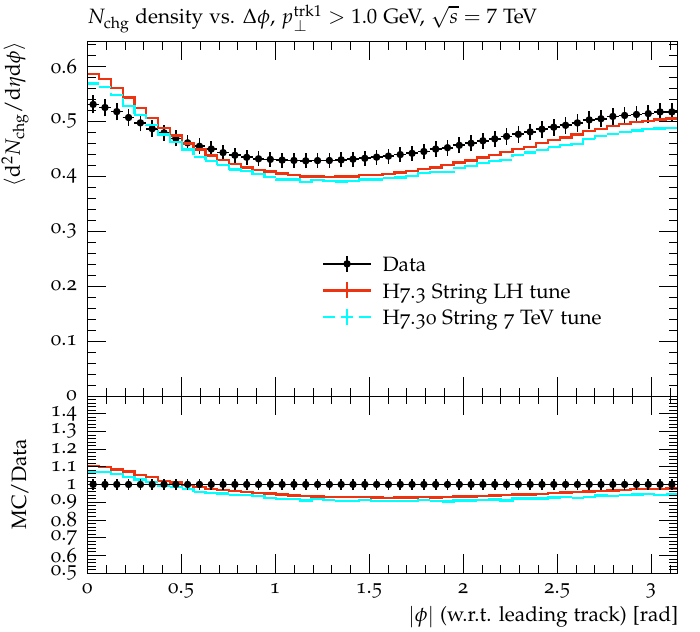}
\end{subfigure}
\begin{subfigure}{0.3\textwidth}
    \captionsetup{labelformat=empty}
    \caption[]{\bf $\mathbf{\sqrt{s}}$ = 13 TeV}
    \includegraphics[width=1\textwidth]{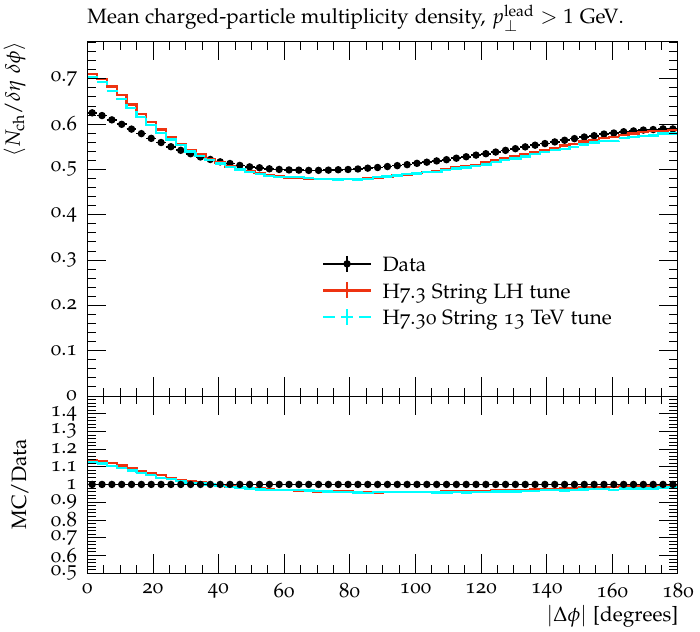}
\end{subfigure}
\caption{\UE measurements for charged particle multiplicity as a function of azimuthal angle related to the leading particle for three energies of the proton collision. These distributions correspond to Fig.~\ref{fig:UE_1}.}
\label{fig:EEvsIndep-UE}
\end{figure}

\begin{figure}[h!]
\centering
        \includegraphics[width=0.3\textwidth]{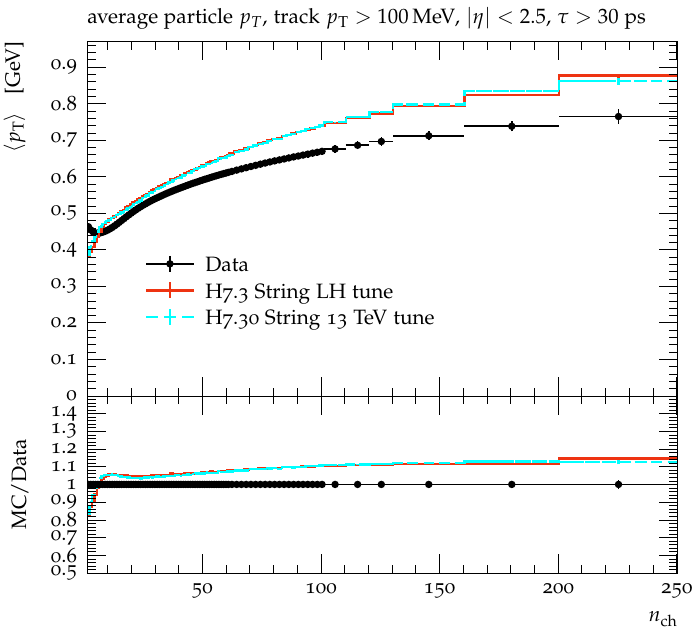}
        \includegraphics[width=0.3\textwidth]{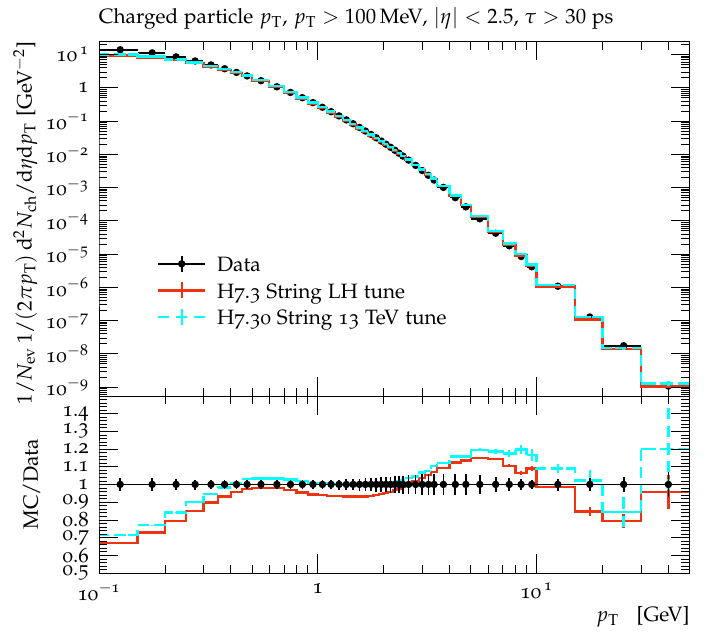}
        \includegraphics[width=0.3\textwidth]{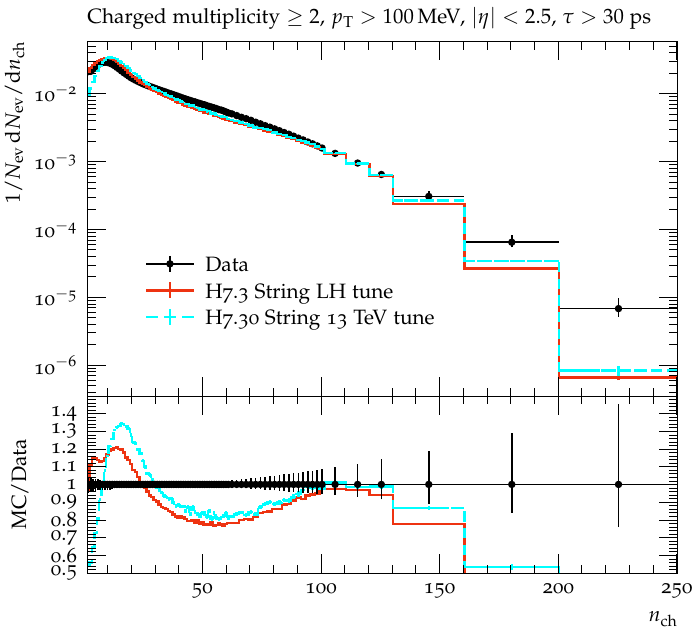}
\caption{\MB measurements for the three different views on distributions of charged particle multiplicity and their transverse momenta for proton collision at $\sqrt{s}$ = 13 TeV. These distributions correspond to Fig.~\ref{fig:MB_1}.}
\label{fig:EEvsIndep-MB}
\end{figure}  

\newpage
\clearpage

\section{LHC Analysis - additional observables}

\begin{figure}[h!]
        \includegraphics[width=.32\textwidth]{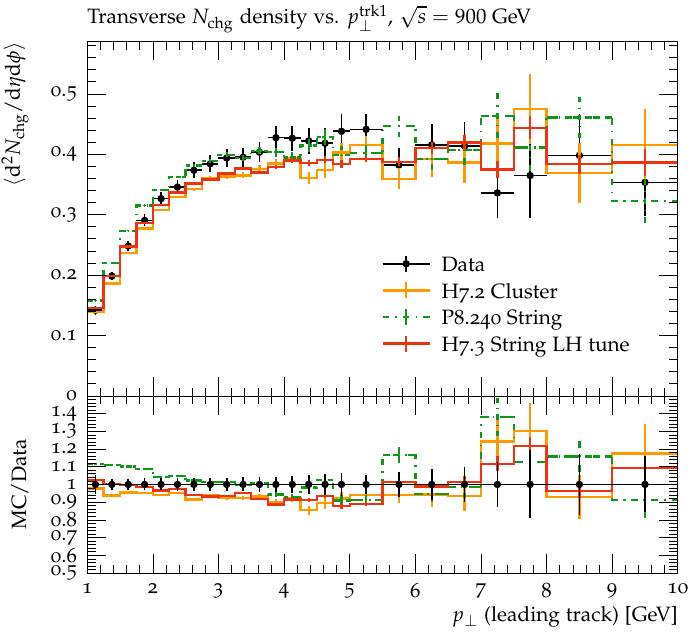}
        \includegraphics[width=.32\textwidth]{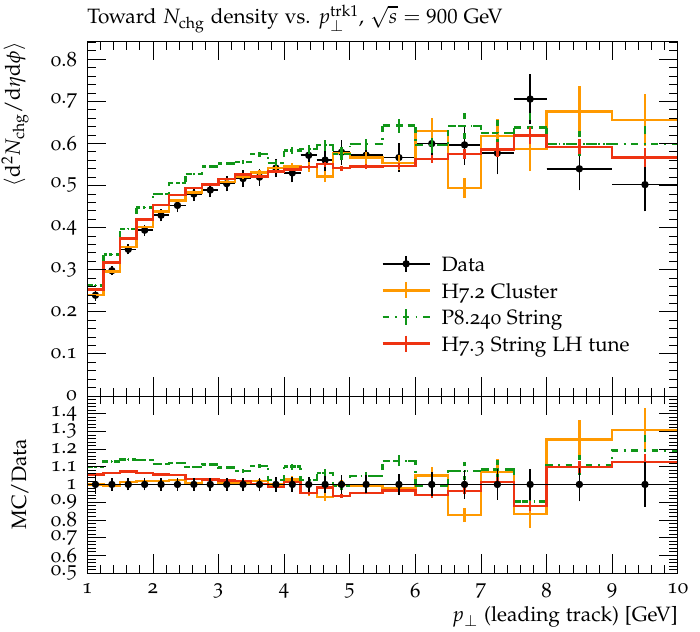}
        \includegraphics[width=.32\textwidth]{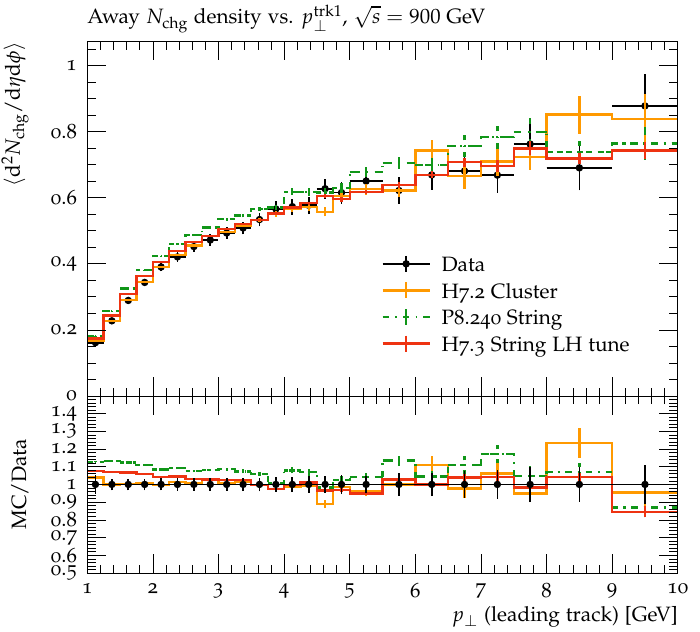}\\
        \includegraphics[width=.32\textwidth]{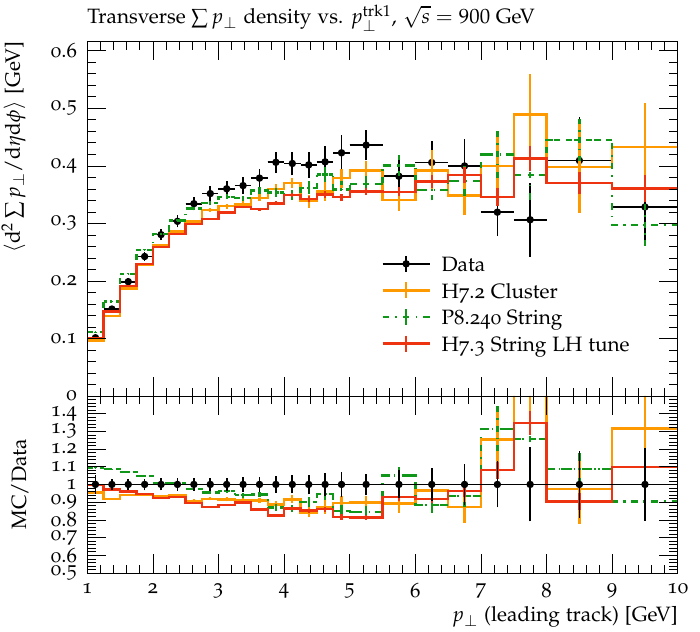}
        \includegraphics[width=.32\textwidth]{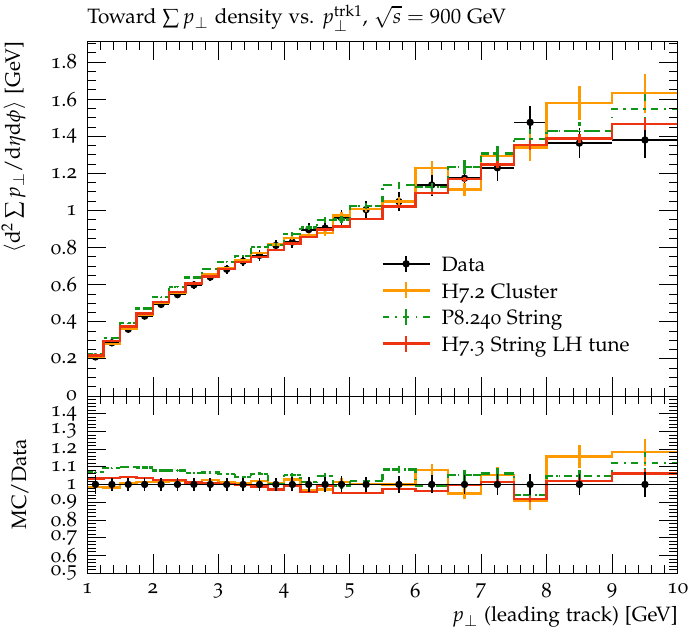}
        \includegraphics[width=.32\textwidth]{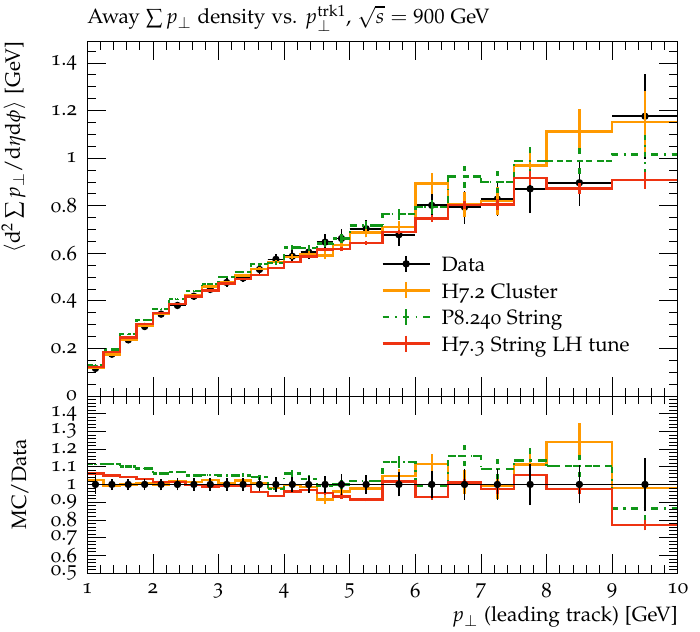}\\
        \includegraphics[width=.32\textwidth]{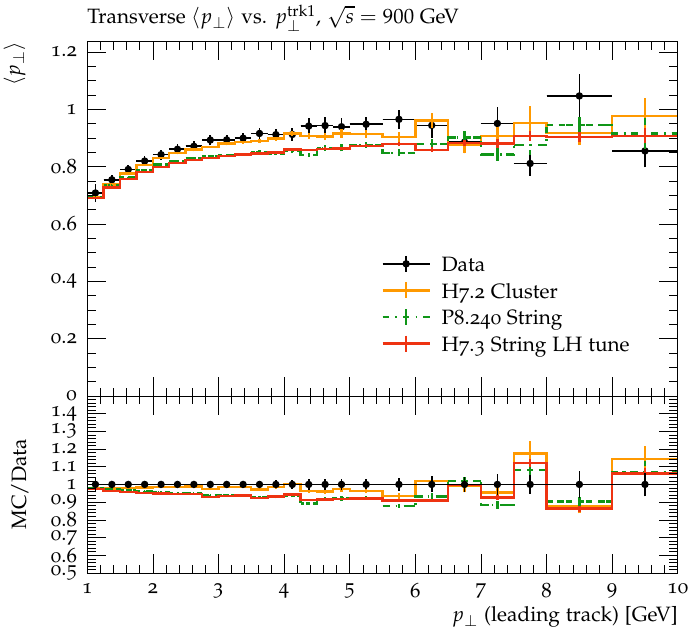}
        \includegraphics[width=.32\textwidth]{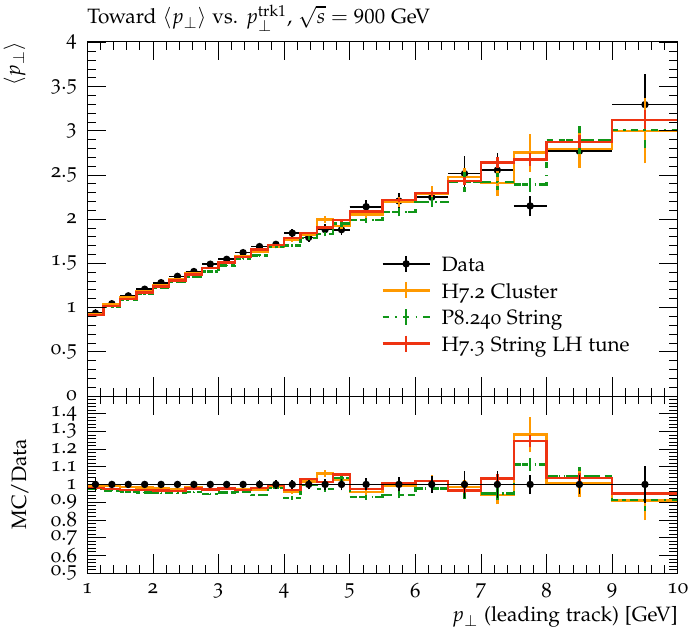}
        \includegraphics[width=.32\textwidth]{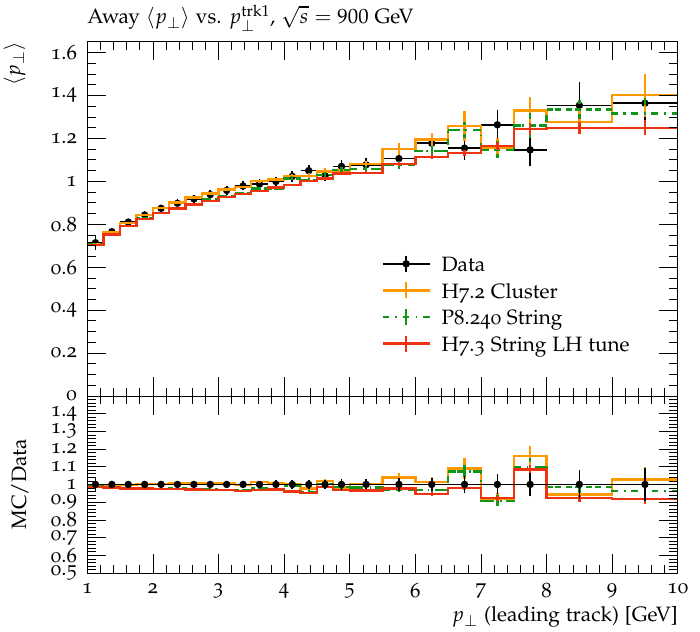}\\
        \includegraphics[width=.32\textwidth]{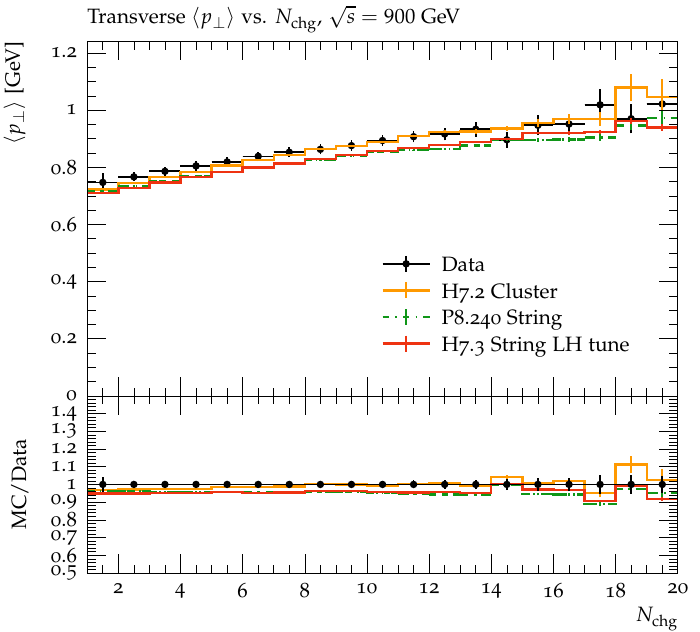}
        \includegraphics[width=.32\textwidth]{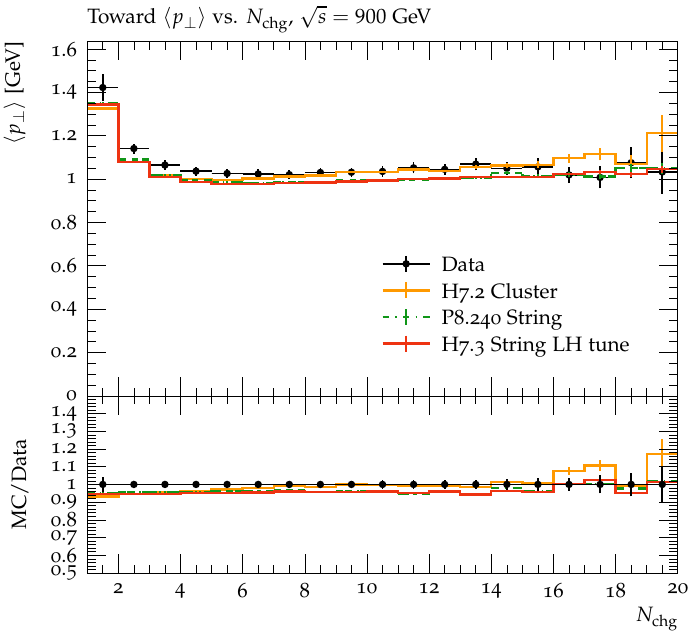}
        \includegraphics[width=.32\textwidth]{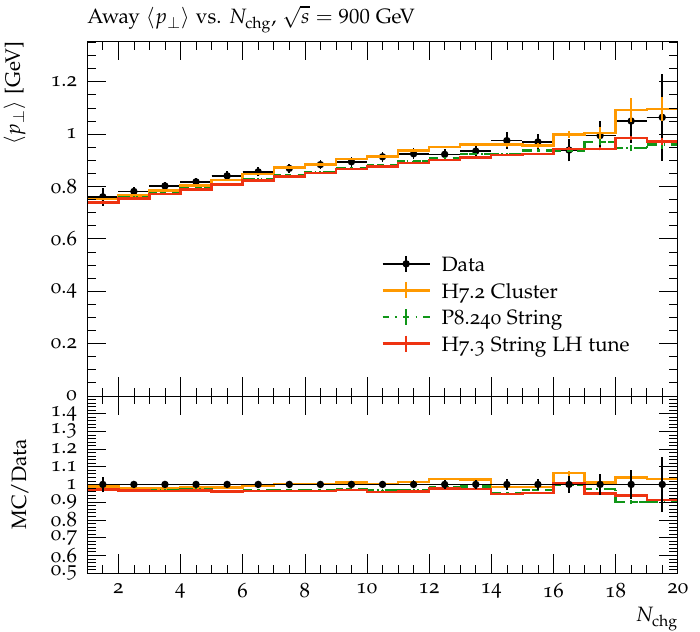}
        \caption{\UE observables for proton collisions at $\sqrt{s} = $ 0.9 TeV \cite{ATLAS:2010kmf}. These observables were included in the tuning interpolation.}
\label{fig:UE_aux_09}
\end{figure}

\begin{figure}[h]
\centering
        \includegraphics[width=.32\textwidth]{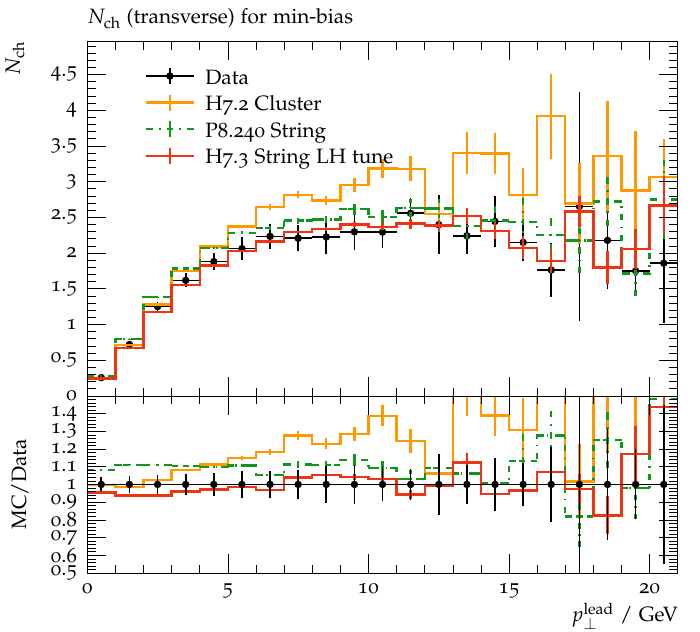}
        \includegraphics[width=.32\textwidth]{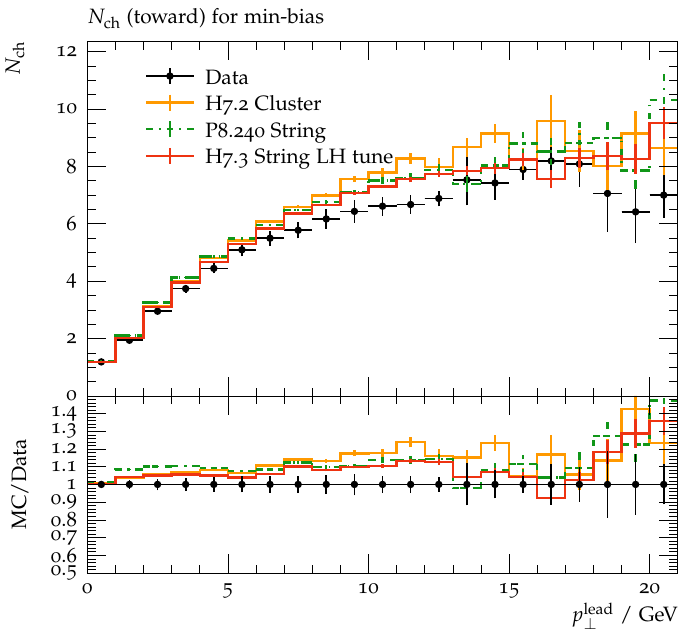}
        \includegraphics[width=.32\textwidth]{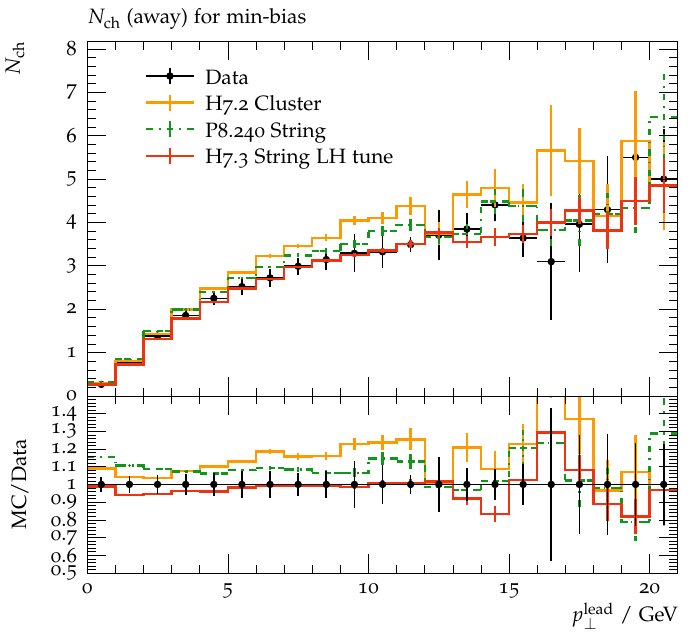} \\
        \includegraphics[width=.32\textwidth]{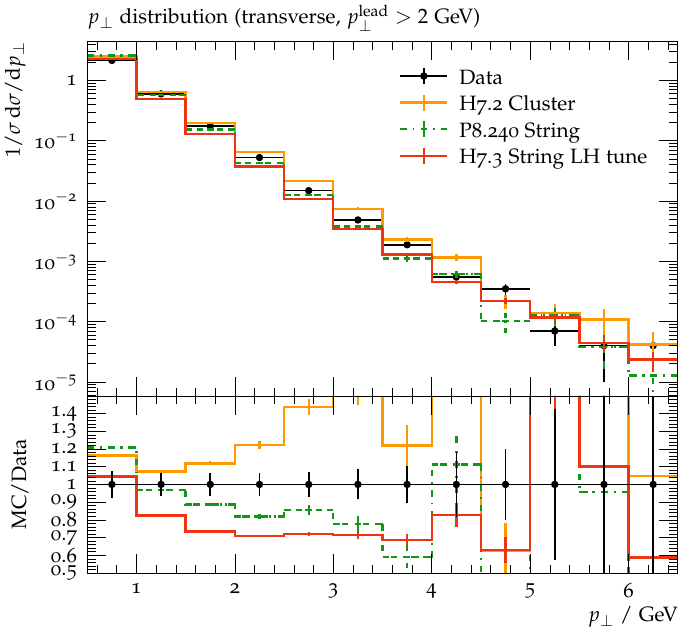}
        \includegraphics[width=.32\textwidth]{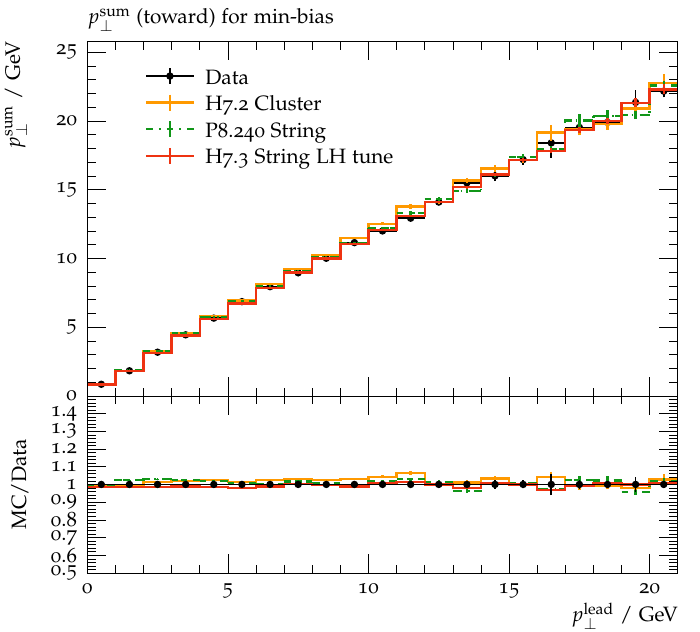}
        \includegraphics[width=.32\textwidth]{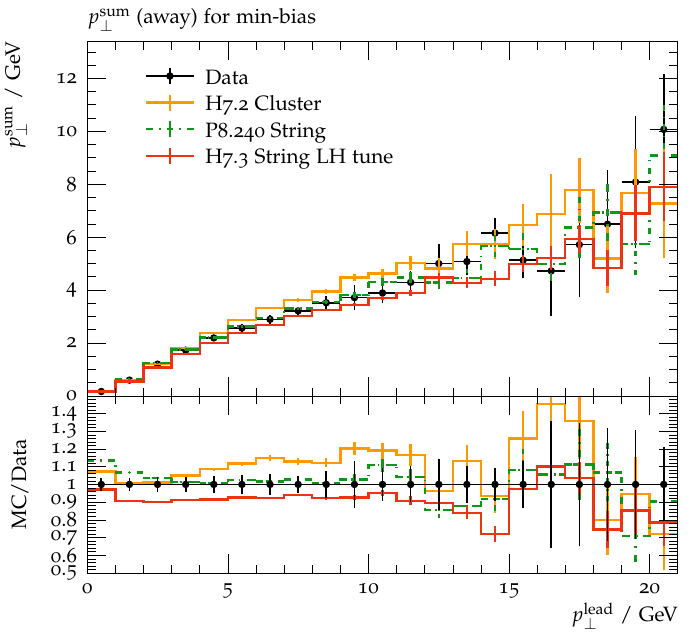} \\
        \includegraphics[width=.33\textwidth]{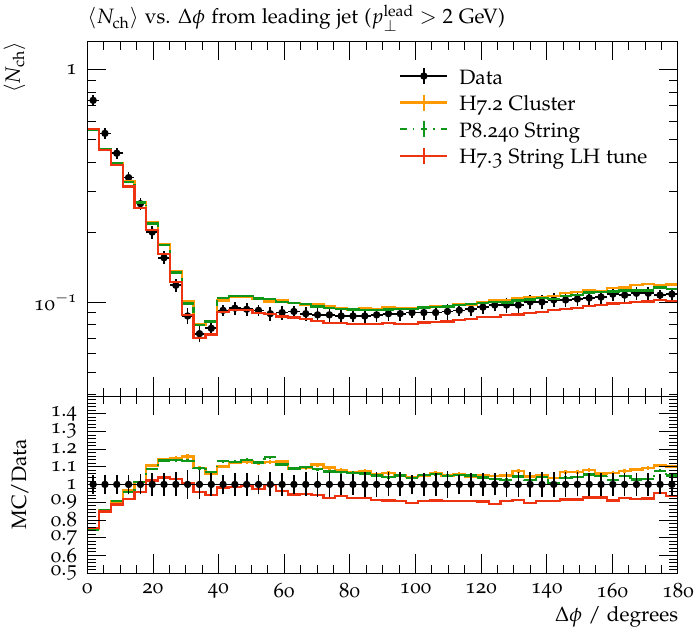}
        \includegraphics[width=.33\textwidth]{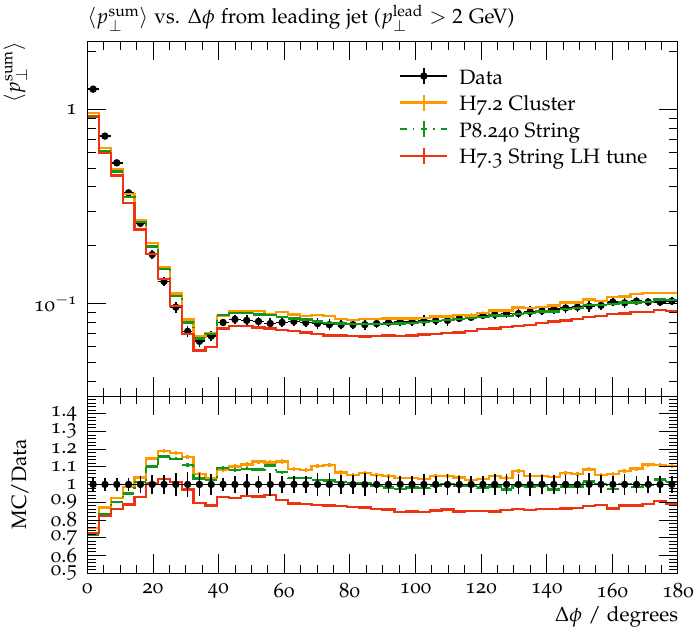}
        \caption{\UE observables for proton anti-proton collisions at $\sqrt{s} =$ 1.8 TeV \cite{CDF:2001onq}. These observables were included in the tuning interpolation.}
\label{fig:UE_aux_18}
\end{figure}   

\begin{figure}[h]
        \includegraphics[width=.33\textwidth]{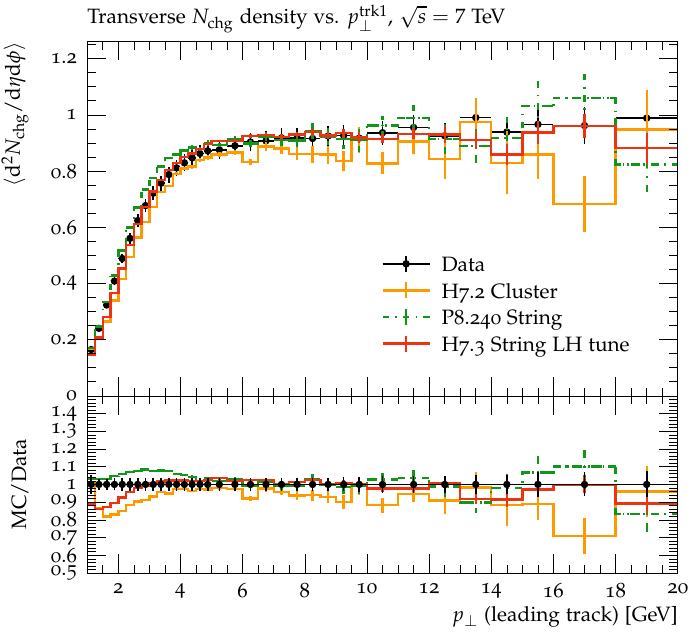}
        \includegraphics[width=.33\textwidth]{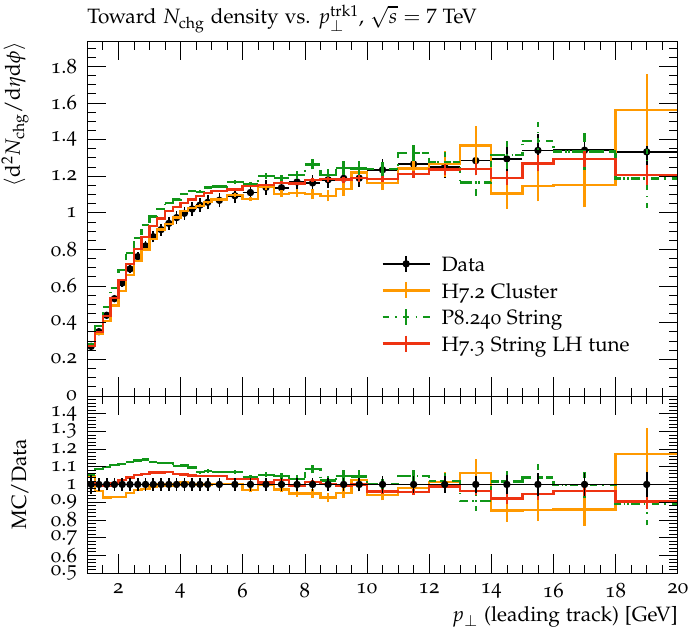}
        \includegraphics[width=.33\textwidth]{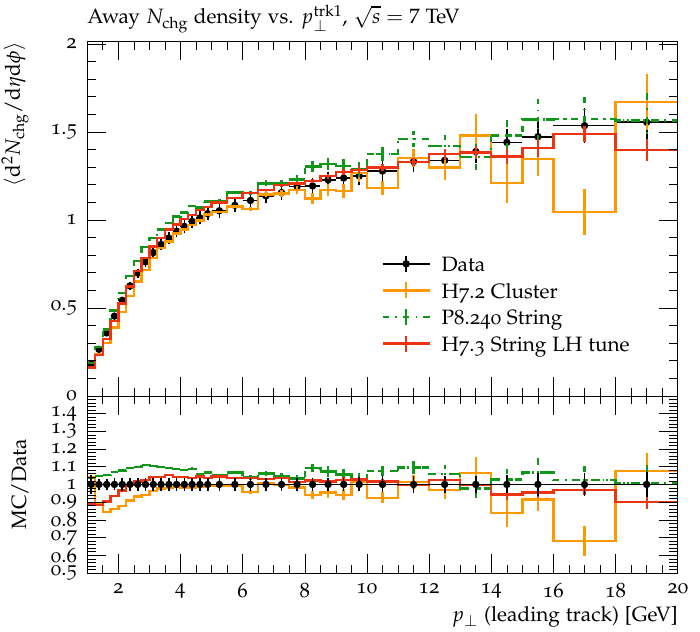}\\
        \includegraphics[width=.33\textwidth]{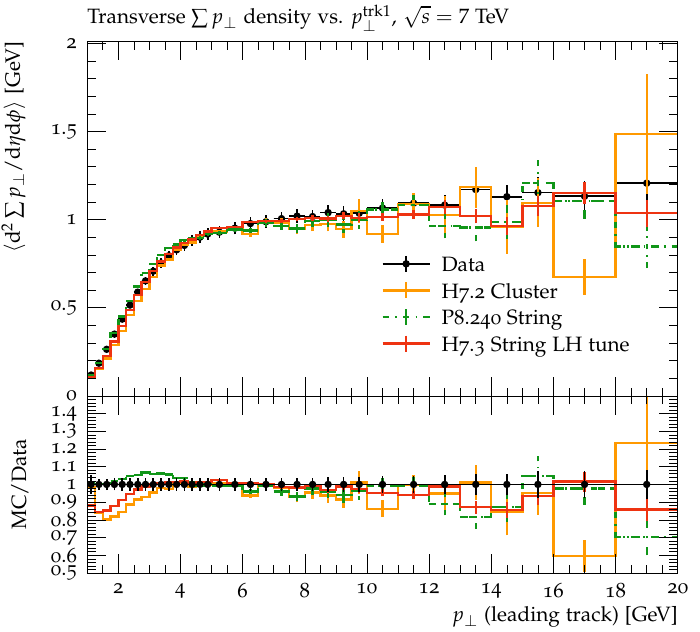}
        \includegraphics[width=.33\textwidth]{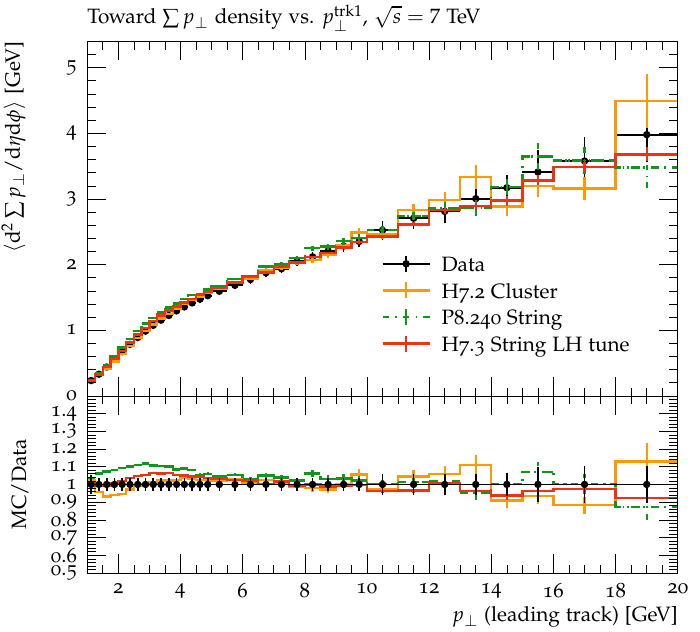}
        \includegraphics[width=.33\textwidth]{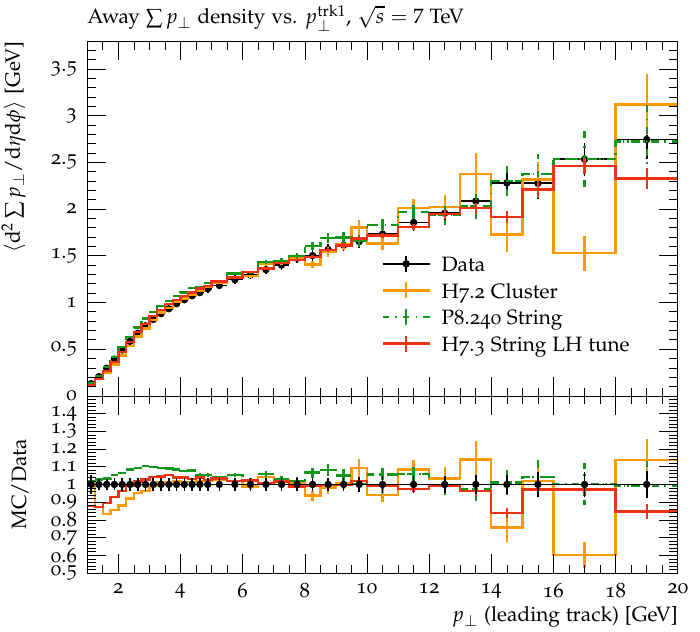}\\
        \includegraphics[width=.33\textwidth]{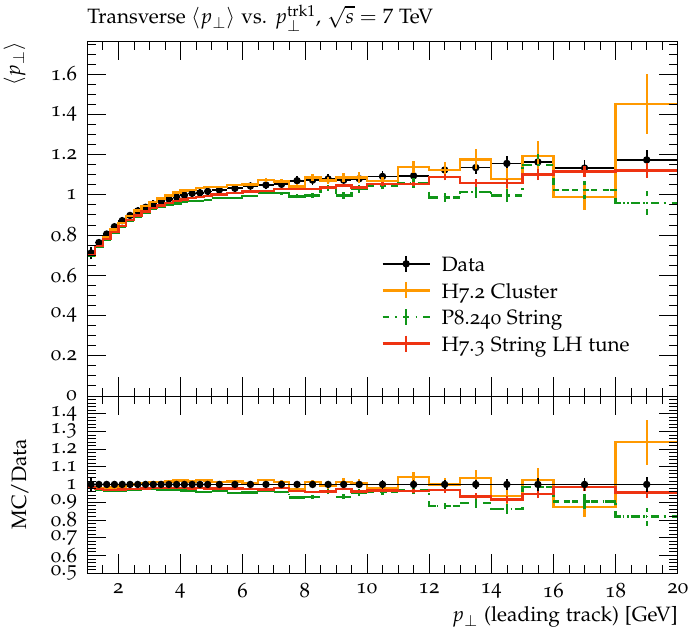}
        \includegraphics[width=.33\textwidth]{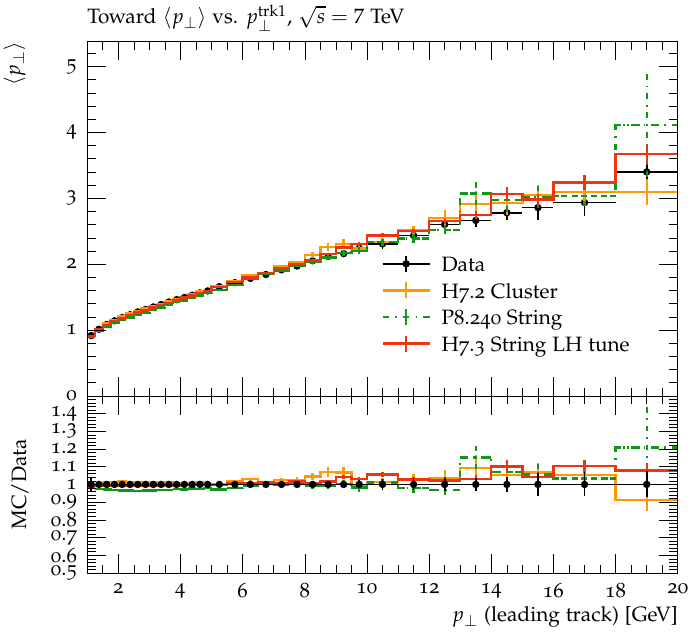}
        \includegraphics[width=.33\textwidth]{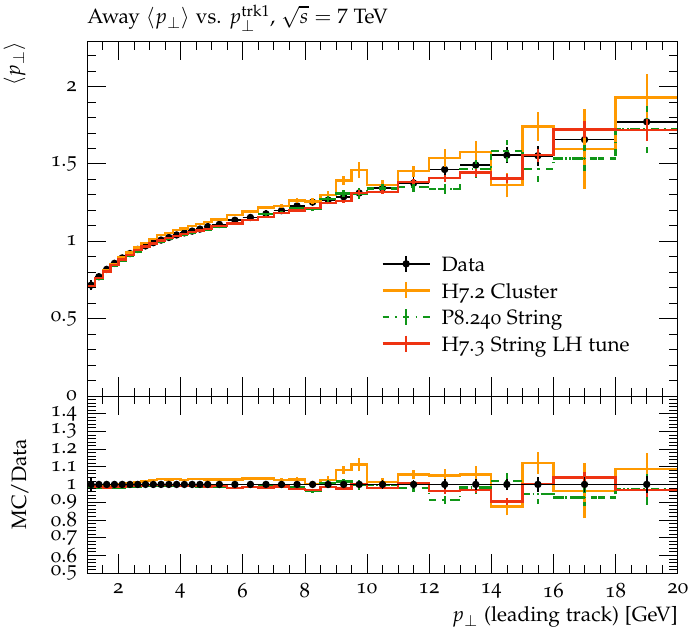}\\
        \includegraphics[width=.33\textwidth]{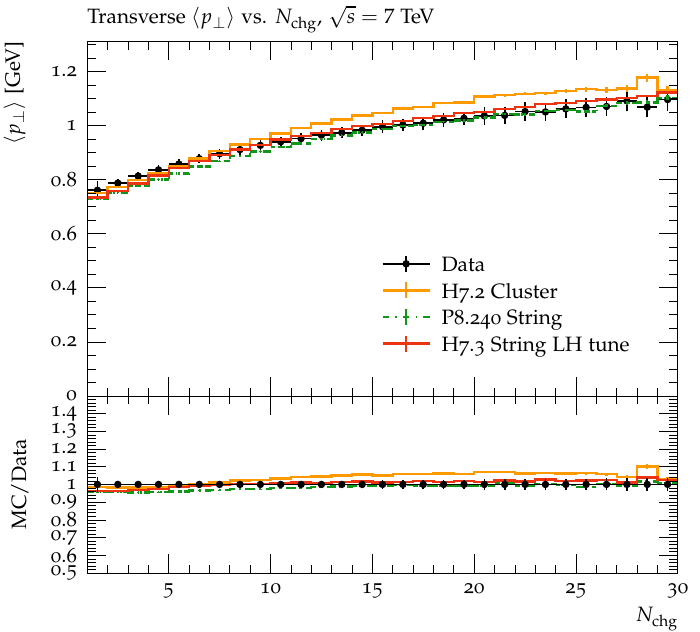}
        \includegraphics[width=.33\textwidth]{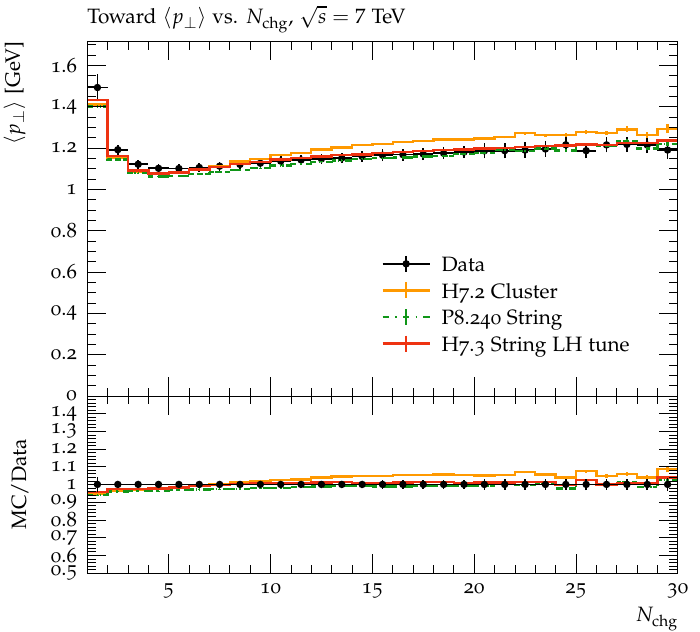}
        \includegraphics[width=.33\textwidth]{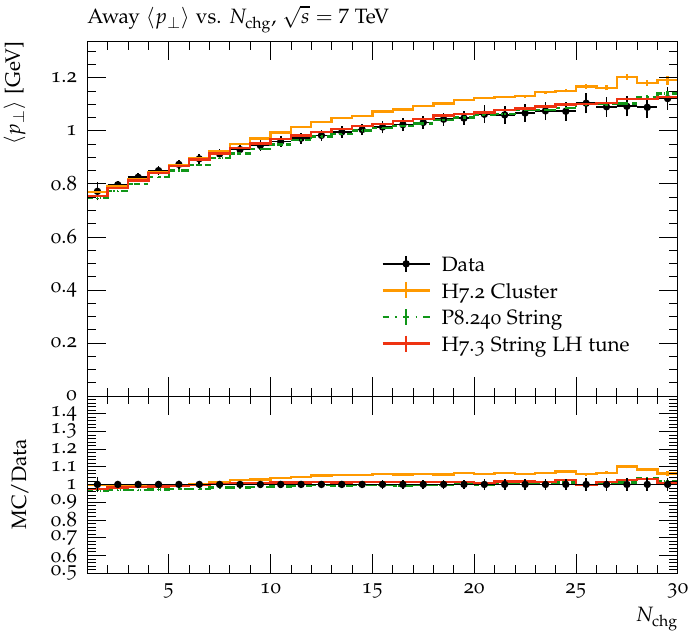}
        \caption{\UE observables for proton collisions at $\sqrt{s} =$ 7 TeV \cite{ATLAS:2010kmf}. These observables were included in the tuning interpolation.}
\label{fig:UE_aux_7}
\end{figure}

\begin{figure}[h]
        \includegraphics[width=.33\textwidth]{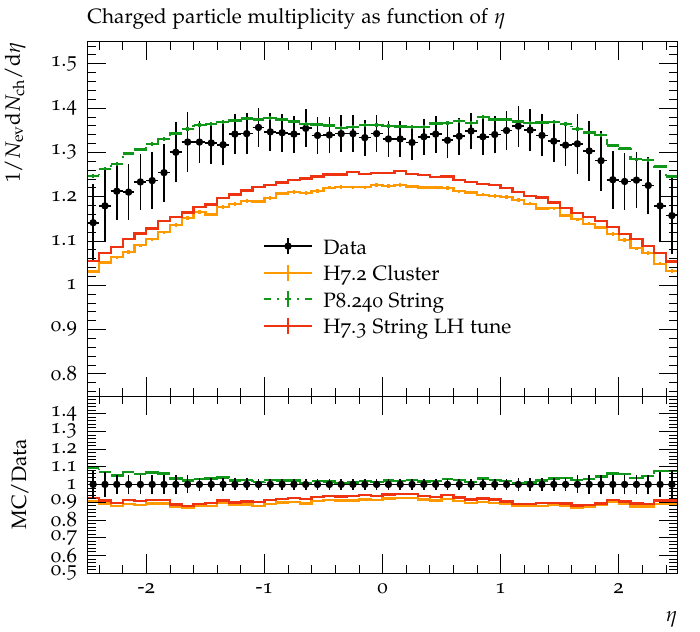}
        \includegraphics[width=.33\textwidth]{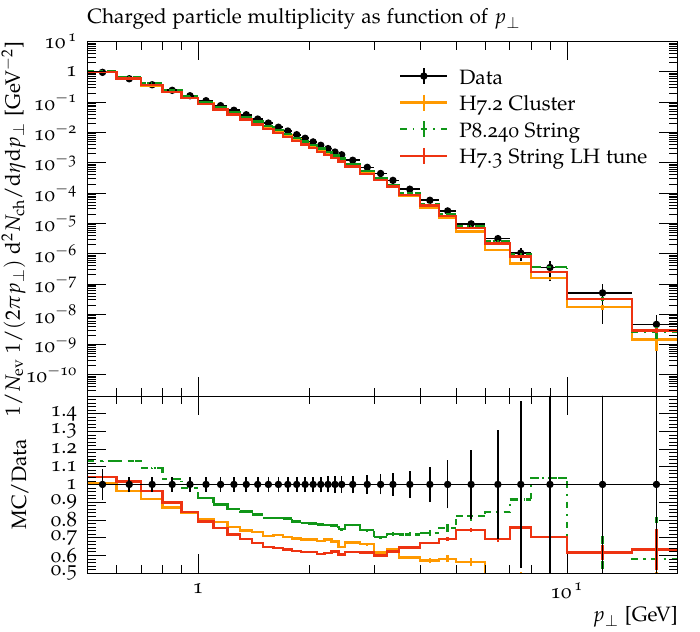}
        \includegraphics[width=.33\textwidth]{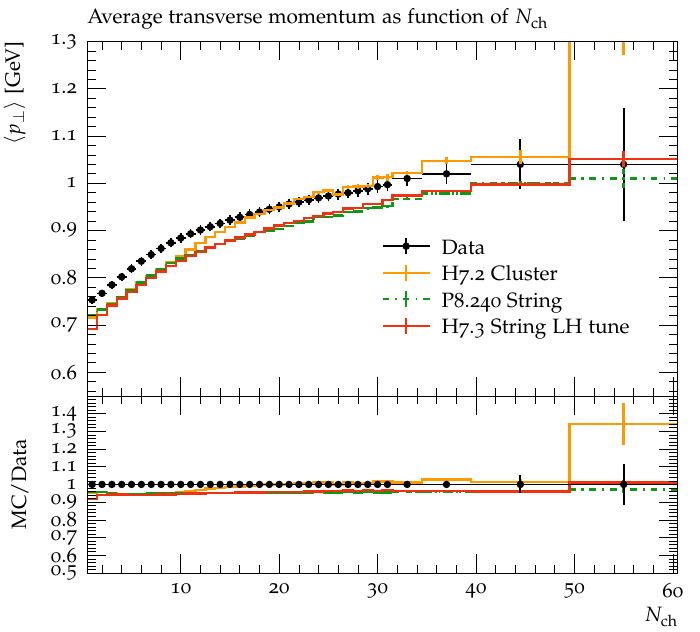}\\
        \includegraphics[width=.33\textwidth]{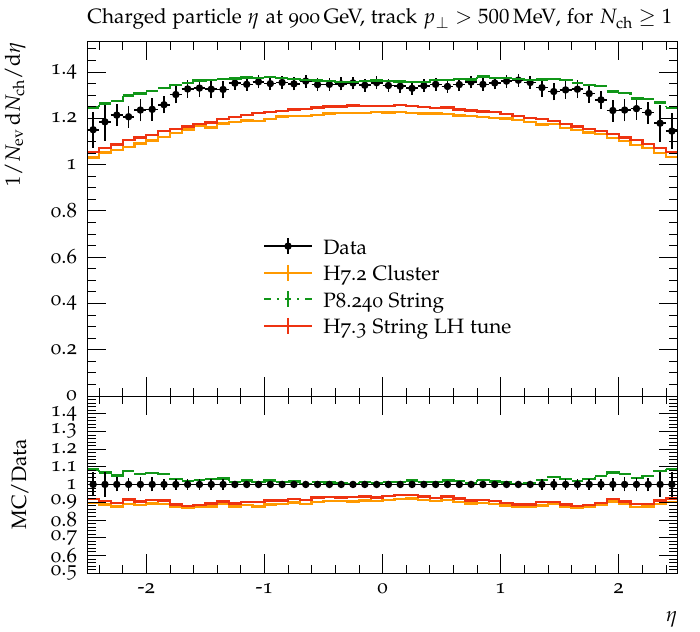} 
        \includegraphics[width=.33\textwidth]{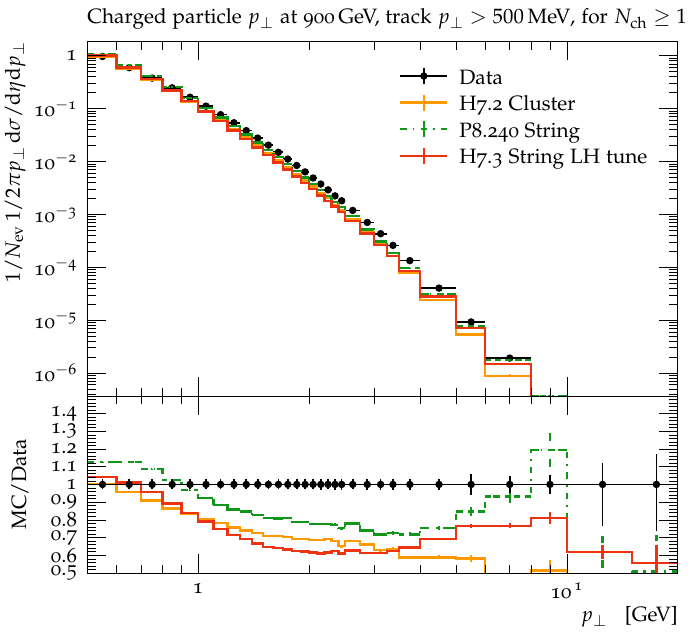}
        \includegraphics[width=.33\textwidth]{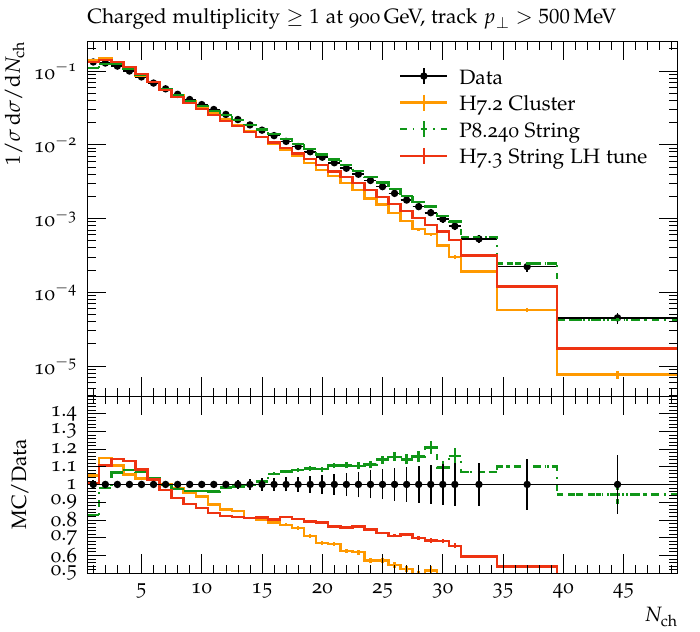}
        \caption{\MB observables for proton collisions at $\sqrt{s} =$ 0.9 TeV \cite{ATLAS:2010zmc, ATLAS:2010jvh}. These observables were included in the tuning interpolation.}
\label{fig:MB_aux_09}
\end{figure}

\begin{figure}[h]
\centering
\begin{subfigure}{0.32\textwidth}
    \includegraphics[width=\linewidth]{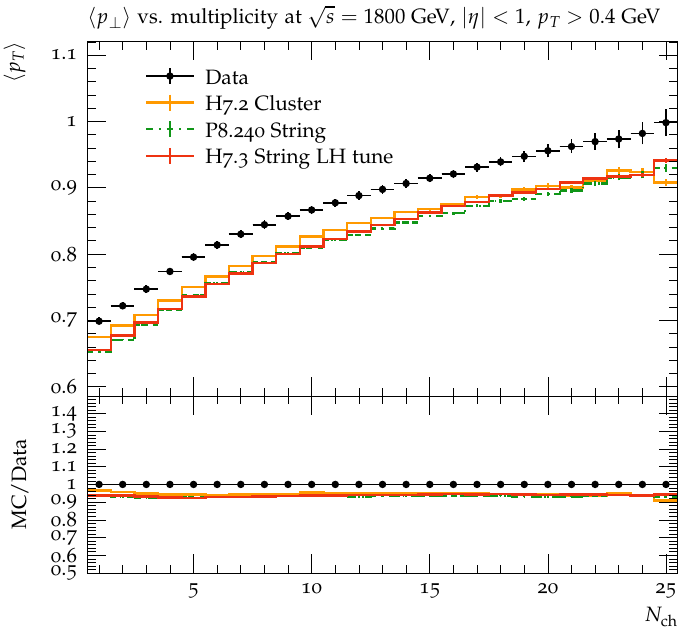}
    \label{fig:MB_aux_18:a}
\end{subfigure}
\begin{subfigure}{0.32\textwidth}
    \includegraphics[width=\linewidth]{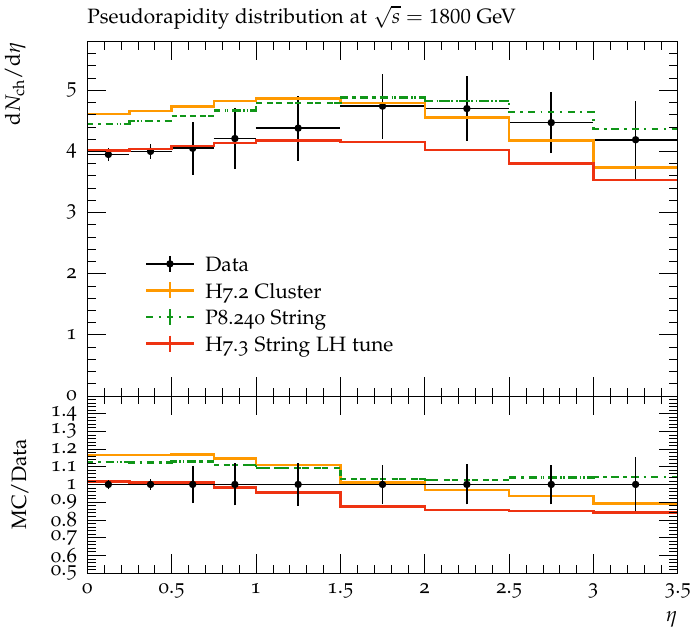}
    \label{fig:MB_aux_18:b}
\end{subfigure}
\begin{subfigure}{0.32\textwidth}
    \includegraphics[width=\linewidth]{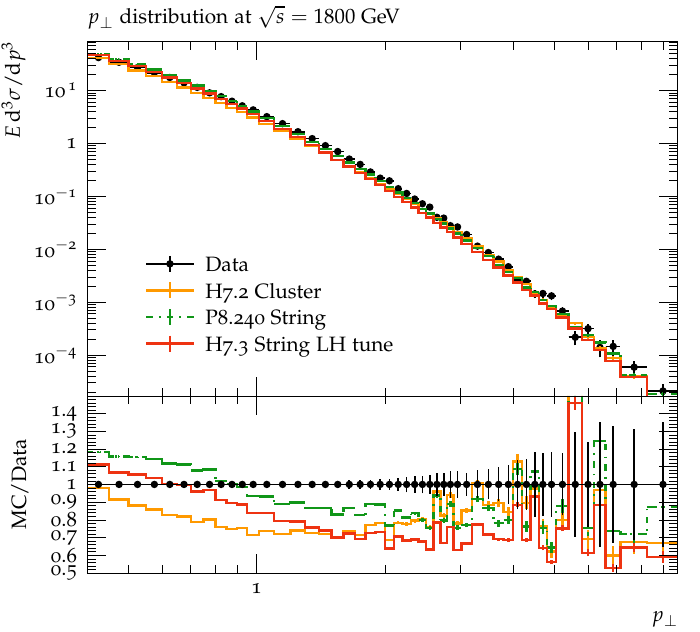}
    \label{fig:MB_aux_18:c}
\end{subfigure}
\caption{\MB observables for proton anti-proton collisions at $\sqrt{s} =$ 1.8 TeV \cite{CDF:2001hmt,CDF:1989nkn,CDF:1988evs,Alexopoulos:1998bi}. These observables were included in the tuning interpolation.}
\label{fig:MB_aux_18}
\end{figure}

\begin{figure}[h]
        \includegraphics[width=.33\textwidth]{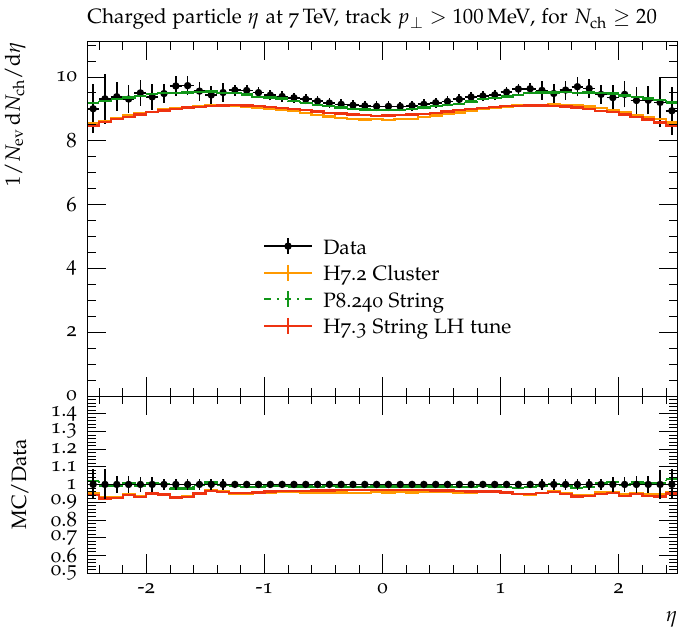}
        \includegraphics[width=.33\textwidth]{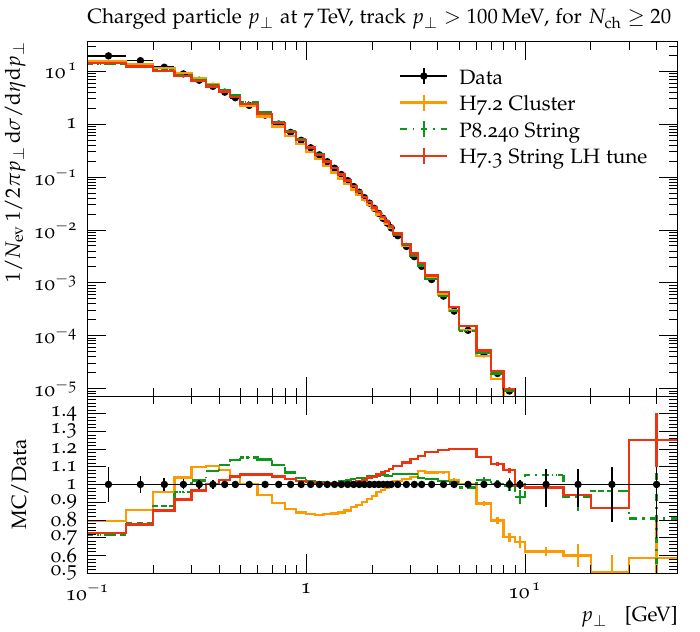}
        \includegraphics[width=.33\textwidth]{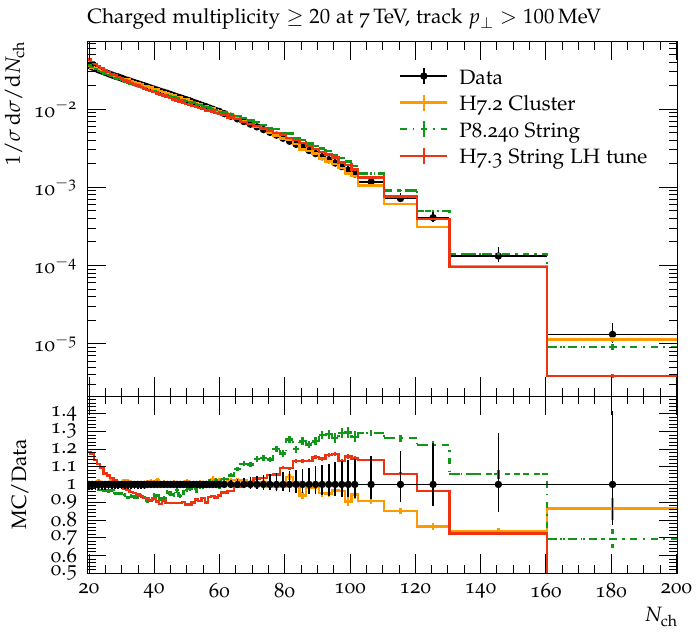}\\
        \includegraphics[width=.33\textwidth]{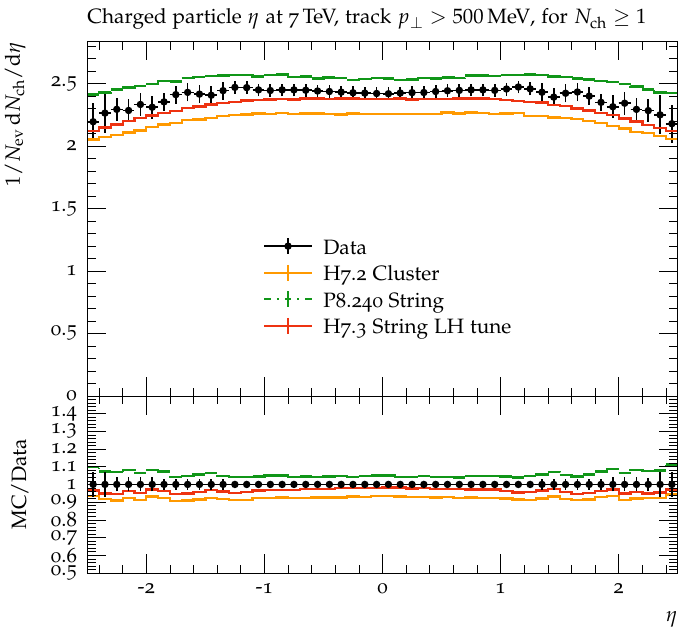}
        \includegraphics[width=.33\textwidth]{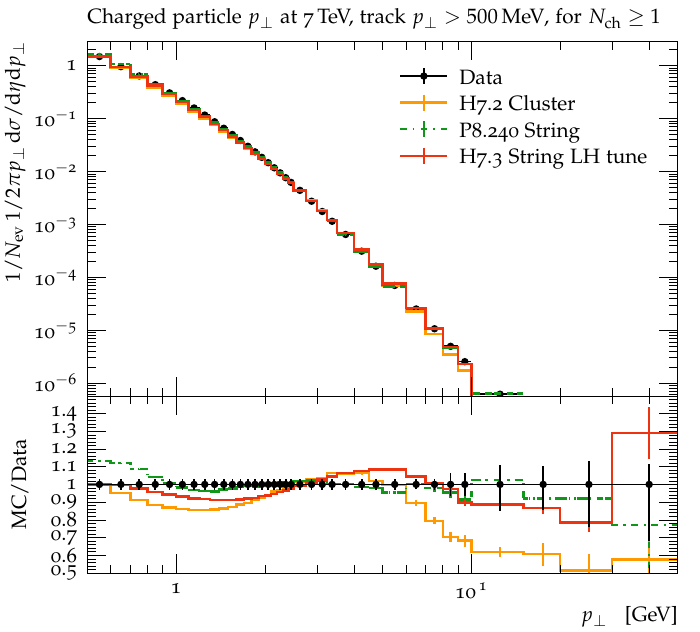}
        \includegraphics[width=.33\textwidth]{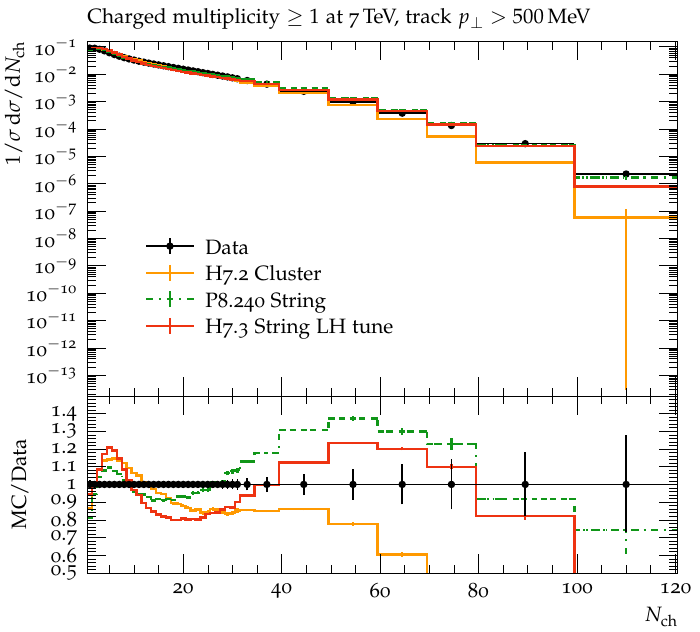}\\
        \includegraphics[width=.33\textwidth]{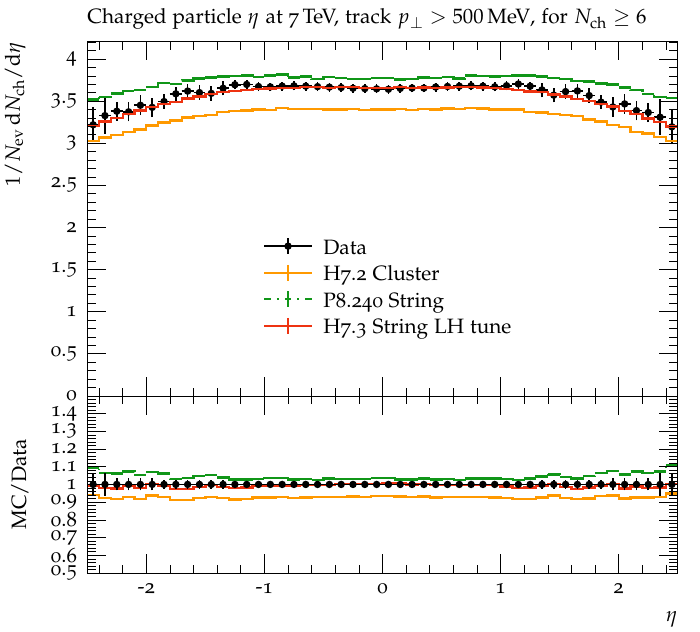}
        \includegraphics[width=.33\textwidth]{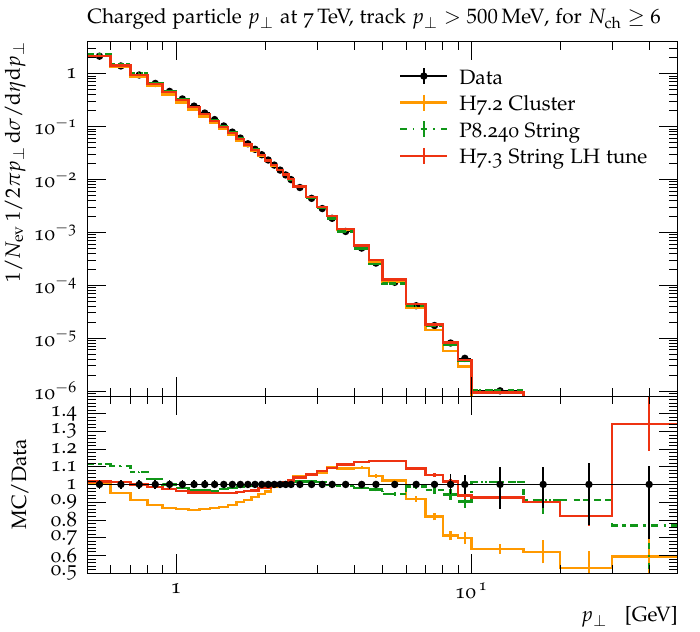}
        \includegraphics[width=.33\textwidth]{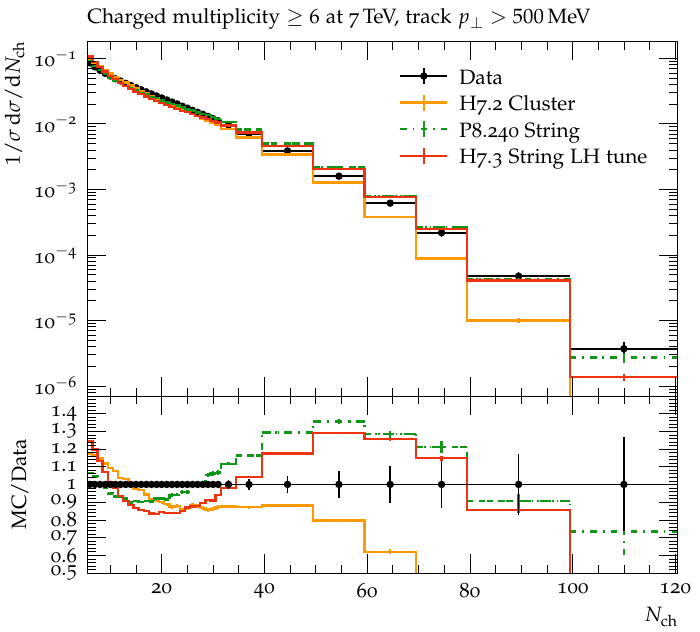}\\
        \includegraphics[width=.33\textwidth]{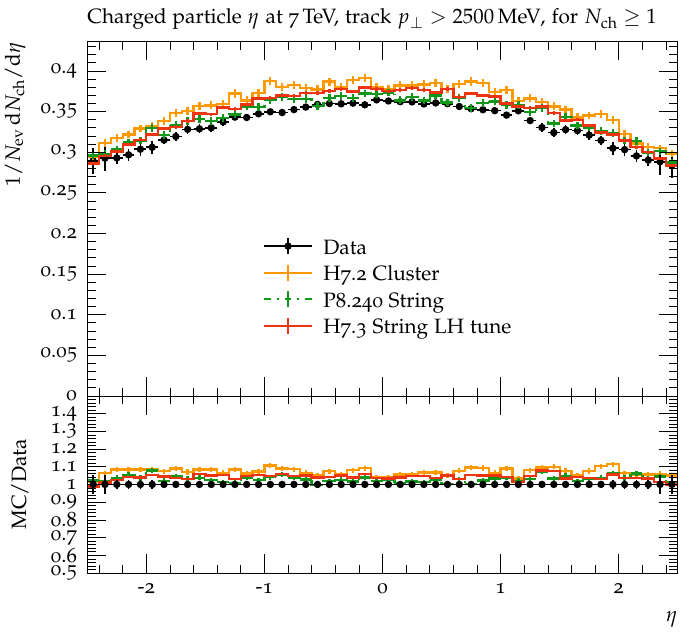}
        \includegraphics[width=.33\textwidth]{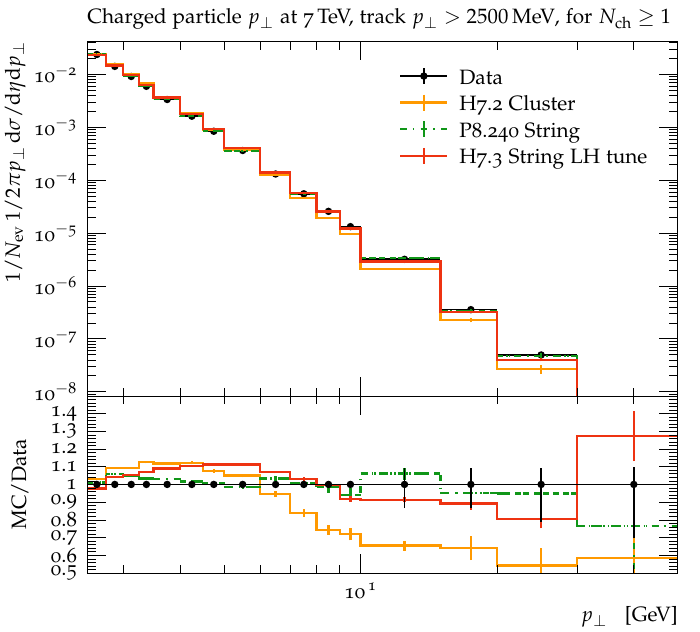}
        \includegraphics[width=.33\textwidth]{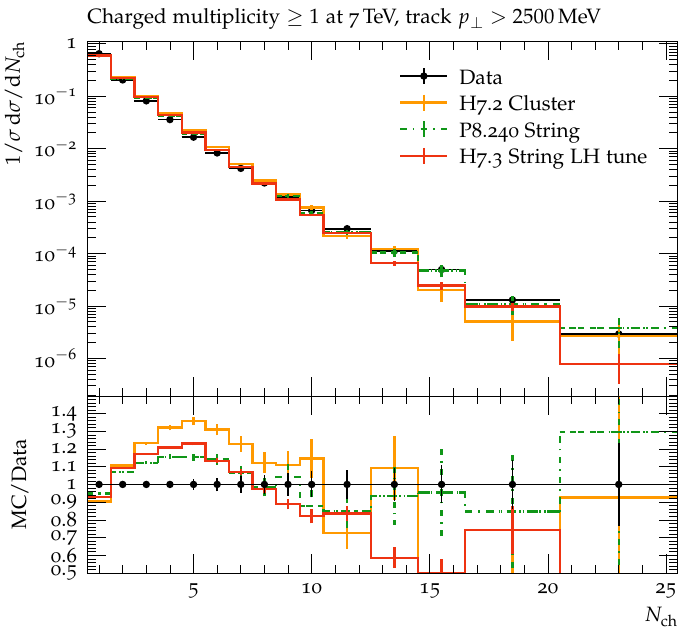}
        \caption{\MB observables for proton collisions at $\sqrt{s} =$ 7 TeV \cite{ATLAS:2010jvh}. These observables were included in the tuning interpolation.}
\label{fig:MB_aux_7}        
\end{figure}

\begin{figure}[h]
\centering
\captionsetup[subfigure]{labelformat=empty}
\begin{subfigure}{0.28\textwidth}
\caption{\bf CMS, $\mathbf{\sqrt{s}}$ = 0.9 TeV}
        \includegraphics[width=1\textwidth]{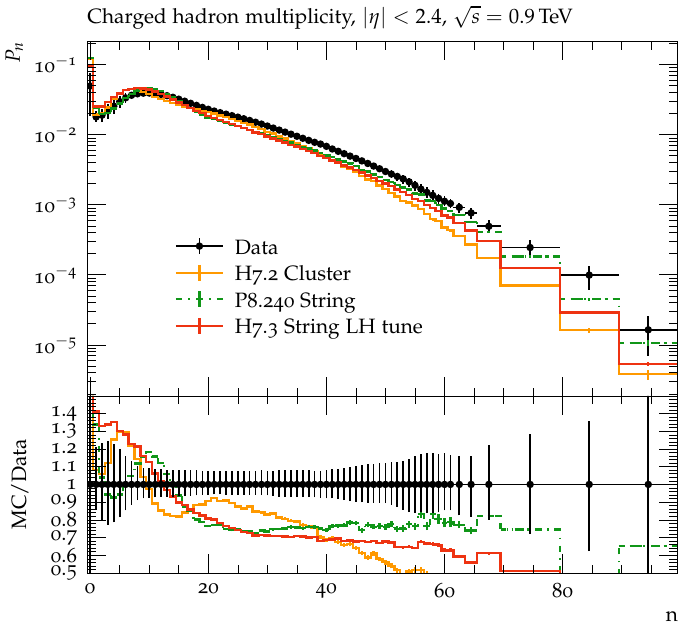}
\end{subfigure}
\begin{subfigure}{0.28\textwidth}
\caption{\bf CMS, $\mathbf{\sqrt{s}}$ = 0.9 TeV}
        \includegraphics[width=1\textwidth]{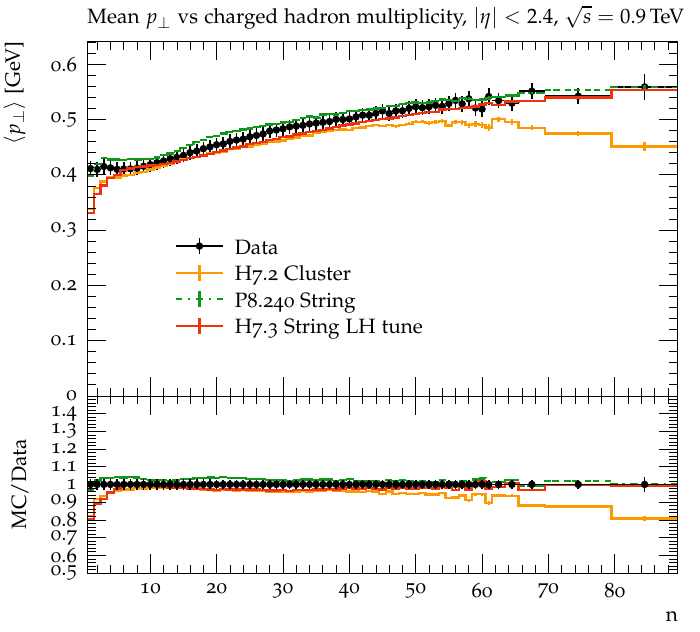}
\end{subfigure}
\begin{subfigure}{0.28\textwidth}
\caption{\bf CMS, $\mathbf{\sqrt{s}}$ = 0.9 TeV}
        \includegraphics[width=1\textwidth]{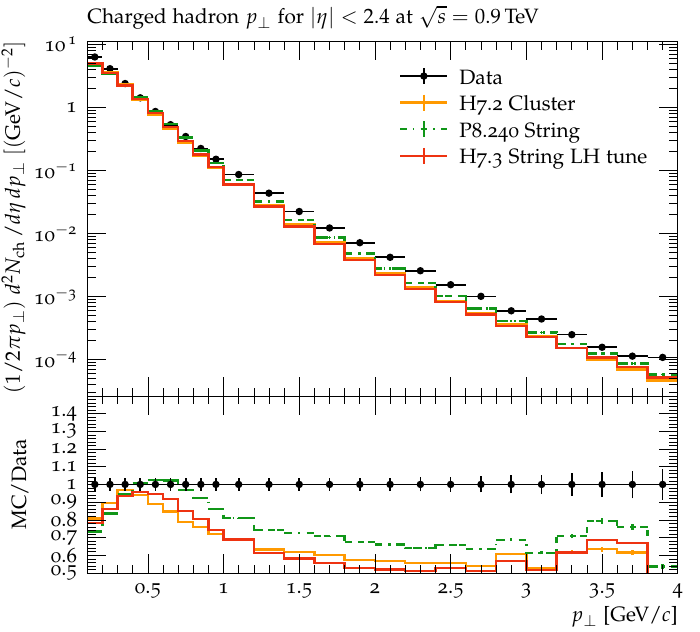}
\end{subfigure}
\begin{subfigure}{0.28\textwidth}
\caption{\bf ALICE, $\mathbf{\sqrt{s}}$ = 0.9 TeV}
        \includegraphics[width=1\textwidth]{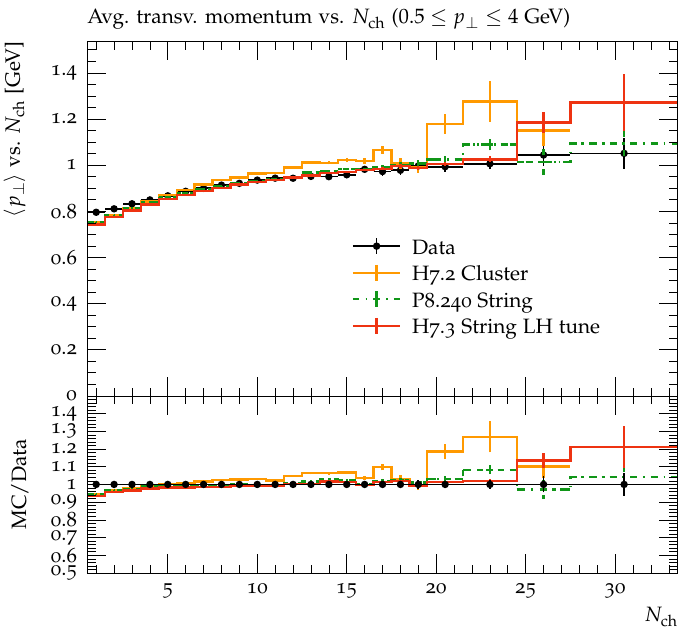}
\end{subfigure}
\begin{subfigure}{0.28\textwidth}
\caption{\bf ATLAS, $\mathbf{\sqrt{s}}$ = 0.9 TeV}
        \includegraphics[width=1\textwidth]{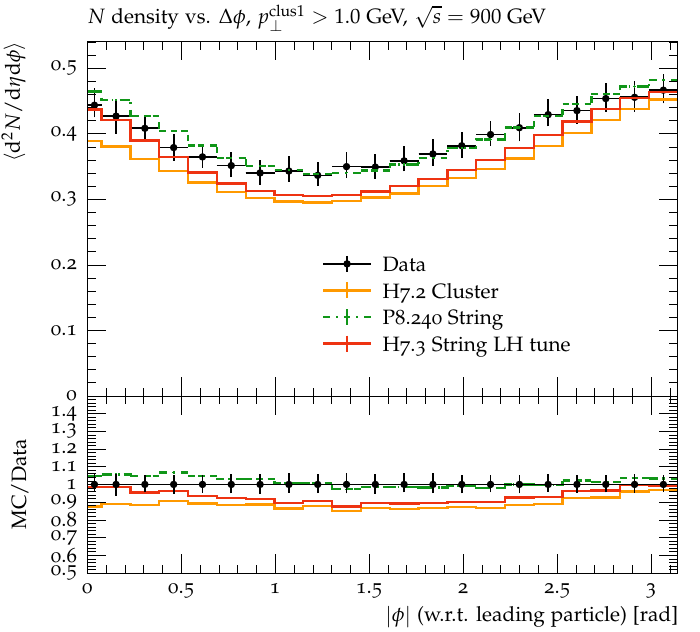}
\end{subfigure}
        \caption{A selection of \UE and \MB observables for proton collisions at $\sqrt{s} =$ 0.9 TeV. These distributions were not included in the tuning interpolation.}
\label{fig:UEMB_aux_09}
\end{figure}

\begin{figure}[h]
\centering
\captionsetup[subfigure]{labelformat=empty}
\begin{subfigure}{0.28\textwidth}
\caption{\bf CMS, $\mathbf{\sqrt{s}}$ = 7 TeV}
        \includegraphics[width=1\textwidth]{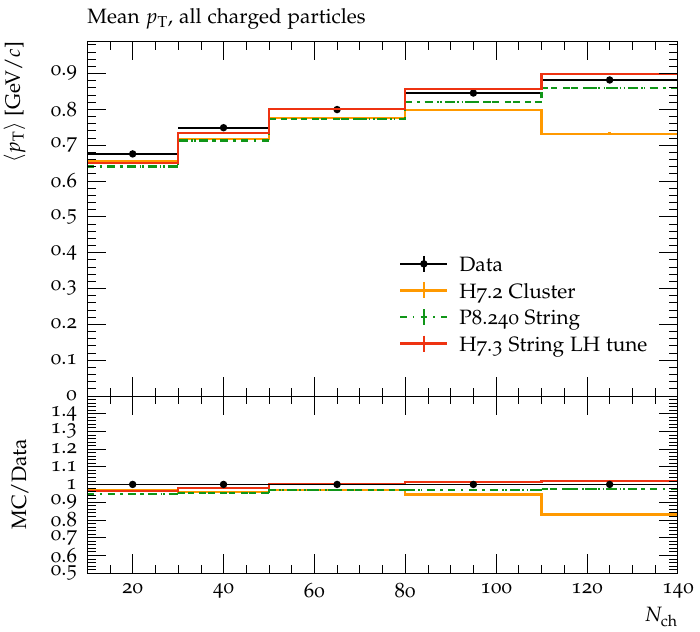}
\end{subfigure}
\begin{subfigure}{0.28\textwidth}
\caption{\bf CMS, $\mathbf{\sqrt{s}}$ = 7 TeV}
        \includegraphics[width=1\textwidth]{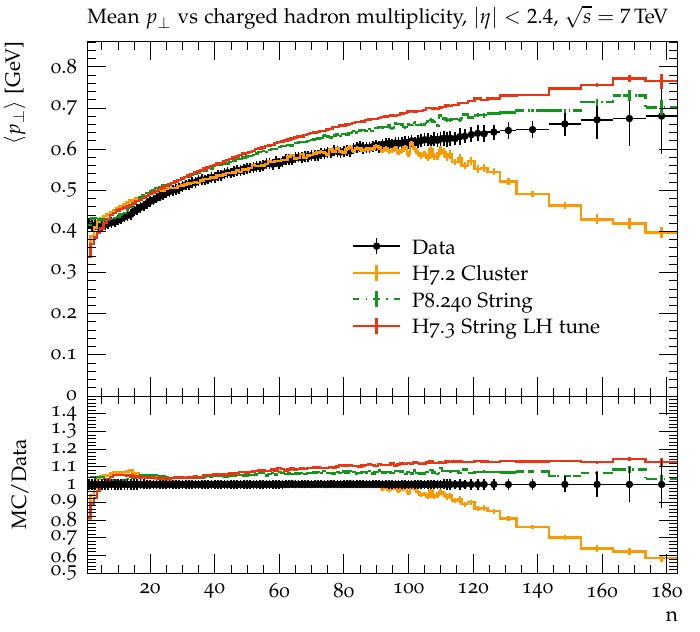}
\end{subfigure}
\begin{subfigure}{0.28\textwidth}
\caption{\bf CMS, $\mathbf{\sqrt{s}}$ = 7 TeV}
        \includegraphics[width=1\textwidth]{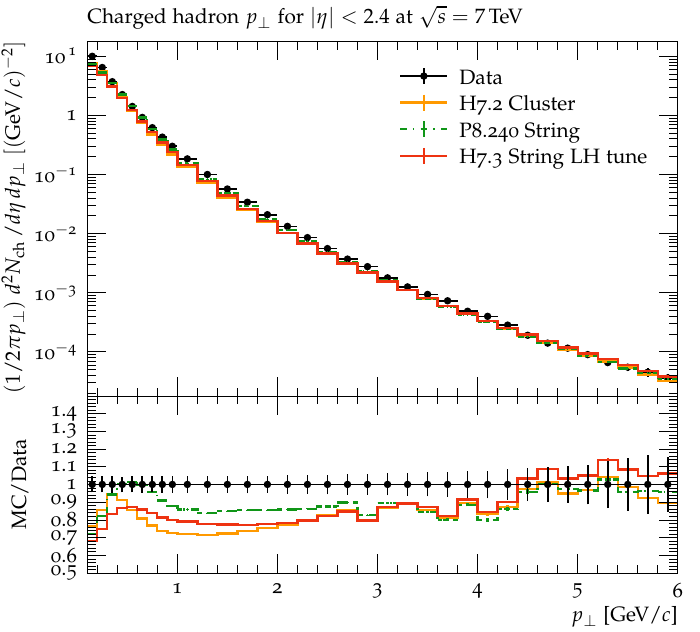}
\end{subfigure}\\
\begin{subfigure}{0.28\textwidth}
\caption{\bf LHCb, $\mathbf{\sqrt{s}}$ = 7 TeV}
        \includegraphics[width=1\textwidth]{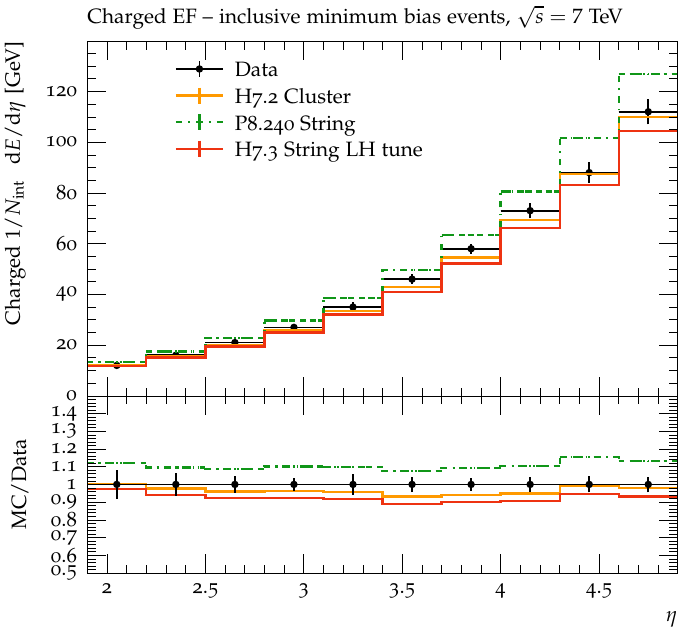}
\end{subfigure}
\begin{subfigure}{0.28\textwidth}
\caption{\bf LHCb, $\mathbf{\sqrt{s}}$ = 7 TeV}
        \includegraphics[width=1\textwidth]{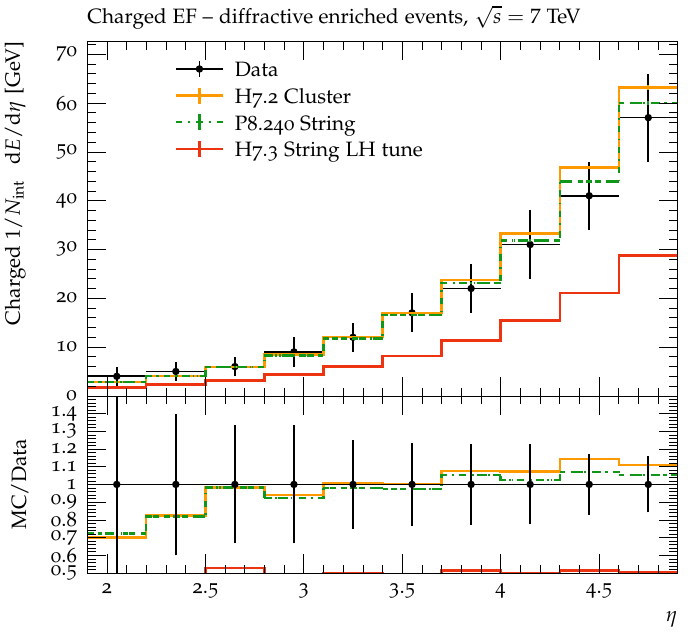}
\end{subfigure}
\begin{subfigure}{0.28\textwidth}
\caption{\bf TOTEM, $\mathbf{\sqrt{s}}$ = 7 TeV}
        \includegraphics[width=1\textwidth]{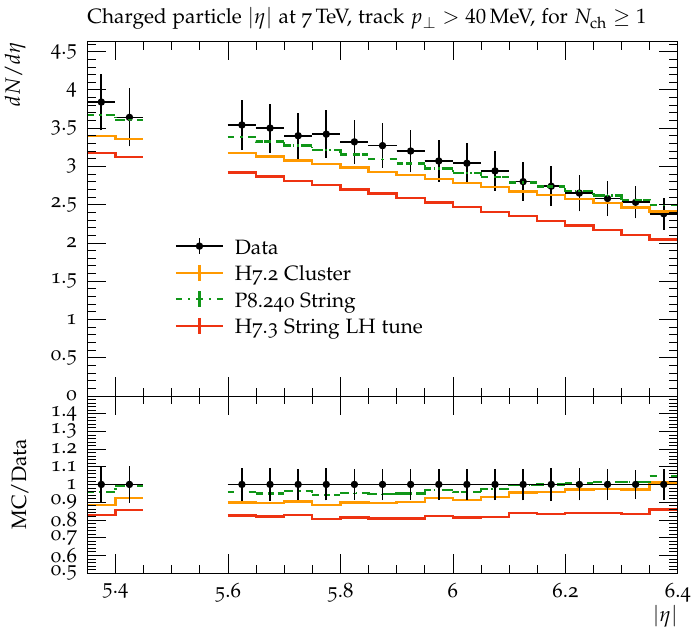}
\end{subfigure}
\caption{A selection of \UE and \MB observables for proton collisions at $\sqrt{s} =$ 7 TeV. These distributions were not included in the tuning interpolation.}
\label{fig:UEMB_aux_7}
\end{figure}

%%%%%%%%%%%%%%%%%%%%%%%%%%%%%%%%%%%%%%%%%%%%%%%%%%%%%%%%%%%%%%%%%%%%%%%%%%%%%%%%%%%%%%%%%%%%%%%%%%%%%%%%%%%

\newpage
\clearpage

\bibliographystyle{JHEP}
\bibliography{references.bib}

\end{sloppypar}
\end{document}